\documentclass[fleqn,usenatbib]{mnras}

\usepackage{newtxtext,newtxmath}

\usepackage[T1]{fontenc}
\usepackage{ae,aecompl}

\usepackage{graphicx}	
\usepackage{amsmath}	
\usepackage{siunitx}
\usepackage{booktabs}
\usepackage{multirow}
\usepackage{color, colortbl}
\usepackage[pdftex,dvipsnames]{xcolor}
\usepackage[ruled,vlined]{algorithm2e}
\usepackage{lineno}

\DeclareMathOperator*{\argmax}{argmax}
\let\vec\mathbf

\usepackage{xcolor}
\usepackage{ulem}
\usepackage{xspace} 
\usepackage{hyperref}

\newcommand{\hyperrank}{\textsc{hyperrank}\xspace}
\newcommand{\redmagic}{RedMaGiC\ }

\newcommand{\chat}{\hat{c}}
\newcommand{\bhat}{\hat{b}}
\newcommand{\W}{\mathcal{W}}
\newcommand{\B}{\mathcal{B}}
\newcommand{\R}{\mathcal{R}}
\newcommand{\D}{\mathcal{D}}
\newcommand{\vf}{\boldsymbol{f}}

\newcommand{\zbar}{\bar{z}}
\newcommand{\zbaravg}{\langle\zbar^{\sdir}\rangle}
\newcommand{\sompz}{\textsc{SOMPZ}\xspace}
\newcommand{\sdir}{\textsc{3sDir}\xspace}
\newcommand{\sdiralt}{\textsc{3sDir-Alt}\xspace}
\newcommand{\sdirmfwz}{\textsc{3sDir-MFWZ}\xspace}
\newcommand{\balrog}{\textsc{Balrog}\xspace}
\newcommand{\buzzard}{\textsc{Buzzard}\xspace}
\newcommand{\snr}{\texttt{SNR}\xspace}
\newcommand{\metacal}{\textsc{metacalibration}\xspace}
\newcommand{\angstrom}{\text{\normalfont\AA}} 
\newcommand{\SPC}{\textsc{SPC}\xspace}
\newcommand{\PC}{\textsc{PC}\xspace}
\newcommand{\SC}{\textsc{SC}\xspace}
\newcommand{\C}{\textsc{C}\xspace}
\newcommand{\SPCMB}{\textsc{SPC-MB}\xspace}

\DeclareMathOperator{\Dir}{Dir}

\usepackage{threeparttable}
\usepackage{eso-pic}

\AddToShipoutPictureBG*{%
  \AtPageUpperLeft{%
    \hspace{0.75\paperwidth}%
    \raisebox{-3.5\baselineskip}{%
      \makebox[0pt][l]{\textnormal{DES-2019-0485}}
}}}%

\AddToShipoutPictureBG*{%
  \AtPageUpperLeft{%
    \hspace{0.75\paperwidth}%
    \raisebox{-4.5\baselineskip}{%
      \makebox[0pt][l]{\textnormal{FERMILAB-PUB-20-659-AE}}
}}}%

\title[DES Y3 Redshift Calibration]{Dark Energy Survey Year 3 Results:\\ Redshift Calibration of the Weak Lensing Source Galaxies}

\author[J. Myles, A. Alarcon et al.]{
\parbox{\textwidth}{
\Large{J.~Myles,$^{1,2,3}$\thanks{E-mail: jmyles@stanford.edu},
A.~Alarcon,$^{4, 65, 66}$\thanks{E-mail: alexalarcongonzalez@gmail.com},
A.~Amon,$^{1,2,3}$
C.~S{\'a}nchez,$^{5}$
S.~Everett,$^{6}$
J.~DeRose,$^{7,6}$
J.~McCullough,$^{2}$
D.~Gruen,$^{1,2,3}$
G.~M.~Bernstein,$^{5}$
M.~A.~Troxel,$^{8}$
% ...
\begin{center} (DES Collaboration) \end{center}
}
\vspace{0.2cm}
\parbox{\textwidth}{ \small
\textit{The authors' affiliations are shown at the end of this paper.}}}}

\date{Accepted 2021 May 18. Received 2021 May 04; in original form 2020 December 17}

\pubyear{2020}
\begin{document}
\label{firstpage}
\pagerange{\pageref{firstpage}--\pageref{lastpage}}
\maketitle

\begin{abstract}

Determining the distribution of redshifts of galaxies observed by
  wide-field photometric experiments like the Dark Energy Survey is an
  essential component to mapping the matter density field with
  gravitational lensing. In this work we describe the methods used to
  assign individual weak lensing source galaxies from the Dark Energy
  Survey Year 3 Weak Lensing Source Catalogue to four tomographic bins
  and to estimate the redshift distributions in these bins. As the
  first application of these methods to data, we validate that the
  assumptions made apply to the DES Y3 weak lensing source galaxies
  and develop a full treatment of systematic uncertainties. Our method
  consists of combining information from three independent likelihood
  functions: Self-Organizing Map $p(z)$ (\sompz), a method for
  constraining redshifts from galaxy photometry; clustering redshifts
  (WZ), constraints on redshifts from cross-correlations of galaxy
  density functions; and shear ratios (SR), which provide constraints
  on redshifts from the ratios of the galaxy-shear correlation
  functions at small scales. Finally, we describe how these
  independent probes are combined to yield an ensemble of redshift
  distributions encapsulating our full uncertainty. We calibrate redshifts with combined effective uncertainties of $\sigma_{\langle z \rangle}\sim 0.01$ on the mean redshift in each tomographic bin.

\end{abstract}




\section{Introduction}
\label{sec:intro}

The matter density fluctuations present in the Universe, and their evolution over time under the impact of gravity and cosmic expansion, are sensitive to cosmological physics, including the nature of dark energy, neutrino masses, and the nature of dark matter. Galaxy surveys like the Dark Energy Survey \citep[DES,][]{Troxel2018,desy1_3x2pt}, 
the Kilo-Degree Survey  \citep[KiDS,][]{kids1000_3x2pt}, Hyper Suprime-Cam survey 
\citep[HSC,][]{HSC_hikage_19}, the Legacy Survey of Space and Time \citep[LSST,][]{lsst_desc}, or the \textit{Euclid} mission \citep{euclid}
use this to achieve competitive constraints on cosmological parameters from observable proxies of the matter density field. In particular, the DES first three years of observation data are used, among other purposes, to measure three two-point ($3\times2$pt.) correlation functions \citep{y3-3x2ptkp}:

\begin{enumerate}
\item Cosmic shear: the correlation function of the shapes of "source" galaxies divided into four tomographic bins \citep*{y3-shapecatalog,y3-cosmicshear1,y3-cosmicshear2}.
\item Galaxy clustering: the auto-correlation function of the positions of luminous red "lens" galaxies selected by the \redmagic algorithm \citep{Rozo2016,y3-galaxyclustering}, or alternatively the positions of an optimized magnitude-limited sample \citep{y3-2x2maglimforecast,y3-2x2ptaltlensresults}.
\item Galaxy-galaxy lensing: the cross-correlation function of source galaxy shapes around lens galaxy positions \citep{y3-gglensing}. 
\end{enumerate}

The use of gravitational lensing signals is indispensable in this approach: In a photometric survey, while the positions of galaxies can be used as tracing matter density, the only direct connection to the underlying density field is through its effect on the images of distant galaxies by means of gravitational lensing. In order to draw conclusions on the physical density fluctuations from observations of gravitational lensing, however, the distances to the lensed background sources must be known.

Any gravitational lensing measurement, including the interpretation of the cosmic shear and galaxy-galaxy lensing correlation functions, therefore relies on a robust characterization of the distribution $n(z)$ of redshifts $z$ of the respective source galaxy samples \citep{Huterer2006,cfht_hilde,Benjamin2013,Huterer2013,Samuroff2017,KidsDEScomb,Tessore2020}. While ideally this could be accomplished by measuring the spectrum of each galaxy in a given catalogue, it is so far only feasible to gather spectra for small, possibly non-representative subsets of galaxies. As a consequence, large optical imaging surveys with measurements of tens or hundreds of millions of galaxies must rely on relatively few, noisy photometric bands to constrain redshifts. The key challenge in doing this is the presence of degeneracies in the statistical colour-redshift relation, making it commonly impossible to uniquely determine the redshift of any given galaxy from wide-band photometry. One can address this challenge by determining a prescription for reweighting the $n(z)$ of a sample with credible, known redshifts according to those galaxies' relative abundance in the overall sample detected and selected by a photometric survey (e.g. \citealt{Lima2008, Cunha2012, Bonnett2016,euclid_photoz_i,euclid_photoz_ii}; \citealt*{Hoyle2018}; \citealt{Tanaka2018,kids1000_pzpaper, KidsSOM20, euclid_photoz_specz_preparation,lsst_photoz}). The problem of degeneracies in the statistical colour-redshift relation in this case manifests as uncertainty on the measured redshift distribution, often quantified in terms of uncertainties on the moments of the measured $n(z)$. Much of the work in estimating redshift distributions is dedicated to understanding how measured $n(z)$ are biased due to sample variance and selection biases in the sample of galaxies with credible redshifts \citep*{GruenBrimioulle2017,specz_incompletenessHartley2020}. In this work, we describe the analysis used to characterize the redshift distributions of the DES Year 3 (the first 3 seasons of observations) source galaxy sample from their photometry, validate this methodology on realistic simulations of the survey data, and present the results of the analysis on the DES data. 

A challenge to determination of $n(z)$ is the combination of incompleteness in the spectroscopic samples and inaccuracies in many-band photometric redshifts used to calibrate the colour-redshift maps.  Our work ameliorates this challenge by weighting the redshift-calibration sample to match the abundance of the target sample in a high-dimensional colour space \citep*{y3-sompzbuzzard}. Differences in reweighting procedures are known to result in scientifically meaningfully different constraints on the matter clustering parameter $\sigma_8$ \citep{Troxel2018, KidsDEScomb}, highlighting the critical importance of properly accounting for the impact of selection biases on redshift distribution measurement.

A robust redshift analysis should be validated on simulations, rely on multiple independent data sets and methodologies, and have well-characterized uncertainties. Besides the work presented in this paper on photometric redshifts, we accomplish this by combining photometric information with galaxy clustering and shear ratios to constrain redshift distributions. Clustering redshifts (WZ) and shear ratios (SR) play the essential role of providing additional, independent constraining power to validate and further constrain photometric redshift distributions \citep*{y3-sourcewz, y3-shearratio}.

We describe this overall DES Year 3 redshift inference scheme in \S\ref{sec:scheme}. In \S\ref{sec:data} we describe the data used in this analysis. We develop the methodology for determining $n(z)$ from galaxy magnitude and colours and the uncertainty on those $n(z)$ in \S\ref{sec:methodology} and \S\ref{sec:uncertainty}, respectively. We present our results in \S\ref{sec:results} and discuss their implications in \S\ref{sec:discussion}.
\section{DES Y3 Redshift Scheme}
\label{sec:scheme}

\begin{figure*}
\centering
\includegraphics[width=\textwidth]{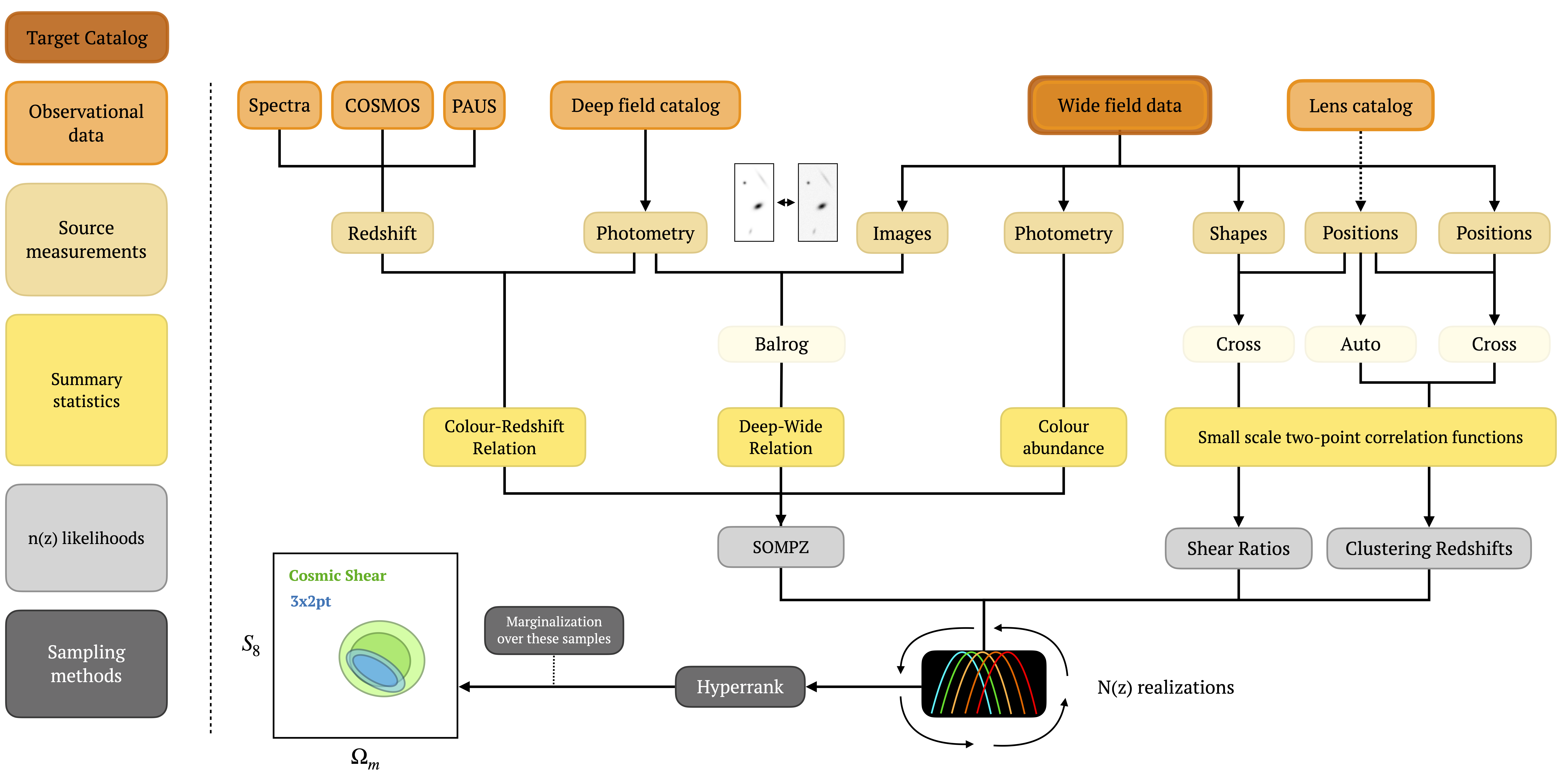}
\caption{Flowchart illustrating the weak lensing redshift distributions calibration scheme. The three main $n(z|\mathrm{model})$ likelihood functions of the analysis, shown in gray, are \sompz, clustering redshifts, and shear ratios. Note that the parameter constraint plot is only an illustration and is not a result from real measurements.}
\label{fig:flowchart}
\end{figure*}

The overarching DES Year 3 redshift inference scheme uses multiple, independent analyses to robustly characterize the weak lensing source galaxy redshift distributions. As illustrated in Fig. \ref{fig:flowchart}, the three likelihood functions computed rely on three independent methods and data: \sompz, clustering redshifts, and shear ratios. 

\begin{enumerate}
  \item Self-Organizing Map $p(z)$ (\sompz) leverages the Y3 DES Deep Fields \citep*{y3-deepfields} to accurately determine the number density of galaxies in deep $ugrizJHK_s$ colour space. Since redshifts are well-constrained at a given $ugrizJHK_s$ colour, this number density can be used to properly weigh galaxies within a sample of credible redshifts in a way that is not subject to selection biases. In brief, this method relies on determining the $p(z)$ at a given cell in 8-band colour space from galaxies with deep 8-band coverage, the probability of each cell in 8-band colour space contributing to the galaxies in a given cell in noisy 3-band colour-magnitude space, and the abundance of galaxies in 3-band colour-magnitude space, to compute the overall redshift distribution of the Year 3 lensing source galaxy sample. The validation of this method and the characterization of its sources of uncertainty are outlined in detail in this work. 
  \item Clustering redshifts constrain the distances to source galaxies from their angular galaxy clustering with samples of reference galaxies within narrow redshift ranges (\citealt{Newman2008,Menard2013,mcquinn2013,Johnson2017,Morrison2017}; \citealt*{Davis2017,Gatti2018}; \citealt{Cawthon2017,kids1000_wzsims,kids1000_pzpaper}). This method is based on the fact that the amplitude of this correlation function is proportional to the fraction of source galaxies in physical proximity to those reference galaxies. Clustering redshifts validate and refine photometric $n(z)$ with the key benefit of avoiding any reliance on the statistical colour-redshift relation and bypassing the completeness issues associated with spectroscopic survey coverage. The details of this analysis are described fully in \cite*{y3-sourcewz}.
  \item Shear ratios (\citealt{Jain2003,Mandelbaum_2005,Heymans_2012}; \citealt*{Y1GGL,DES_spt_lensingratio}; \citealt{Hildebrandt2018}) provide additional constraining power and validation by measuring the galaxy-galaxy lensing signal of a lens galaxy redshift bin at small scales. The ratio of this signal from two source bins reflects the ratio of mean lensing efficiencies of objects in those source bins with respect to the lens bin redshift. This, in turn, depends on the redshift distribution of the sources. Because this methodology utilizes lensing signals, it is virtually independent from \sompz and clustering redshifts. The methodology of this analysis is described fully in \cite*{y3-shearratio}. Both the clustering and shear ratio redshift constraints are derived from data on small angular scales, which allows the redshift constraints to remain largely statistically independent of cosmological constraints based on larger-scale signals.
  \end{enumerate}
  
  In summary, we use galaxy photometry to constrain $n(z)$ with \sompz, galaxy positions to constrain $n(z)$ with clustering redshifts, and galaxy shapes to constrain $n(z)$ with shear ratios. As in past work, we assess consistency of these measurements. We further subsequently \textit{combine} these measurements. The final result of this analysis is an ensemble of redshift distributions whose variation encodes the combined uncertainties on the $n(z)$ due to all sources of information. Any DES Y3 lensing likelihood that uses the same redshift bins can be estimated by sampling from this ensemble. Specifically, the $n(z)s$ in this ensemble are ordered with an algorithm called \hyperrank, which facilitates sampling and marginalization over the $n(z)$ ensemble within the cosmological likelihood Markov chains \citep{y3-hyperrank}. 
\section{Data}
\label{sec:data}
\subsection{DES Wide Field Survey}  
This work presents tomographic redshift distributions for the DES Year 3 weak lensing source catalogue, described in \cite*{y3-shapecatalog}. The source catalogue is a subset of the DES Year 3 Gold catalogue of photometric objects \citep{y3-gold}. After the applied selections, it consists of 100,208,944 galaxies with measured $r$, $i$, and $z$ \metacal photometry and shapes \citep{Sheldon2017}. We note that a subset of the selections defined in \cite*{y3-shapecatalog} were motivated by achieving a more homogeneous photometric catalogue, and therefore more accurate redshift calibration. These cuts on \metacal photometry are as follows:

\begin{enumerate}
    \item $18 < m_i < 23.5$
    \item $15 < m_r < 26$
    \item $15 < m_z < 26$
    \item $-1.5 < m_r - m_i < 4$
    \item $-4 < m_z - m_i < 1.5$
\end{enumerate}

The bright limits of selections (i), (ii), and (iii) remove nearby galaxies for which no lensing signal is expected. They also remove some remaining stars that were incorrectly included in the source galaxy sample. The faint limit of these selections excludes the region of magnitude space where \texttt{COSMOS-30} thirty-band photometric redshifts are found to be more biased \citep{Laigle2016,KidsDEScomb}. Selections (iv) and (v) remove unphysical colors which are assumed to be caused by catastrophic flux measurement failures. 

In this work, we frequently refer to this sample and its photometry as \textit{wide (field)} data. For further details on this catalogue, we refer the reader to \cite*{y3-shapecatalog}.

For the DES Y3 weak lensing analysis, we exclude DES wide-field $g$-band data due to biases caused by difficulties in modeling the $g$-band point-spread function (PSF). In particular, \metacal requires an accurate PSF model to deconvolve (and subsequently reconvolve) a galaxy image from the PSF in order to determine how a galaxy image responds to artificial shear. Inadequate modeling of the PSF would lead to an imprecise constraint on the shear response $R_s$ of each galaxy. In the $g$-band, such model inaccuracies are expected to result e.g.~from chromatic effects on the PSF \citep{Plazas2012}. Our diagnostics indeed show that PSF models are significantly less accurate in the $g$-band than in the redder DES filters. As a result, we do not use $g$-band data for any purpose that requires accurate PSF deconvolution, including the \metacal correction for selection biases. This problem precludes the use of $g$-band for defining redshift bins, since selection biases can only be corrected within \metacal when all selections (including the selection into a redshift bin) are made based on properties also measured on artificially sheared images, which are not available in the $g$-band. For further details on this challenge, see \citet*{y3-shapecatalog}.

\subsection{DES Deep Field Survey and Artificial Wide Field Photometry} \label{sec:data_DF_and_balrog}

The DES Y3 Deep Fields and mock wide-field photometry for the deep-field detections are the cornerstone of \sompz. Full characterization of these data products are provided in \citet*{y3-deepfields} and \citet{y3-balrog}, respectively, and we summarize requisite details here. Our inference method relies on extracting source density information from four \textit{deep} fields named E2, X3, C3, and COSMOS (COS) covering areas of 3.32, 3.29, 1.94, and 1.38 square degrees, respectively, as shown in Fig. \ref{fig:RS_deep_footprint}. After masking regions with artefacts such as cosmic rays, artificial satellites, meteors, asteroids, and regions of saturated pixels, 5.2 square degrees of overlap with the UltraVISTA and VIDEO near-infrared (NIR) surveys \citep{McCracken2012,Jarvis2013} remain. This yields 2.8M detections with measured $ugrizJHK_s$ photometry with limiting magnitudes 24.64, 25.57, 25.28, 24.66, 24.06, 24.02, 23.69, and 23.58, substantially fainter than the faintest galaxies in the sample of source galaxies. In this work we frequently refer to this sample and its photometry as \textit{deep (field)} data.

In order to relate galaxies with given deep photometry to observed lensing sources with wide photometry, we rely on the \balrog \citep{Suchyta2016} software which injects simulated galaxies, based on the deeper photometry from the DES deep fields, into real images. For this analysis, \balrog was used to inject model profiles fit to deep-field galaxies into the broader wide-field footprint \citep{y3-balrog}. After injecting galaxies into images, the output is passed into the DES Y3 photometric pipeline. Each deep-field galaxy is injected multiple times at different positions, and injected galaxies are detected equivalently to real galaxies, yielding multiple realisations of each deep-field galaxy. The output matched catalogue of 2,417,437 injection-realisation pairs containing both deep and wide photometric information is a key part of our redshift calibration inference method. This catalogue is called the \textit{Deep/\balrog} Sample. Note that this sample contains a total
of 267,229 unique deep-field galaxies having $\ge1$ \balrog realisation that passes the wide-field selection criteria.
    
With respect to the consistency of \balrog and Y3 GOLD fluxes, we highlight that the Y3 GOLD catalogue \citep{y3-gold} accounts for photometric effects including reddening due to the interstellar medium, achromatic (i.e. `gray') zero-point recalibrations relative to an original DES Y3 calibration, and chromatic corrections for the SED-dependent effects of differential optical throughput as a function of focal plane location and variable environmental conditions at the telescope site. The work of \citet{y3-balrog} captures corrections for reddening as described above in the injections, but does not model the other two effects at injection time (thus eliminating any need to apply the corrections to detections). We emphasize that \citet{y3-balrog} verify that the mock wide-field fluxes generated by \balrog are more than sufficiently robust for all Y3 calibration purposes. Our findings discussed in \S \ref{sec:balrogunc} reinforce this conclusion in the context of redshift calibration. For details on the origins of these photometric calibration procedures see \citep{y3-gold, Burke2018}.

\subsection{Redshift samples} \label{sec:redshift_samples}

Our analysis relies on the use of galaxy samples with known redshift and deep-field photometry. To this end, we use catalogues of both high-resolution spectroscopic and multi-band photometric redshifts and develop an experimental design that allows us to test uncertainty in our redshift calibration due to biases in these samples. The spectroscopic catalogue we use contains both public and private spectra from the following surveys: zCOSMOS \citep{Lilly09zcosmos}, C3R2 \citep{Masters2017,C3R2_DR2}, VVDS \citep{vvds}, and VIPERS \citep{vimos}. We use two multi-band photo-$z$ catalogues from the COSMOS field \citep{Scoville2007_COSMOS}:  the \texttt{COSMOS2015} 30-band photometric redshift catalogue \citep{Laigle2016}, which includes 30 broad, intermediate, and narrow bands covering the UV, optical, and IR regions of the electromagnetic spectrum, and the \texttt{PAUS+COSMOS} 66-band photometric redshift catalogue \citep{Alarcon2020} from the combination of PAU Survey data \citep{paucam,eriksen2019} in 40 narrow-band filters and 26 COSMOS2015 bands excluding the mid-infrared.

\begin{figure}
\centering
\includegraphics[width=\linewidth]{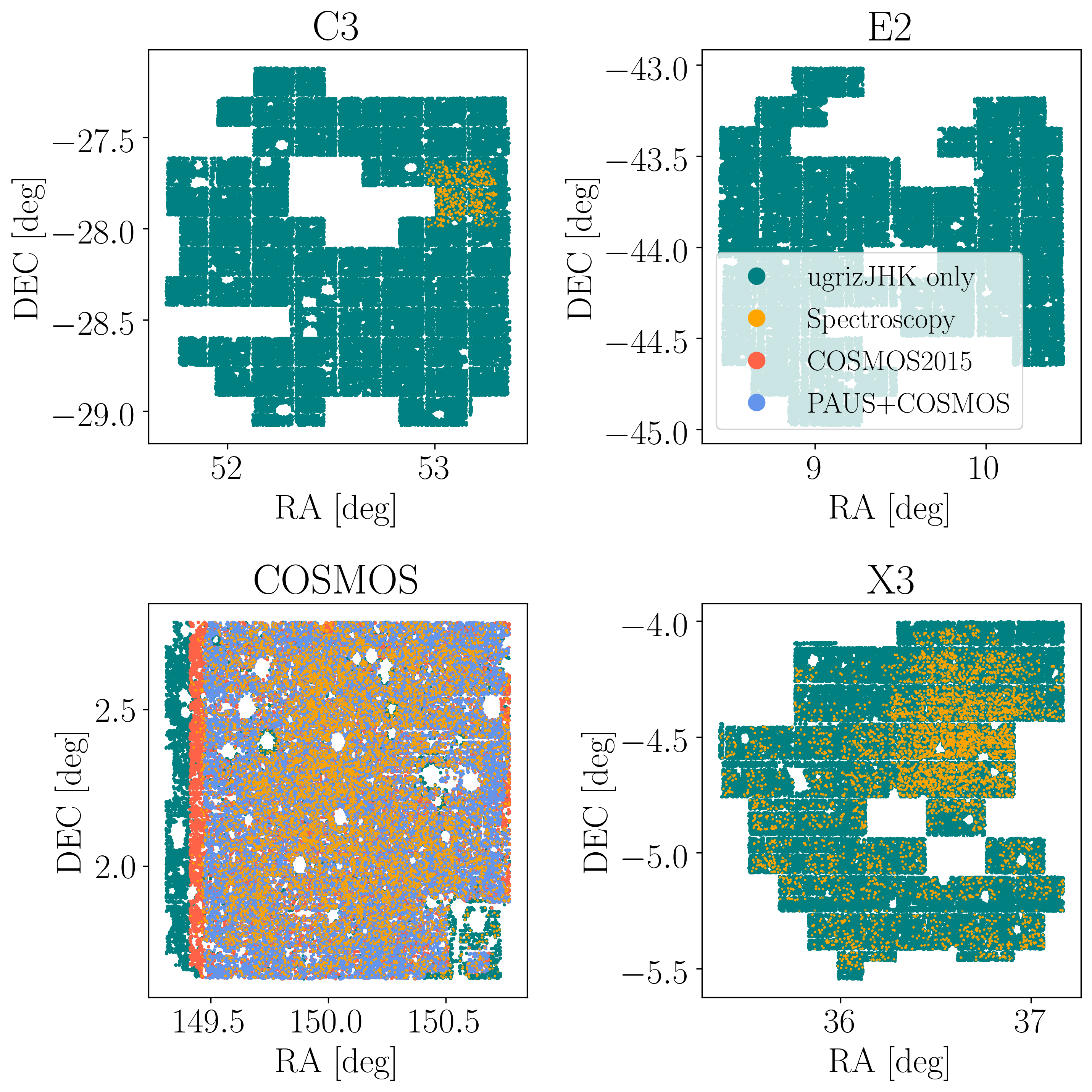}
\caption{The four DES deep fields used for our redshift analysis. Each field has overlapping deep DES \textit{ugriz} bands and archival \textit{JHK} bands from the VIDEO or UltraVISTA surveys. Green points indicate DES deep-field galaxies with no spectroscopic or many-band photometric redshifts. Yellow (S), blue (C), and red (P) indicate deep-field galaxies with redshifts from spectroscopy, \texttt{COSMOS2015}, or \texttt{PAUS+COSMOS}, respectively. Missing rectangular regions are DECam CCDs on which scattered light hampered precision deep photometry.}
\label{fig:RS_deep_footprint}
\end{figure}

\begin{figure}
\centering
\includegraphics[width=\linewidth]{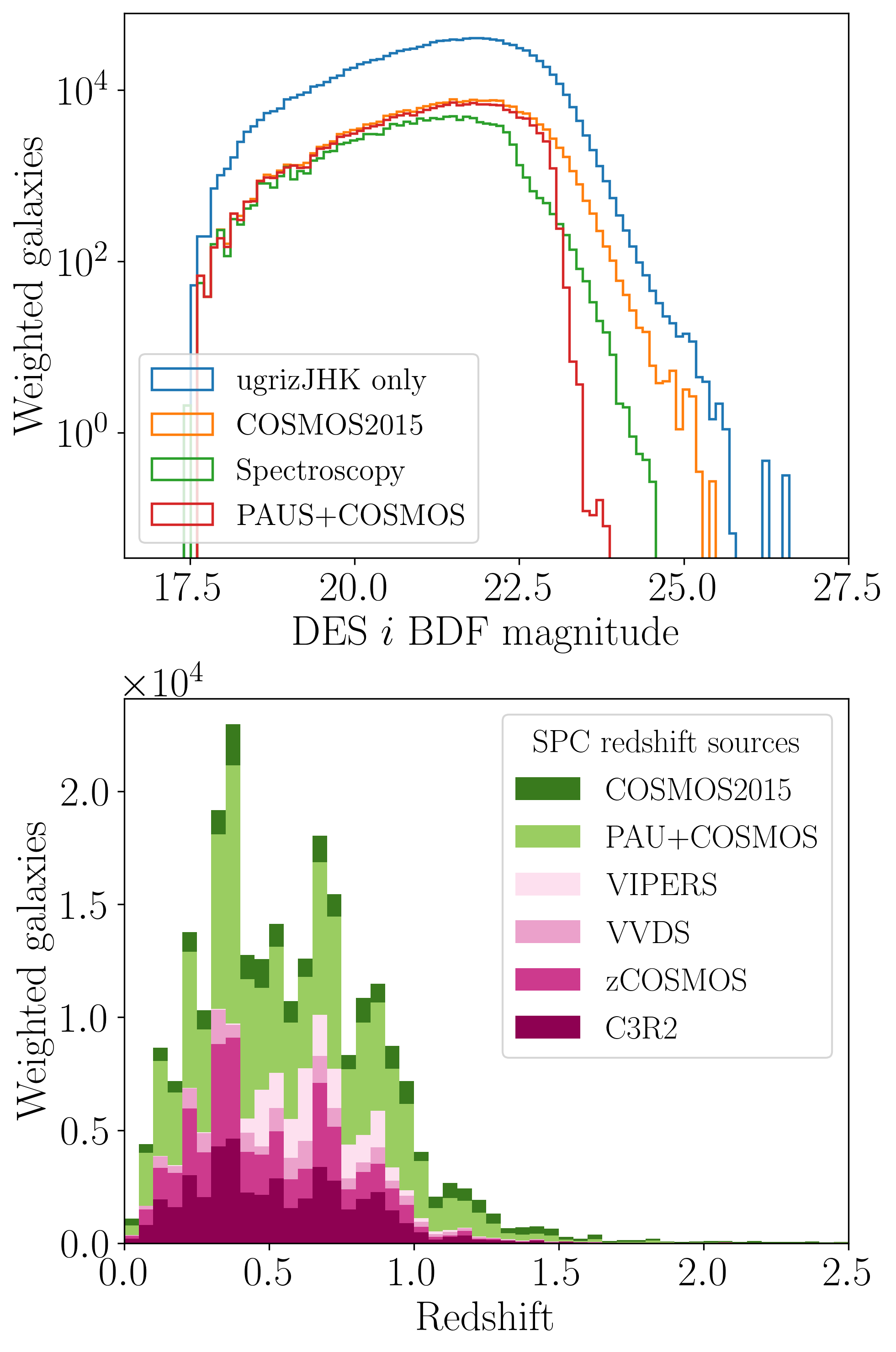}
\caption{\textit{Top}: Distribution of redshift samples as a function of DES $i$-band magnitude. Each galaxy in this histogram is weighted by all weights used in the cosmological analyses: probability of detection from \balrog, \metacal response, and lensing weight (see \S~\ref{sec:sompz_formalism}). For details on the definition of the `Bulge Plus Disk, Fixed Ratio' (BDF) galaxy profile see \citet*{y3-deepfields}. \textit{Bottom}:  Distribution of redshifts used in our analysis, for one of our redshift samples \SPC. This sample is defined to preferentially use redshift from spectroscopy, then \texttt{PAUS+COSMOS}, then \texttt{COSMOS2015}. Each galaxy in this stacked histogram is weighted by all weights used in the analysis: probability of detection from \balrog, \metacal response, and lensing weight (see \S~\ref{sec:sompz_formalism}).}
\label{fig:RS_deep_imaghist}
\end{figure}

Fig.~\ref{fig:RS_deep_footprint} shows the DES deep-field footprints \citep*{y3-deepfields} and highlights the footprint of each of the different redshift catalogues. While the two photo-z catalogues are limited to the COSMOS field, our spectroscopic compilation partially covers the COSMOS, X3 and C3 fields. Fig.~\ref{fig:RS_deep_imaghist} shows the DES $i$-band magnitude distribution for all galaxies with $ugrizJHK_s$ photometry and for each of the redshift samples \citep*[for a definition of BDF magnitude see][]{y3-deepfields}. Each galaxy has been weighted by the same weight used in the cosmological analysis, which includes the galaxy detection probability from \balrog, the \metacal response and a lensing weight (see section~\ref{sec:sompz_formalism} for more details on these weights). While the spectroscopic compilation spans the largest area among the redshift catalogues, it is also the shallowest. The \texttt{COSMOS2015} catalogue is the deepest, but also has the lowest redshift precision. Finally, the \texttt{PAUS+COSMOS} catalogue is more precise than \texttt{COSMOS2015} and, unlike spectroscopic samples, is nearly complete in the highly relevant magnitude range of up to $i\approx23$ but has the lowest areal coverage at faint magnitudes.

To estimate the redshift distribution of each tomographic bin, we compose three main redshift samples for which we rank the redshift information differently, meaning that for an object with redshift information from multiple origins, we choose the estimation from the highest ranked one. These redshift samples are:
\begin{itemize}
    \item \textit{SPC}: This sample ranks first the spectroscopic catalogue  (S), then \texttt{PAUS+COSMOS} (P), and finally \texttt{COSMOS2015} (C). This sample is designed to inform an understanding of cosmological results that is minimally reliant on the \texttt{COSMOS2015} data without introducing potential selection biases such as those discussed by \citet{GruenBrimioulle2017}.
    \item \textit{PC}: This sample ranks first the \texttt{PAUS+COSMOS} catalogue  before \texttt{COSMOS2015}, and does not include spectroscopic redshifts. This sample is designed to inform an understanding of cosmological results that are maximally reliant on many-band photometric redshifts, and thus not affected by selection effects resulting from spectroscopic survey selection functions.
    \item \textit{SC}: This sample ranks first the spectroscopic catalogue  before \texttt{COSMOS2015}, and does not include the \texttt{PAUS+COSMOS} catalogue. This sample is designed to inform an understanding of cosmological results that are not  reliant on PAU multi-band photometric redshifts.
\end{itemize}

The fiducial ensemble of redshift distributions is generated by marginalizing over all three of these redshift samples (SPC, PC, SC) with equal prior, which in practice is achieved by simply concatenating the $n(z)$ samples produced from these three redshift samples. In addition to the three samples used for our fiducial analysis, we define the following alternative redshift samples that we deem less reliable. These samples are used to test the robustness of our redshift information:
\begin{itemize}
\item \C:  This sample includes only information from the \texttt{COSMOS2015} catalogue and would therefore suffer most strongly from systematic biases in these photometric redshifts.
\item \SPCMB: This sample (\SPC, Magnitude-Biased) is artifically constructed to enable an additional robustness test of our dependence on the \texttt{COSMOS2015} catalogue. The motivation for constructing this sample is that the redshift information used in \SPC still preserves 10 per cent of the effective information from \texttt{COSMOS2015}, primarily at the faintest magnitudes, due to the paucity of spectroscopic redshifts for galaxies at these fainter magnitudes. We thus construct \SPCMB to test the impact on our $n(z)$ of including these redshifts from primarily fainter galaxies in \texttt{COSMOS2015}. In order to assess the potential impact of biases in these faint \texttt{COSMOS2015} galaxies without removing them, which would introduce selection effects such as those discussed by \cite{GruenBrimioulle2017}, we must define some prescription for producing realistic de-biased redshifts for these galaxies. We achieve this with the following prescription: we bin galaxies for which we have a spectroscopic/PAU redshift and a \texttt{COSMOS2015} photometric redshift into magnitude-redshift bins with lower magnitude bin limits $[18, 21, 22.4]$ and redshift bin widths of $0.01$. For each of these galaxies, we compute the redshift bias $\Delta = z_{\SPC} - z_{\texttt{COSMOS2015}}$. We remove all outlying galaxies with $|\Delta| > 0.15$. For each magnitude-redshift bin, we compute the mean bias $\langle \Delta \rangle$. We then add this mean bias in each bin to the \texttt{COSMOS2015} galaxies in that bin for which we \textit{do not} have a spectroscopic/PAU redshift, thus yielding a realistic mock spectroscopic/PAU redshift for them. In this way, we generate a sample of realistically corrected \texttt{COSMOS2015} redshifts without being subject to selection effects that would be introduced by removing these galaxies entirely.
\end{itemize}

These variant samples are detailed in Table \ref{tab:specz}. The impact of using these respective samples to produce redshift distributions is discussed in \autoref{sec:redshiftsampleunc}. Note that we do not attempt a calibration of the DES Y3 lensing source redshift distribution that is solely informed by spectroscopic redshifts. The sample of available spectroscopic redshifts in the deep fields does not span the full $ugrizJHK_s$ color-space of the DES data. If any cell in deep color space were to only include the subset of galaxies with successful spectroscopic redshift, we expect the resulting estimates of the redshift distributions would suffer from unquantified selection biases. However, comparisons of redshift calibration between the samples used, some of which are almost a 1:1 mixture of spectroscopic and high-quality photometric redshifts, should provide robust indications of any relevant biases in the PAU or \texttt{COSMOS2015} photometric redshift samples.

\begin{table}
    \centering
    \begin{tabular}{|l|c|c|c|}
        \hline
         Name &  \% Spectra & \% \texttt{PAUS+COSMOS} & \% \texttt{COSMOS2015} \\
         \hline
         \SC & 47 & 0 & 53\\
        \PC & 0 & 87 & 13\\
         \SPC & 47 & 43 & 10 \\
         \hline
         \C & 0 &  0 & 100 \\
         \SPCMB & 47 & 43 & $10^{*}$\\
         \hline
    \end{tabular}
    \caption{Redshift samples used in our analysis, both in the fiducial case (\SC, \PC, and \SPC) and in the less reliable cases (\C and \SPCMB) and their relative contribution from spectroscopic data, \texttt{PAUS+COSMOS} and \texttt{COSMOS2015}. The relative contribution includes all galaxy weights used in the analysis: probability of detection from \balrog, \metacal response, and lensing weight (see \S~\ref{sec:sompz_formalism}). $^{*}$Note, as described in \S~\ref{sec:redshift_samples}, we artificially bias the \texttt{COSMOS2015} redshifts when constructing the \SPCMB sample to enable the robustness test for which this sample is designed.}
    \label{tab:specz}
\end{table}

\subsection{Simulated Galaxy Catalogues} \label{sec:buzzard}
We use the \buzzard cosmological simulations to validate aspects of our analysis. These simulations are briefly described here, and discussed comprehensively in \citet{y3-simvalidation}, as well as additional validation tests of the photometry in these simulations in \citet{DeRose2019}. 

The \buzzard simulations are galaxy catalogues that have been populated in $N$-body lightcones by applying the \textsc{Addgals} algorithm.  They make use of a set of 3 independent $N$-body lightcones with box sizes of $[1.05,\, 2.6,\, 4.0]\, (h^{-3}\, \rm Gpc^3)$, with mass resolutions of $[0.33,\, 1.6, \, 5.9] \, \times10^{11}\, h^{-1}M_{\odot}$, and spanning redshift ranges in the intervals $[0.0,\, 0.32,\, 0.84, \,  2.35]$ respectively. This produces a simulation that spans $10,313$ square degrees. We use the \textsc{L-Gadget2} $N$-body code, a memory-optimized version of \textsc{Gadget2} \citep{Springel2005}, with initial conditions generated using \textsc{2LPTIC} at $z=50$. 

\textsc{Addgals} provides simulated galaxy positions, velocities, absolute magnitudes, spectral energy distributions (SEDs), ellipticities and half-light radii for each galaxy. Positions and absolute magnitudes are assigned such that the simulated galaxies reproduce projected clustering measurements in the Sloan Digital Sky Survey Main Galaxy Sample (SDSS MGS). Likewise, SEDs are assigned from SDSS MGS using a conditional abundance-matching model \cite{DeRose2020b}, that reproduces the color-and-luminosity-dependent clustering in SDSS MGS. Broad band photometry is produced from these SEDs by k-correcting them to each galaxy's rest frame, and integrating over the DES and VISTA bandpasses to produce $ugrizJHK_s$ photometry. While we find reasonably good agreement between the \buzzard photometry and that observed in our deep and wide fields, the match is by no means perfect, particularly in bluer bands and for redshifts $z>1.2$, as illustrated in fig. 1 of \citet{y3-simvalidation}.

The simulations are ray-traced using \textsc{Calclens} using an $N_{\rm side}=8192$ \textsc{HealPix} grid \citep{Becker2013}, and angular deflections, shear, and magnification quantities are computed for each galaxy. The DES Y3 footprint mask is applied to the ray-traced simulations, resulting in a footprint with an area of 4305 square degrees. We apply a photometric error model to the mock wide-field photometry in our simulations based on a relation measured from \balrog. A weak lensing source selection is applied to the simulations using the PSF-convolved sizes and $i$-band \snr in order to match the non-tomographic source number density, 5.84 arcmin$^{-2}$, in the \metacal source catalogue. In order to simulate a lens galaxy catalogue, we also apply the \textsc{redMaGiC} selection algorithm on the simulations using the same configuration as used in the Y3 data.

\section{\sompz Methodology}
\label{sec:methodology}
We aim to determine 
the redshift distribution $n(z)$ of the weak lensing galaxy sample, proportional to the probability $p(z)$ of a 
galaxy in that sample to be at a given redshift $z$, by reweighting the distribution of redshifts of a sample with reliable redshift information in a suitable way that prevents selection bias and reduces sample variance.
A sample of galaxies with both well-constrained redshift and deep photometry in several bands, 
and an additional, larger sample of galaxies with deep photometry in the same set of bands
provide crucial information on how to accurately perform that weighting. In this section, we provide details of the methodology and, in addition, brief descriptions of the additional steps of DES Y3 redshift distribution calibration related to clustering redshifts \citep*{y3-sourcewz}, image simulations \citep{y3-imagesims}, and shear ratios \citep*{y3-shearratio}.

\subsection{Redshift Distribution Inference Formalism}
\label{sec:sompz_formalism}

Extracting the redshift information from deep, several-band photometry to estimate the redshift of an observed wide-field galaxy amounts to marginalizing over deep photometric information \citep*{y3-sompzbuzzard}. The probability distribution function for the redshift of a galaxy, conditioned on observed wide-field colour-magnitude $\mathbf{\hat{x}}$ and covariance matrix $\hat{\Sigma}$, and on passing a selection function $\hat{s}$, can be written by marginalizing over deep photometric colour $\mathbf{x}$ as follows:
\begin{equation}
\label{eqn:marginal_pz}
    p(z|\mathbf{\hat{x}} , \mathbf{\hat{\Sigma}}, \hat{s}) = \int d \mathbf{x} \ p(z| \mathbf{x}, \mathbf{\hat{x}}, \mathbf{\hat{\Sigma}}, \hat{s}) p(\mathbf{x}|\mathbf{\hat{x}}, \mathbf{\hat{\Sigma}} ,\hat{s}).
\end{equation}

The large number of dimensions of the variables on the right-hand side of Equation \ref{eqn:marginal_pz} make these probabilities unfeasible to evaluate directly. We instead must discretize the smooth colour and colour-magnitude spaces spanned by $\mathbf{x}$ and ($\mathbf{\hat{x}}$, $\mathbf{\hat{\Sigma}}$) into categories $c$ and $\hat{c}$. These $c$ and $\hat{c}$, which we call cells, define a set of galaxy photometric \textit{phenotypes} \citep*{Sanchez2018, y3-sompzbuzzard}. While any of the many existing unsupervised classification or clustering algorithms can be used to categorize galaxies in this way, we use the \textit{Self-Organizing Map} because it allows for a two-dimensional representation of the data set whose continuity facilitates interpolation and easily interpretable visualizations \citep{Kohonen1982, Kohonen2001,CarrascoKind2014SOMz, Greisel2015, Masters2015}. With this compressed information, we can marginalize over deep-field information $c$ to write the $p(z)$ for the ensemble of galaxies associated with a particular cell $\hat{c}$ as:

\begin{equation}
p(z|\hat{c},\hat{s}) = \sum_{c} p(z|c, \hat{c}, \hat{s}) p(c|\hat{c},\hat{s}).
\end{equation}

After associating $\hat{c}$ with tomographic bins according to a given binning algorithm (discussed in detail in \S\ref{sec:binning}), the $n(z)$ in each tomographic bin $\hat{b}$ can be constructed by marginalizing over  (i.e. summing) the constituent cells $\hat{c} \in \hat{b}$ of the tomographic bin:
\begin{align} 
    p(z|\hat{b}, \hat{s}) &= \sum_{\hat{c} \in \hat{b}} p(z|\hat{c}, \hat{s}) p(\hat{c}|\hat{s}, \hat{b}) \\
    &= \sum_{\hat{c} \in \hat{b}} \sum_{c} p(z|c, \hat{c}, \hat{s}) p(c|\hat{c},\hat{s})p(\hat{c}|\hat{s}, \hat{b}).
\end{align}

Each galaxy is assigned to exactly one wide SOM cell and each wide SOM cell $\hat{c}$ is assigned to exactly one tomographic bin. 

\begin{figure*}
\centering
\includegraphics[width=0.5\textwidth]{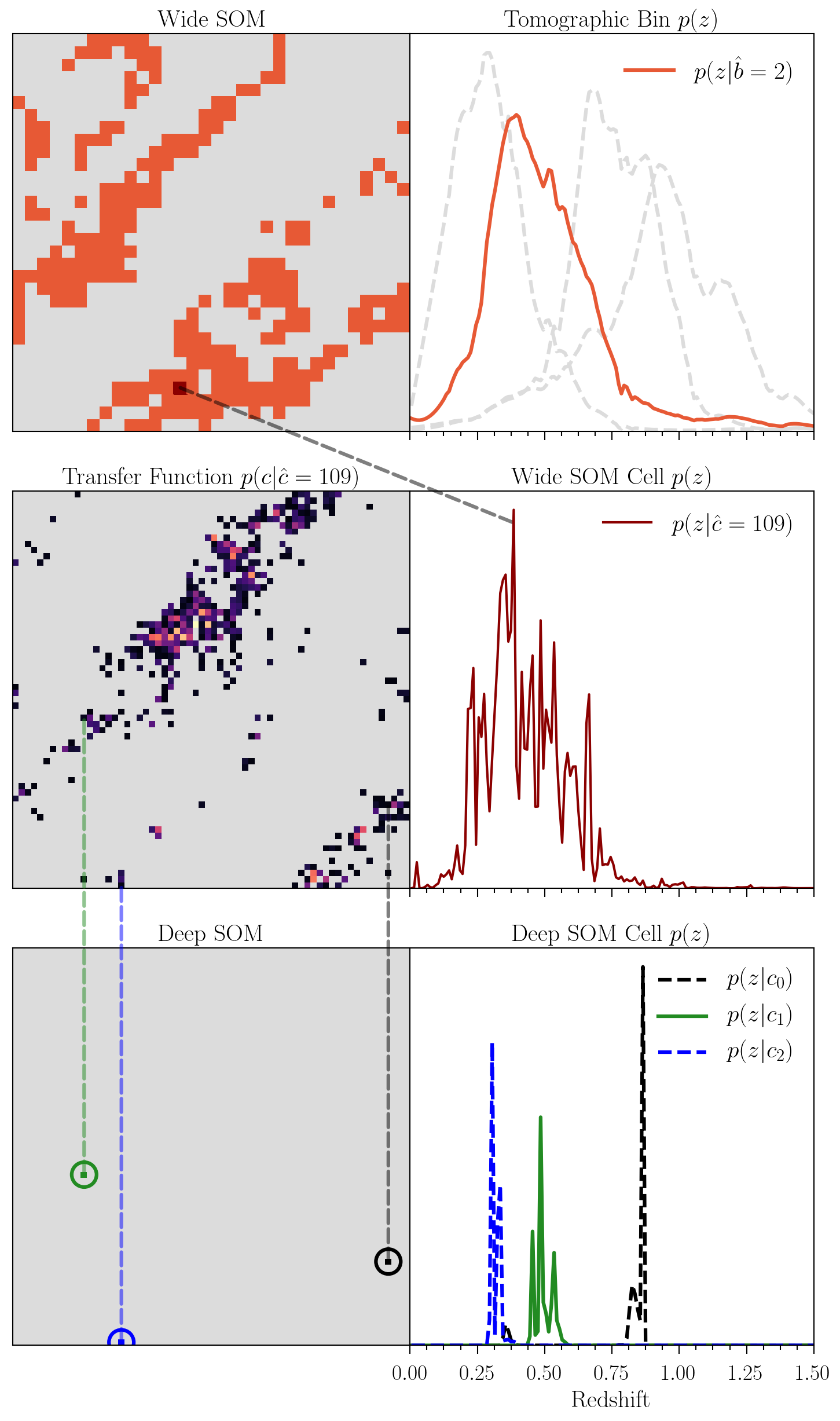}
\caption{Visual representation of each term in the \sompz inference methodology. \textit{Top left}: Wide SOM cells assigned to the second tomographic bin. \textit{Middle left}: Transfer Function $p(c|\hat{c})$ for the selected wide SOM cell $\hat{c}$. Lighter color indicates higher values of $p(c|\hat{c})$, which corresponds to deep SOM cells with a larger number of \balrog draws in the selected $\hat{c}$. \textit{Bottom left}: Three selected deep SOM cells $c$ with non-zero $p(c|\hat{c})$. Different colors indicate different deep SOM cells. \textit{Top right}: The redshift distribution of a tomographic bin. \textit{Middle right}: One wide SOM cell in that bin. \textit{Bottom right}: Three deep SOM cells associated with the highlighted wide SOM cell.}\label{fig:method}
\end{figure*}

\begin{figure*}
\centering
\includegraphics[width=\linewidth]{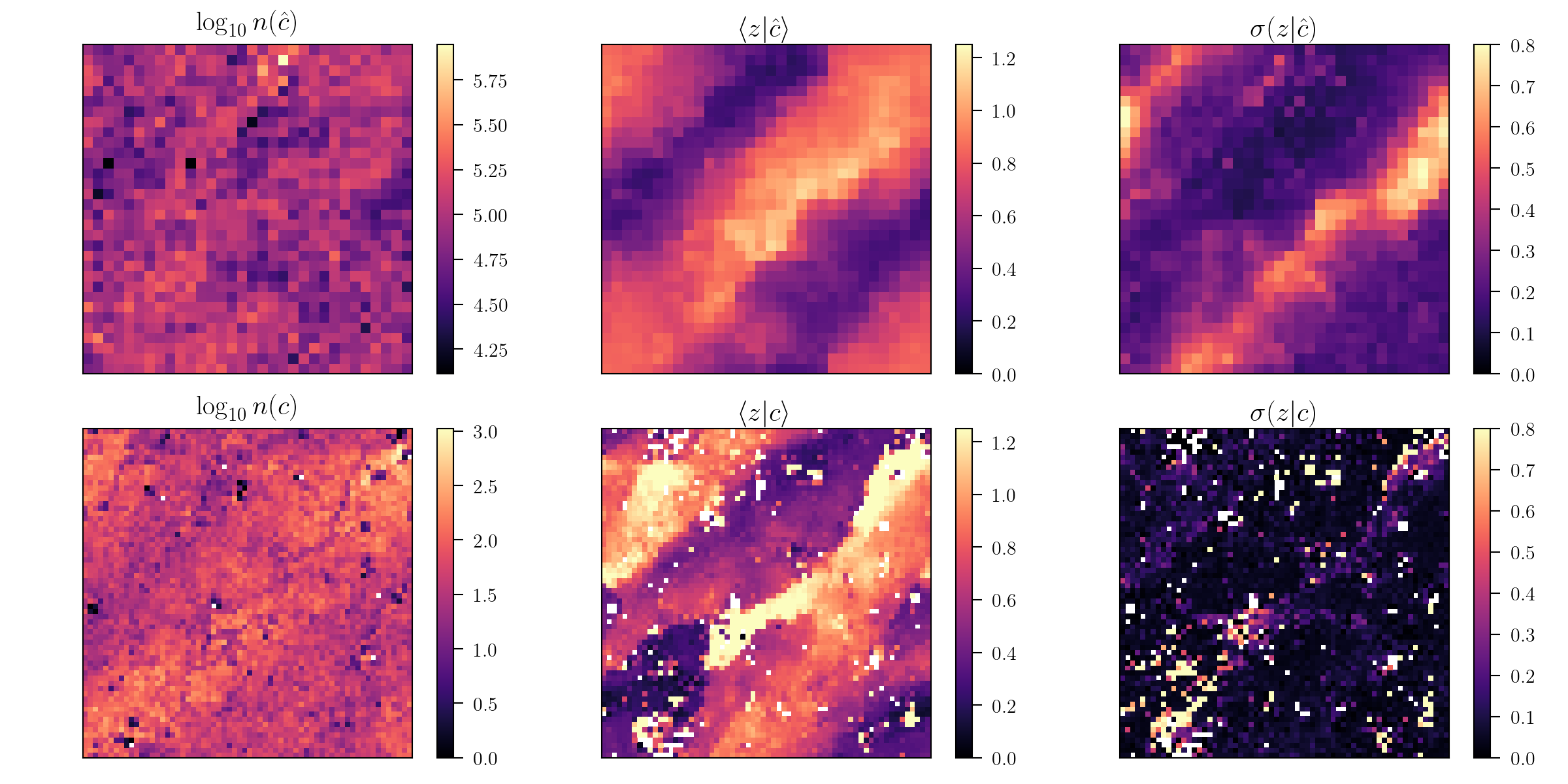}
\caption{Visualization of the wide (top panel) and deep (bottom panel) field Self-Organizing Maps. Shown here are the total number of unique galaxies assigned to each SOM (left), the mean redshift of each cell (middle), and the  standard deviation of the redshift distribution of each cell (right). White cells in the deep SOM are parts of color space for which there are no galaxies in the \texttt{COSMOS2015} sample.
  \label{fig:soms}}
\end{figure*}

The redshift probability conditioned on both c and $\hat{c}$ is statistically difficult to estimate because very few galaxies will meet both conditions simultaneously. In other words, because the number of pairs $(c,\hat{c})$ is so large, each pair will have very few, if any, galaxies. However, under the assumption that the $p(z)$ for galaxies assigned to a given deep photometric cell $c$ should not depend sensitively on the noisy wide photometry of that galaxy, we can relax the selection condition $\hat{c}$ to $\hat{b}$ (as in Equation \ref{eqn:bincond}) or remove this selection entirely (as in Equation \ref{eqn:nobincond}):
\begin{align} 
    p(z|\hat{b}, \hat{s}) &\approx \sum_{\hat{c} \in \hat{b}} \sum_{c} p(z|c, \hat{b}, \hat{s}) p(c|\hat{c},\hat{s}) p(\hat{c}|\hat{s})\label{eqn:bincond}\\
    &\approx \sum_{\hat{c} \in \hat{b}} \sum_{c} p(z|c, \hat{s}) p(c|\hat{c},\hat{s})p(\hat{c}|\hat{s}). \label{eqn:nobincond}
\end{align}
We use the approximations in Equations \ref{eqn:bincond} and \ref{eqn:nobincond} for our fiducial measurement on the Y3 weak lensing source catalogue. In particular, for each tomographic bin, we use Equation \ref{eqn:bincond} when possible (i.e. in cases for which at least one galaxy satisfies both $c$ and $\hat{b}$), and Equation \ref{eqn:nobincond} otherwise. For our tests on the equivalent simulated catalogue, we use Equation \ref{eqn:bincond} exclusively, discarding cases for which there is no galaxy satisfying both $c$ and $\hat{b}$. We illustrate each factor in this equation in Fig. \ref{fig:method} and show the fiducial Self-Organizing Maps in Fig. \ref{fig:soms}. The validity and impact of these assumptions are discussed in \S\ref{sec:method_validation}. 

The terms in this equation are estimated from the following different samples of galaxies:
\begin{enumerate}
    \item $p(\hat{c}|\hat{s})$ is computed from our Wide Sample, which consists of all galaxies in the DES Year 3 weak lensing source catalogue.
    \item $p(c|\hat{c},\hat{s})$ is computed from our Deep and \balrog Samples, which consist of all detected and selected \balrog realisations of the galaxies in the Deep Sample. We call this term the \textit{transfer function}.
    \item $p(z|c, \hat{b}, \hat{s})$ is computed from the Redshift Sample subset of the Deep Sample, for which we have reliable redshifts, 8-band deep photometry, and wide-field \balrog realisations. \footnote{This term could, in principle, be computed from the overlapping photometry of the deep and wide fields, but is much more well sampled by making use of \balrog.}
\end{enumerate}

\subsection{Weighting Redshift Distributions for Lensing Analyses}
\label{sec:lensingweight_sompz}

Under weak lensing shear $\gamma$, the measured galaxy ellipticity transforms as $e\rightarrow e+ R \gamma$ with a shear response $R$. Average quantities like mean tangential shear or two-point correlation functions are thus implicitly weighted by $R$. 

Additionally, each galaxy has an explicit lensing weight $w$ defined to reduce the variance of the measured shear (for more detail, see \citealt*{y3-shapecatalog}). When predicting any shear signal, the $n(z)$ must be weighted by the product of response and explicit weight, $R\times w$ (see \S 3.3 in \citealt{y3-imagesims} for details and blending-related limitations of this approach).

\subsubsection{Lensing Weighted Wide SOM Cell Occupation}

The contribution of a wide cell $\chat$ to the lensing signal measured by some selection $\hat{s}$ of galaxies needs to take into account the response and lensing weights of individual galaxies in $\hat{c}$. Thus, the weight of wide SOM cell $\hat{c}$ is computed with the following sum over all galaxies $i$ assigned to that cell:
\begin{equation}
    p(\hat{c}|\hat{s}) = \sum_{i \in \hat{c}} \frac{w_i R_i}{\sum_{j\in \hat{s}} w_j R_j}.\label{eqn:pchat_lensing_weighted}
\end{equation}

\subsubsection{Lensing Weighted $p(z|c, \hat{b}, \hat{s})$}
In addition to the response and lensing weightings, each selected galaxy in the \balrog Sample must be weighted by the number of times it was detected, passed the selection $\hat{s}$, and was assigned to the same bin $\hat{b}$; this weight must also be normalized by the number of times $N_{\mathrm{inj}}$ it was injected with \balrog. 

The lensing weighted $p(z)$ for a galaxy $i$ in the Deep Sample given its assignment to a deep cell $c$ and a wide bin $\hat{b}$ is:
\begin{equation}
    p(z|c, \hat{b}, \hat{s}) \propto \sum_{i \in (c, \hat{b})} \frac{w_i R_{i} p_i(z)}{N_{i, \mathrm{inj}}}  \; ,\label{eqn:pz_lensing_weighted}
\end{equation}
where the sum runs over \balrog realisations $i$ of Redshift Sample galaxies that are assigned to deep-field cell $c$ and tomographic bin $\hat{b}$, and $p_i(z)$ is either the spectroscopic or many-band photometric redshift posterior for that galaxy.

\subsubsection{Lensing Weighted Transfer Matrix}
Finally, the lensing weighted transfer matrix $p(c|\hat{c},\hat{s})$ is found by similarly weighting the counts of ($c,\hat{c}$) pairs among \balrog realisations:
\begin{align}
\label{eqn:transfer_func}
p(c|\hat{c},\hat{s}) &= \frac{p(c, \hat{c}|\hat{s})}{p(\hat{c}|\hat{s})}.
\end{align}

The respective sums over \balrog realisations $i$ to compute the numerator and denominator of this term are:

\begin{align}
\label{eqn:transfer_num_and_denom}
p(c,\hat{c}|s) &\propto \sum_{i \in \hat{s}} \delta_{c,c_i} \delta_{\hat{c},\hat{c}_i} w_i R_i/N_{i, \mathrm{inj}}\\
p(\hat{c}|\hat{s}) &\propto \sum_{i \in \hat{s}} \delta_{\hat{c}, \hat{c}_i} w_i R_i/N_{i, \mathrm{inj}}.
\end{align}

Note that the transfer function is computed from \balrog realisations, not the full wide galaxy sample, since only for the former are both wide-field and deep-field photometry available.

\subsubsection{Smooth Response Weights}

As a consequence of using response to weight on a per-galaxy basis, the derived redshift distribution can carry the noise inherent in the responses themselves. This may even generate a non-physical negative distribution at some redshifts. To remedy this, the response weights are smoothed over a grid of galaxy size and signal-to-noise according to the treatment in \citet[][see their appendix D]{y3-imagesims}. As demonstrated there on the simulated sample, this introduces an error in mean redshift (per tomographic bin) of the order of $|\Delta \bar{z}| \approx 10^{-3}$. By contrast, the effect of response weighting overall is an order of magnitude larger at $|\Delta \bar{z}| \approx 0.01$. Therefore, we can conclude that the uncertainty introduced due to smoothing the response weights is negligible with respect to the other effects at play, and that the resulting redshift distributions benefit from the reduced noise in response.

\subsection{Construction of Tomographic Bins}
\label{sec:binning}
Once galaxies have been categorized into phenotypes based on their photometric observations, we construct tomographic bins and assign each phenotype $\chat$ to a bin. For our fiducial result, we construct these bins according to the following procedure:

\begin{enumerate}
\item{To construct a set of $n$ tomographic bins $\hat{b}$, begin with an arbitrary set of $n+1$ bin edge values $e_j$.}

\item{Assign each galaxy in the Redshift Sample to the tomographic bin $\hat{b}$ in which the best-estimate median redshift value of its $p(z)$ (or its spectroscopic redshift $z$) falls. This yields an integral number of galaxies $N_{\mathrm{spec}, (\hat{c},\hat{b})}$ satisfying the dual condition of membership in a wide SOM cell $\hat{c}$ and a tomographic bin $\hat{b}$. This can be written as a sum over \balrog realizations $i$ of redshift galaxies:}

\begin{equation}
    N_{\mathrm{spec}, (\hat{c},\hat{b})} = \sum_{i} \delta_{\hat{c}, \hat{c}_i} \delta_{\hat{b}, \hat{b}_i}.
\end{equation}

\item{Assign each wide cell $\hat{c}$ to the bin $\hat{b}$ to which a plurality of its constituent Redshift Sample galaxies are assigned:}

\begin{equation}
    \hat{b} = \{ \hat{c} | \argmax_{\hat{b}} N_{\mathrm{spec}, (\hat{c},\hat{b})}\}.
\end{equation}

\item{Adjust the edge values $e_j$ post hoc such that the numbers of galaxies in each tomographic bin $\hat{b}$ are approximately equal and repeat the procedure from step (ii) with the final edges $e_j$.}

This procedure yields bin edges of [0.0, 0.358, 0.631, 0.872, 2.0] for the Y3 weak lensing source catalogue. As an inconsequential result of the slight differences in the Y3 source galaxy catalogue and the simulated equivalent, the bin edges in the equivalent \buzzard catalogue are [0.0, 0.346, 0.628, 0.832, 2.0]. We discuss this choice to homogenize the number of galaxies in each tomographic bin separately for data and simulations in \S \ref{sec:method_validation}.
\end{enumerate}

\subsection{Clustering redshift information}

Fully independent information on the redshift distribution of the tomographic bins of our source sample is provided by its angular cross-correlation with galaxy samples of known redshift \citep{Newman2008,Menard2013}. Previous experiments have used this type of information to validate and/or further constrain the mean redshift of their sources \citep[e.g.][]{Hildebrandt2017,Davis2017,kids1000_pzpaper}. A dominant confounding factor in this approach is the redshift evolution, within the tomographic bin, of the clustering bias of the source galaxies, which is highly degenerate with the mean redshift of a tomographic bin \citep[e.g.][]{Gatti2018,kids1000_wzsims}.

The full description of the DES Y3 source galaxy clustering redshift analysis is given by \citet*{y3-sourcewz}. In brief, as reference galaxies we use the combination of redMaGiC luminous red galaxies with high quality photometric redshifts \citep{Rozo2016,y3-galaxyclustering} and spectroscopic galaxies from BOSS and eBOSS \citep{Smee2013,Dawson2013,Dawson2016,Ahumada2019} where they overlap the DES survey area.

There are two ways in which the clustering redshift data is used to validate and inform the redshift calibration. From comparing the clustering signal to the signal expected for a fiducial redshift distribution within a redshift range where the former exists, and assuming that clustering bias is constant as a function of source redshift, one can determine the best shift $\Delta z$ of the fiducial redshift distribution and compare it to zero within its statistical and systematic uncertainty. This first method is only used as cross-check to validate the photometric estimate of $n(z)$. Alternatively, one can include the clustering redshift information in a likelihood analysis, jointly with sample variance and shot noise, that returns samples of probable redshift distributions, while marginalizing over a flexible model of source clustering bias redshift evolution. This second method is used to generate the ensemble of redshift distributions in this paper (see \autoref{sec:method_validation} and \autoref{sec:3sdir_mfwz}), and it is shown to vastly improve the accuracy of the shape of $n(z)$ derived from photometric data alone. For details of both approaches, we refer the reader to \citet*{y3-sourcewz}. 

\subsection{Image simulations and the effect of blending}
\label{sec:imagesims}
The calibration as described thus far is aimed at recovering the distribution of redshifts of the dominant galaxies associated with an ensemble of detections in the DES Y3 \metacal catalogue, weighted by the individual detections' shear response. However, the measurement of a detection's shape commonly depends not just on the shear of the dominant associated galaxy, but also on the shear applied to galaxies blended with it. As \citet{y3-imagesims} show, this leads to significant response to the shear of light at other redshifts. This is best accounted for by a modification of the redshift distribution to be used for predicting lensing signals. In \citet{y3-imagesims}, such a modification is derived for the DES Y3 source galaxy bins defined here. This modification reduces the mean redshift of the bins (see \S \ref{tab:uncertainty}) and is calibrated with an uncertainty shown in Table \ref{tab:uncertainty}.

We note that this correction to the $n(z)$ calibrated by photometry and clustering is expected to have non-zero shifts on the mean redshift in each tomographic bin. Additionally, several aspects of our photometric calibration strategy are validated in image simulations \citep[see][e.g.~recovered true redshift distributions and their appendix D]{y3-imagesims}.

\subsection{Shear ratio information}
\label{sec:shearratio}
For physically separated pairs of a gravitational lens with sources from two bins, the ratio of the shear signals is indicative of the redshift distributions of the sources for fixed parameters of the cosmological model. In the DES Y3 lensing analyses, we use this information as an additional term in the likelihood of the lensing signals. We provide a brief summary here and refer readers to \citet*{y3-shearratio} for details of the methodology.

Gravitational shear signals on small to moderate scales are calculated for the source bins defined here around samples of redMaGiC lens galaxies. The ratio of these signals between pairs of source bins is used as the data over which likelihoods are calculated. The use of a ratio removes sensitivity of the measured shear signal to the mean matter overdensity profile around the lens galaxies but magnification, intrinsic alignments of sources relative to physically nearby lenses, and a mild dependence of the geometric shear ratio to cosmology require the likelihood to be evaluated alongside the cosmological and nuisance parameters of the Y3 lensing analyses. The shear ratio information provides constraints on this multi-dimensional parameter space in addition to, and somewhat degenerate with, the source redshift information.

For consistency tests in this paper, we use constraints from a shear-ratio-only chain to judge the consistency of the $n(z)$s with the lensing signals, from a free parameter with flat prior for the shift of the fiducial redshift distribution, at fixed cosmological parameters \citep*[see][for details]{y3-shearratio}. Note that for the reasons described in \autoref{sec:imagesims}, perfect agreement of the shear ratio constraint and the redshift distribution derived by means of photometry and clustering is not expected.

\section{Characterization of Sources of Uncertainty in Photometric n(z)}\label{sec:uncertainty}

In this section we will characterize the uncertainties in our measurement of redshift distributions from galaxy photometry. In brief, our method consists in using secure redshifts to determine $p(z)$ in 8-band colour-space, and using the DES deep fields to determine the abundance of galaxies in 8-band color-space in the 3-band magnitude and colour-space of the lensing source galaxy sample. As a result, we must incorporate uncertainties in the redshifts used and in the estimated abundances of galaxies in each region of colour-space. The fully enumerated list of contributing sources of uncertainty is thus:

\begin{enumerate}
    \item Sample Variance: fluctuations in the underlying matter density field determine the abundance of observed deep field galaxies of a given 8-band colour and at a given redshift (\S\ref{sec:SV_SN})
    \item Shot Noise: shot noise in the counts of deep field galaxies of a given 8-band colour and at a given redshift (\S\ref{sec:SV_SN})
    \item Redshift Sample Uncertainty: biases in the redshifts of the secure redshift galaxy samples used (\S\ref{sec:redshiftsampleunc})
    \item Photometric Calibration Uncertainty: uncertainty in the 8-band colour of deep field galaxies (\S\ref{sec:zeropointunc})
    \item \balrog uncertainty: imperfections in the procedure of simulating the wide field photometry of deep field galaxies (\S\ref{sec:balrogunc})
    \item \sompz Method Uncertainty: bias in the estimated redshift distributions relative to truth inherent to the methodology (\S\ref{sec:sompz_method_unc})
\end{enumerate}
 
 We now turn to developing the formalism necessary to describe each of these uncertainties and how they affect our measured $n(z)$.
 
Our ultimate goal is to characterize the uncertainty in our estimation of the redshift distribution of each tomographic bin $p(z|\hat{b},\hat{s})$. It is useful to rewrite this probability (following Equation~\ref{eqn:bincond} and Equation~\ref{eqn:transfer_func}) explicitly as a function of the four galaxy samples involved in its estimation:
\begin{equation} \label{eqn:redshift_prob_samples}
    p(z|\hat{b}, \hat{s}) \approx \sum_{\hat{c} \in \hat{b}} \sum_{c} \underbrace{p(z|c)}_\text{Redshift} \underbrace{p(c)}_\text{Deep} \underbrace{\frac{p(c,\hat{c})}{p(c)p(\hat{c})}}_\text{Balrog}  \underbrace{p(\hat{c})}_\text{Wide}, 
\end{equation}
where the right-hand-side terms are implicitly conditioned on the selections $\hat{b}, \hat{s}$ (not shown in Equation \ref{eqn:redshift_prob_samples} for clarity). Note that the \balrog Sample does not inform the marginal distributions of either the deep nor the wide SOM cells, \textit{i.e.} the \balrog Sample is not used to compute $p(c)$ or $p(\hat{c})$.

First, there is uncertainty because the galaxy samples involved are finite in both number and area. The finite area and size of the Redshift and Deep samples introduce shot noise and sample variance, which we model analytically, as explained in \S~\ref{sec:SV_SN}. Moreover, as mentioned in \S~\ref{sec:sompz_formalism}, the current finite size of the combined Redshift and \balrog Samples makes it difficult to empirically estimate $p(z|c,\chat)$ for all $(c,\chat)$ pairs, so we implement an approximate estimate for this term (Equations~\ref{eqn:bincond} \& \ref{eqn:nobincond}). We describe and explore the effects of this approximation on the $n(z)$ in \S~\ref{sec:method_validation}, where we validate the methodology using simulated mock catalogues.

Second, the $z$ values of the Redshift Sample
carry uncertainty. In \S~\ref{sec:redshiftsampleunc} we compare the redshift information that we have available from different sources in the deep fields (from spectroscopy and many-band photometry) and discuss the limitations of each.
Third, the cell assignments are stochastic and thus their rate estimates are subject to shot noise as well as systematic biases. In \S~\ref{sec:balrogunc}  we test the robustness of the \balrog transfer function against variable observing conditions across the footprint and by comparing to an alternative transfer function estimated directly with actual wide and deep photometry. Finally, in \S~\ref{sec:zeropointunc} we examine the photometric zero-point uncertainty across the deep fields which introduces noise in the deep field colours, and we describe the method used to propagate that noise to each estimated $p(z|\hat{b}, \hat{s})$.

\subsection{Sample Variance and Shot Noise} \label{sec:SV_SN}

The \sompz Bayesian formalism described in \S~\ref{sec:sompz_formalism} makes it very explicit how we estimate the redshift distribution of our four source weak lensing tomographic bins. As highlighted by Equation~\ref{eqn:redshift_prob_samples}, we use the sample with the best statistics to infer each particular probability that is needed to determine the $n(z)$. Therefore, quantifying the $n(z)$ uncertainty means describing the limitations of each sample at determining each of these probabilities. 

In this subsection we discuss some of the limitations of the Redshift and Deep Sample in estimating the redshift and color probability $p(z,c)$. Common limitations in redshift calibration samples are: shot noise due to finite sample size; sample variance due to large scale structure fluctuations; photometric selection effects; photometric calibration errors; spectroscopic selection effects and incompleteness; and photometric redshift errors. We explore systematic errors due to spectroscopic redshift biases or photometric redshift biases in \S~\ref{sec:redshiftsampleunc}, and discuss errors in the deep field photometric zero-point calibration in \S~\ref{sec:zeropointunc}. We match the photometric selection effects from the wide field by injecting deep field galaxies into wide-field images using \balrog \citep{y3-balrog} and calculating the rate at which deep field galaxies would be detected and selected for the weak lensing sample. Since the deep fields are $\sim1.5$ magnitudes deeper than the wide field \citep*{y3-deepfields}, deep-field depth variations are negligible. 

Here we focus on how to estimate the shot noise and sample variance uncertainty in our deep-field samples. Typically this has been achieved by performing the same redshift estimation analysis on mocked realisations of the redshift calibration samples at different line-of-sight positions. Then, the variance and correlations in the mean redshift of the tomographic bin $n(z)$ are obtained from the variations across simulated versions of the data \citep*[\textit{e.g.}][]{Hildebrandt2017,kids1000_pzpaper,Hoyle2018,y3-sompzbuzzard,KidsSOM20}. While we also run all methods in multiple simulated deep field realisations (\S~\ref{sec:method_validation}), we do so as a validation and to verify if there are any remaining systematic uncertainties intrinsic to the methods themselves, but not to get an estimate of sample variance for real data. Instead, we build an analytical model of sample variance that predicts the distribution of the redshift-colour distribution in the deep fields given the data that we have observed, \textit{i.e.} we write the distribution of a distribution: $ P(p(z,c)|\text{data})$. Given this, one can propagate this distribution of uncertainties with Equation~\ref{eqn:redshift_prob_samples} and calculate the distribution of plausible $n(z)$ shapes allowed by sample variance and shot noise.

To analytically model sample variance we use a model involving \textit{three-step Dirichlet} sampling, labelled  \sdir in this work. This approximate model of sample variance was introduced in \citet{Sanchez2020}, and is the product of three independent Dirichlet distributions. \citet{Sanchez2020} showed in simulations that \sdir predicted well the levels of uncertainty due to sample variance and shot noise in the first two moments of the $n(z)$ for a non-tomographic galaxy sample. We explore its performance at describing the sample variance of our four tomographic bins using the Buzzard simulations and discuss the results in \S~\ref{sec:method_validation}. We give extensive technical details of the model's mathematical formalism and application to DES Y3 in Appendix~\ref{app:3sdir_details}. 

In short, the \sdir model describes the probability that galaxies belong to a redshift bin $z$ and colour phenotype $c$, given that a number of galaxies have been observed to be at redshift bin $z$ and colour phenotype $c$. We describe the probability in redshift and deep color $p(z,c)$ with a finite set of coefficients $\{f_{zc}\}$ indicating the probability in redshift bin $z$ and color phenotype $c$, where $\sum_{zc}f_{zc}=1$ and $0\leq f_{zc} \leq 1$. If each Redshift Sample galaxy were representative and independently drawn, then a Dirichlet distribution parameterized by
  the Redshift Sample counts $N_{zc}$ would fully characterize $p(\{f_{zc}\}).$ Sample variance correlates the redshifts, however, and the more complex \sdir model incorporates this, \textit{i.e.} $p(\{f_{zc}\}|\{N_{zc}\})\approx \sdir$.

  An alternative approach to estimate the sample variance and shot noise present in our calibration fields would be to perform spatial bootstrap or jackknife resampling of the calibration samples (see e.g. \citealt{Hildebrandt2017} and \citealt{kids1000_pzpaper}). This technique could be used separately to estimate the variance in the deep field color distribution $p(c)$ from all four Deep Fields, and the variance in $p(z|c)$ in the COSMOS field (the only calibration field where we have complete redshift information). Such a procedure would correctly estimate the shot noise contribution to our uncertainty and would additionally account for the variance from the density fluctuations present within the calibration fields and variance on scales comparable to the angular distance between the field locations on the sky.  We note, however, that to efficiently combine this information with our WZ likelihood function or in a hierarchical Bayesian model one would need an analytic model describing the bootstrap-resampled calibration samples. Chief among the reasons we implement \sdir for the DES Y3 redshift calibration is that, as an analytic model, it can be readily jointly sampled with the WZ likelihood function.
  
\subsubsection{Methodology Validation} \label{sec:method_validation}

\begin{figure*}
\centering
\includegraphics[width=\linewidth]{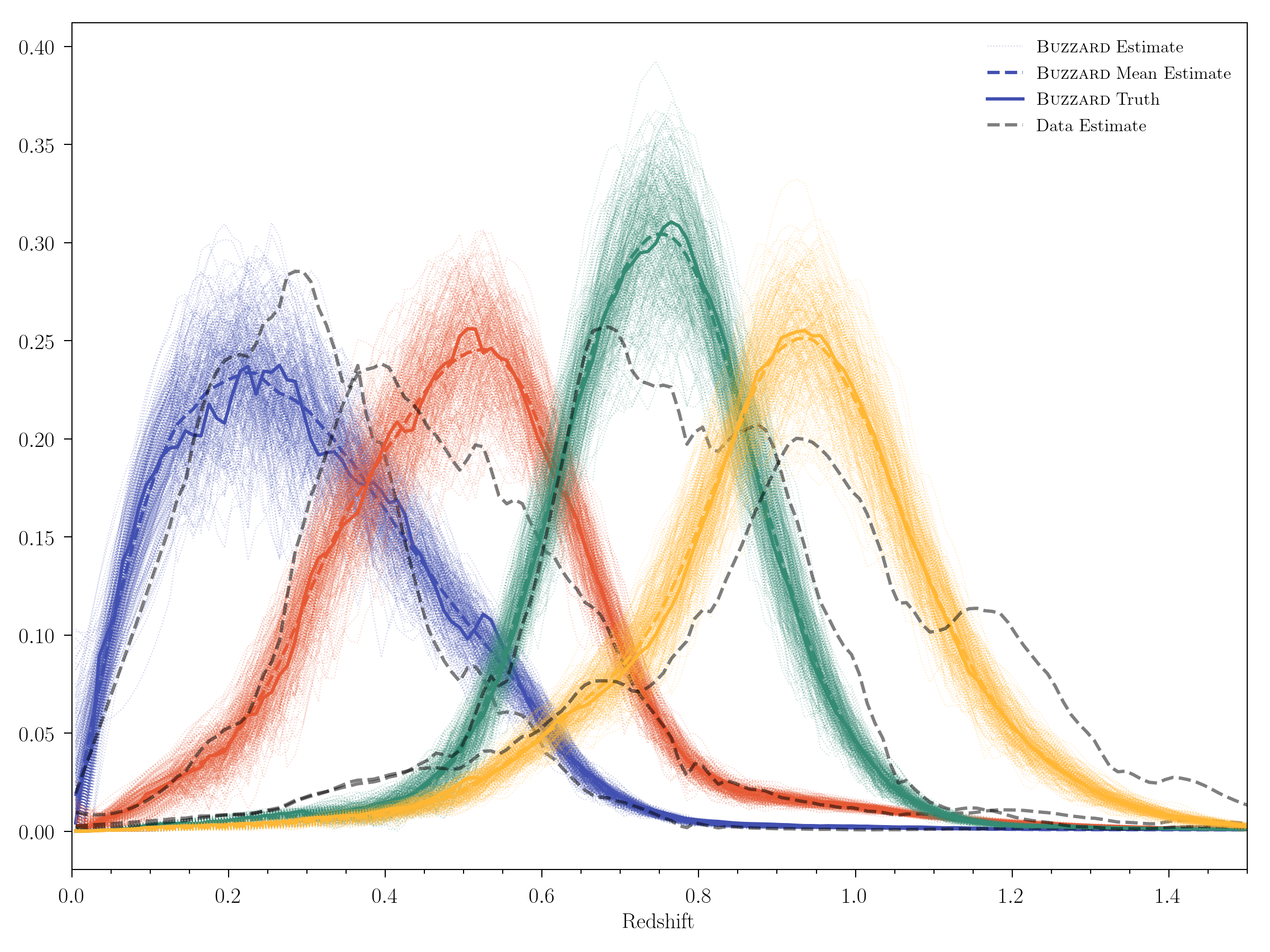}
\caption{Estimated $n(z)$ in four tomographic bins from the \buzzard simulations using an ensemble of 300 different sets of deep fields on the \buzzard sky (colourful fine dashed lines). The similarity of the mean of the estimated $n(z)$ (colourful broad dashed lines) relative to the truth (colour broad solid lines) is a basic illustrative validation of the method. The Redshift Sample used here has 100000 galaxies drawn from 1.38 deg$^2$, the Deep Sample in each realisation is drawn from three fields of size  $3.32$, $3.29$, and $1.94$ deg$^2$, respectively from the \buzzard simulated sky catalogue. The variation in estimated $n(z)$ reflects the uncertainty of the \sompz method primarily due to sample variance in the deep fields. The similarity of the $n(z)$ from simulations to the fiducial result in data (gray broad dashed line) reflects the similarity of the simulated catalogue to the data.}
\label{fig:nz_ensemble_buzz}
\end{figure*}

In order to validate the methodology we use the suite of \buzzard simulations (\S~\ref{sec:buzzard}), where we simulate the DES Y3 Wide, Deep, \balrog and Redshift Samples. First, we want to test the accuracy of the \sompz methodology in estimating the wide field $n(z)$ using the  8-band colour and complete redshift information available in the DES deep fields, in the same spirit as the \citet*{y3-sompzbuzzard} and \citet{KidsSOM20} analyses. Secondly, we want to test the accuracy of the \sdir method in describing the sample variance uncertainty in the estimated $n(z)$ within the \sompz framework. \citet*{y3-sompzbuzzard} validated the \sompz methodology in the context of DES, but here we use a more realistic set of simulated samples, and we also introduce `bin conditionalization' (see Equation~\ref{eqn:bincond}), which reduces the intrinsic bias in mean redshift of the estimated $n(z)$. \citet{Sanchez2020} validated the \sdir method but in a different context: for a non-tomographic sample with a different selection than the DES Y3 source sample, where all galaxies in the deep fields had redshift information, and without a transfer function to reweight the colours in the deep field.

We first turn to discussion of testing the accuracy of our methodology with the \buzzard simulated galaxy catalogue. The goal of this analysis is to calibrate the inherent uncertainty of our method in simulated conditions that are as realistic as possible. This uncertainty can be quantified in terms of, for example, uncertainty on the mean redshift in simulations, which serves as an ideal point of comparison to the data and can be propagated as a source of uncertainty in our final ensemble. Because the colour-redshift relation in this simulation is necessarily an imperfect reproduction of the true colour-redshift relation of our observed source galaxy sample, the deliverable uncertainty on the mean redshift is centered on zero. This stands in contrast to an alternative approach exemplified by \cite{kids1000_pzpaper}, where the colour distributions of galaxies in simulations are matched to data, thus enabling tests on simulations to yield an estimate of the residual bias on the mean redshift in simulations relative to the truth and thus correct for the magnitude of that bias in the data. As a consequence of our analysis choice, there are, in general, different colour-redshift degeneracies in simulations than in data, and the colour-edges of tomographic bins are effectively different. While this means our \buzzard SOMs and tomographic bins are of limited use beyond the specific goal of calibrating uncertainty, we view this analysis choice as an appropriate path given the absence of fully forward-modeled galaxy colour distributions.

The effect of sample variance in our estimated $n(z)$ is of particular interest. To this end we generate 300 versions of the four DES Deep Samples (where one of the four has perfect redshift information) at different random line-of-sight positions in the \buzzard simulations. For each of the 300 realisations of the deep fields, we run the \sompz algorithm and we obtain a $n(z)$ estimate for each tomographic bin by fixing the probabilities to the observed redshift and colour phenotype number counts. Fig.~\ref{fig:nz_ensemble_buzz} shows the 300 $n(z)$s estimated by \sompz for each tomographic bin (light solid lines), together with their average (dark dashed lines) and the true wide field $n(z)$ in the simulation (dark solid lines with colour). We find the average simulated $n(z)$ to be extremely close to the truth. For comparison, we show the estimated $n(z)$ from data (grey dashed lines), which shows a reasonable agreement to the simulated ones. Note that the averaged $n(z)$ in simulations looks much more smooth than that from data as we are averaging out sample variance, while the $n(z)$ from data corresponds to a single realisation observed in the DES deep fields, which is affected by sample variance. In addition, to test the performance of the \sdir method, we calculate multiple samples of $n(z)$ in each \buzzard realisation by drawing from the \sdir likelihood; with the range of $n(z)$ samples spanning the sample variance uncertainty allowed by the \sdir model in the redshift-colour probability.

We show technical details and specific figures of the methods validation in Appendix~\ref{app:validation}, and highlight the main findings here. We find the average mean redshift (average $\zbar$ or $\langle \zbar \rangle$) across the 300 Buzzard realisations to be consistent between the \sompz and \sdir methods. However, when compared to the truth we find a residual offset of $\Delta_z=[0.0051, 0.0024, -0.0013, -0.0024]$ in each bin, where $\Delta_z\equiv\langle \zbar^{\text{SOMPZ}} \rangle - \zbar^{\text{true}}$.

We take this nonzero offset as a systematic error intrinsic to the method and due to the assumption of bin conditionalization (Equation \ref{eqn:bincond}); we describe how we propagate this uncertainty in \S~\ref{sec:sompz_method_unc}. 

Using the \sdir model one can compute, in each Buzzard realisation, a distribution of mean redshift values, or $\zbar$, with the values allowed by sample variance and shot noise uncertainty. We find the expected value of that distribution to be unbiased with the mean redshift value from \sompz in individual \buzzard realisations, and in each tomographic bin. We also compare the width of the $\zbar$ distribution from \sdir in each Buzzard realisation, and the width across the 300 $\zbar$ from \sompz in all realisations. We find the width predicted by \sdir to be within $10$ per cent from the width estimated with \sompz in the  3 lower redshift tomographic bins, but $50$ per cent wider in the last tomographic bin. This is a feature of the \sdir model, which gives an unbiased likelihood at the expense of slightly underestimating the uncertainty due to sample variance at lower redshifts, and overestimating it at higher redshifts.

We have taken great care to validate that \sdir provides a likelihood of $n(z)$ whose mean redshift is fully compatible with the mean redshift from \sompz. The mean redshift serves as the leading order statistic of the $n(z)$ affecting the cosmological constraints of cosmic shear analysis, and historically the $n(z)$ has been parameterized with a fiducial $n(z)$ fixed from galaxy counts and a shift parameter incorporating the uncertainty information \citep[\textit{e.g.}][]{Bonnett2016,Hoyle2018,Troxel2018,Tanaka2018, KidsSOM20, Kids450_somcosmology,kids1000_pzpaper}. However, here we present a change of paradigm and write a full likelihood function for the redshift distribution. Therefore, we want to make sure that no intrinsic biases are introduced in $\zbar$ with respect to the mean redshift of the \sompz methodology. There are a number of advantages to preferring a full likelihood function to a fixed $n(z)$ with a shift to its mean: it more accurately represents our uncertainty from photometry and the redshift-colour relation, it propagates higher order moment uncertainties of the redshift distribution, and it is more suitable to be combined with other sources of redshift information like clustering redshifts. As shown in \cite{kids1000_wzsims} and \cite{kids1000_pzpaper}, combining a clustering redshift likelihood function with a fixed $n(z)$ from photometry parametrized with a shift can introduce a bias in the values of the shift parameter when the $n(z)$ is inaccurate. However, having a full likelihood over $n(z)$ presents the full set of possible $n(z)$ distributions spanning our uncertainty from photometry, with variable shapes, which can be combined, for example, with a clustering redshifts likelihood function.

In order to combine the \sdir likelihood with a clustering redshifts (or WZ) likelihood, one can draw \sdir $n(z)$ samples and importance sample them by the value of their WZ likelihood with each $n(z)$ draw. Even though drawing from \sdir is very fast, this is an extremely inefficient process as the drawn $n(z)$ samples very often contain sample variance fluctuations that deliver a low WZ likelihood. By contrast, a Hamiltonian Monte-Carlo (HMC) sampler has the ability to draw from the joint combination of both likelihoods and, although drawing individual samples is slower, sampling the joint space becomes much more efficient and fast. We have defined a modified version of the \sdir likelihood that we use in a HMC chain to sample together with the WZ likelihood. For further details on this HMC chain, see \citet{hmcpaper_WZ}.
This modified likelihood, or \sdirmfwz, is defined in Appendix~\ref{sec:3sdir_mfwz}, and is by construction more sensitive to sample variance. In short, it is using less information of the colour distribution observed in the deep fields. As a result, the width of $\zbar$ values from \sdirmfwz is larger in all redshift bins -- $[78, 31, 23, 39]$ per cent larger than \sdir in each bin, respectively (see Appendix~\ref{app:validation} for further details).

\subsection{Redshift Sample Uncertainty} \label{sec:redshiftsampleunc}

If all galaxies with redshift information were selected independently and representatively from the source population, with no systematic uncertainties on $z$, then we could simply merge them all into a single sample regardless of their origin. In reality, we have overlapping redshift information from several surveys, each with unique selection criteria and biases, as described in \S~\ref{sec:redshift_samples}. We label the different redshift surveys (or combinations thereof) with $R$. There are different ways that we could combine information from multiple surveys.  One limit is to state that one combination $R$ is correct, but we only have some prior guess $p(R)$ about which one it is. Sampling the Bayesian posterior for $n(z)$ under this assumption is simple: we simply produce samples of $f_{zc}$ from each survey independently by the methods of the previous subsections; and then make a final set of samples for which a fraction $p(R)$ comes from each survey.  In our case we do not know that any of $R$ is correct, but we none the less execute this marginalization over $R$ under the principle that it is still likely to now contain the truth and also span the range of uncertainty that we have from our ignorance of the quantitative errors in different surveys.

As each of P$\equiv$\texttt{PAUS+COSMOS}, C$\equiv$\texttt{COSMOS2015}, S$\equiv$\texttt{SPEC} do not span the same region of colour space (or deep SOM cells $c$), as detailed in  \S~\ref{sec:redshift_samples}, we define three redshift samples (\SPC, \PC, \SC) to maximize the completeness of the redshift coverage in any sample by combining information from different sources, but as a consequence the different samples also become correlated. We still sample them separately, assigning them an equal prior probability, $p(R)=\frac{1}{3}$. We note that for those cells $c$ that only have redshift information from one catalogue, we assume that information to be correct. Although the spectroscopic samples \texttt{SPEC} technically span a larger area than the COSMOS field, and are therefore not completed by photometric data outside this area, they are comprised of several catalogues with different selection functions in redshift and different footprints. For simplicity, we use a sample variance theory prediction which assumes an area equal to the COSMOS area in all redshift samples, which is a conservative approach.

Both \texttt{COSMOS2015} and \texttt{PAUS+COSMOS} multi-band photometric redshift catalogues report an individual redshift function $q(z)$ for each galaxy, which is not a proper posterior, but a marginalized likelihood function for different templates of galaxies. If the full likelihood of redshift, templates, and $c$ were known, one could simultaneously and hierarchically infer the underlying $f_{zc}$ and the redshift posterior for each galaxy (note that $f_{zc}$ is at the same time the prior for each galaxy). However, we have found that the width of these $q(z)$ is so small compared to the redshift resolution that we have with the DES $riz$ bands that the \sompz mean redshift changes by less than $10^{-3}$ in all tomographic bins if we treat $q(z)$ as a delta function centred at the mode of the distribution. This is also a much smaller effect than both the uncertainty from different redshift samples $R$ and that from sample variance, so we decide to completely neglect it and treat $q(z)$ as a delta function when generating the \sdir $f_{zc}$ samples.

\begin{figure}
\centering
\includegraphics[width=\linewidth]{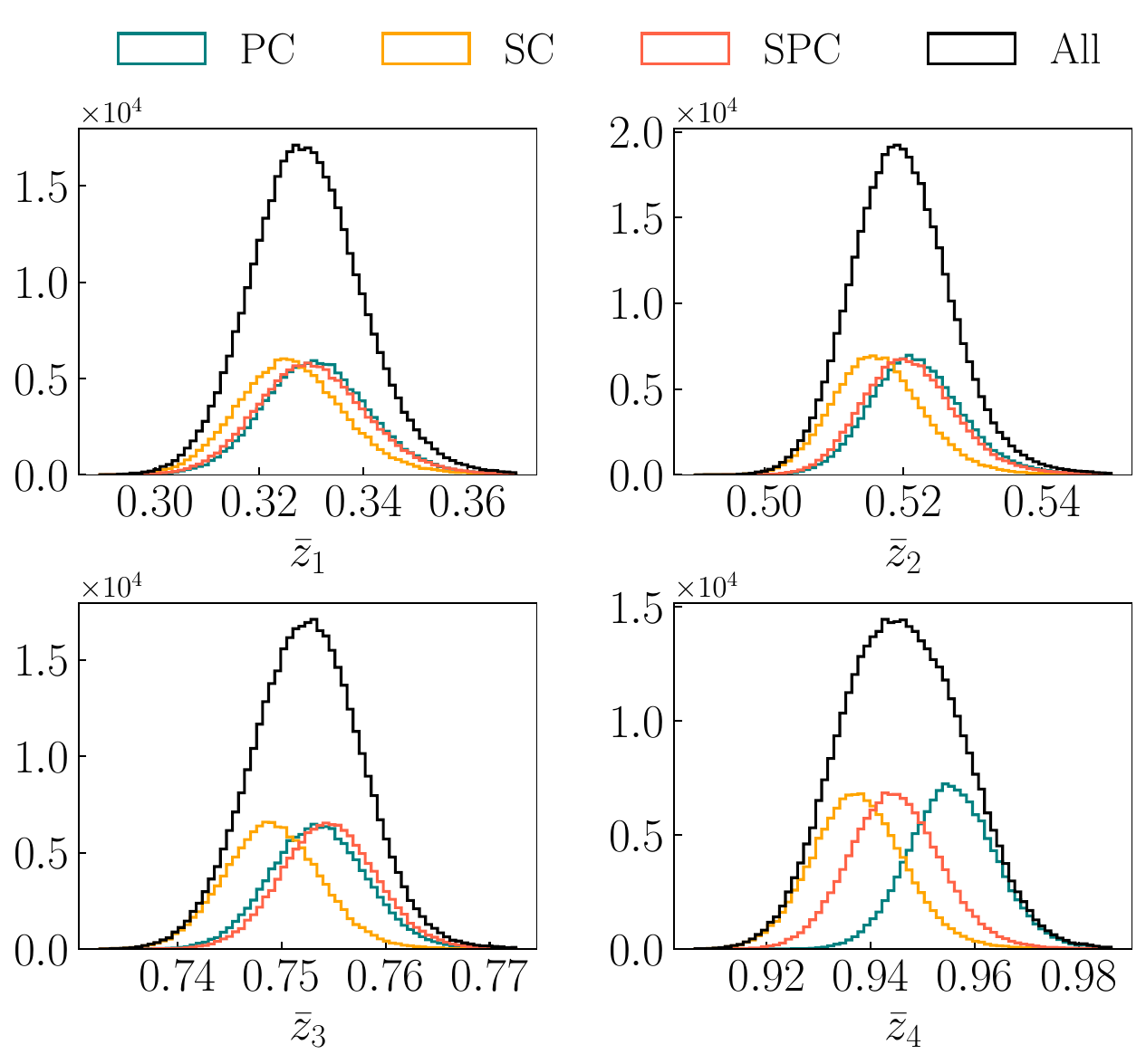}
\caption{The distribution of mean redshift values $\zbar$ from \sdirmfwz for each of the three redshift samples -- \SPC, \PC and \SC -- on real data. We assume that their combination (shown as black histograms) contains the truth and also spans the range of uncertainty that we have from biases of the redshift samples.}
\label{fig:redshiftsamples_meanz3sdir}
\end{figure}

Fig.~\ref{fig:redshiftsamples_meanz3sdir} shows the distribution of mean redshift values predicted by \sdirmfwz for each of the samples \SPC, \PC, \SC; which we find to be generally in agreement. We find a lower mean redshift for samples coming from \SC, while samples from \SPC and \PC agree very well with each other. This is in agreement with \cite{Alarcon2020}, which finds \texttt{PAUS+COSMOS} to be unbiased compared to spectra, but finds \texttt{COSMOS2015} to be systematically biased towards lower redshifts.

\begin{figure}
\centering
\includegraphics[width=\linewidth]{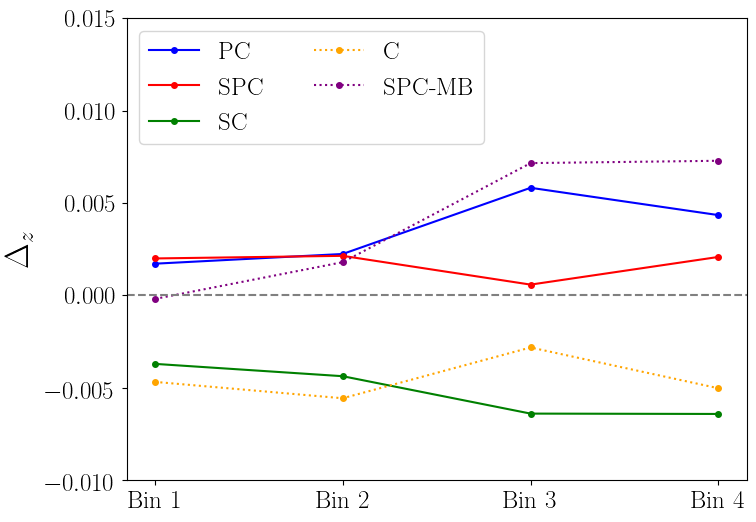}
\caption{Mean redshift difference $\Delta_z$ in each tomographic bin for each Redshift Sample being tested: \SPC, \SC, \PC, \SPCMB, C, relative to  the average mean redshift of \SPC, \SC and \PC.  Spectroscopic catalogues are labelled as S, the \texttt{PAUS+COSMOS} catalogue as P and the \texttt{COSMOS2015} as C. \SPC, \SC and \PC (solid lines) are the three redshift samples used in this work. SPC-MB shows the effect of extrapolating the bias between S and C to all galaxies that are still using redshift information from C in \SPC (mainly faint galaxies) (see section~\ref{sec:redshift_samples} for a definition of the different samples used in this figure). The mean redshift is obtained by computing the $n(z)$ from \sompz using each sample.}
\label{fig:redshiftsamples_sompz}
\end{figure}

The small differences between SPC and PC (Fig.~\ref{fig:redshiftsamples_meanz3sdir} and \ref{fig:redshiftsamples_sompz}) show that our best photometric redshift and spectroscopic redshift information produce $n(z)$ samples that are largely in agreement.
We check for additional robustness using the \SPCMB sample, to test the impact of the faintest galaxies whose redshift information is dominated by \C. We find the $\zbar$ shift between \SPCMB and \SPC to be smaller than $\sim 0.006$, adding confidence that our faintest galaxies, for which we do not have redundant redshift information, are not significantly biasing our mean redshift. This test is limited in that the applied bias as a function of magnitude (described in \S \ref{sec:data}) is inferred from the available overlap between S, P and C, which is limited for faint galaxies. Fig. \ref{fig:redshiftsamples_sompz} shows the difference in $\zbar$ values between several samples: \SPC, \PC, \SC, \C, \SPCMB; and the average $\zbar$ value of \SPC, \PC and \SC.
The size of the offset in mean redshift due to using these different underlying redshift samples, as shown in Fig. \ref{fig:redshiftsamples_sompz}, illustrates the value of additional follow-up spectroscopic and narrow-band photometric observational campaigns. As shown in Fig. \ref{fig:barsummary}, this uncertainty is a significant contributor to our overall error budget, however we find a smaller uncertainty due to this effect than \cite{KidsDEScomb}, who report mean offsets due to varying the redshift sample of $[0.014, -0.053, -0.020, -0.035]$ (see their table 1) in a reanalysis of the Dark Energy Survey's Year 1 analysis \citep*{Hoyle2018}. An important difference in analyses that acts as a caveat to any direct comparison of our work with \cite{KidsDEScomb} is the different selection of source galaxies we apply, in particular the faint magnitude cut $i<23.5$ discussed in Section~\ref{sec:data} and motivated largely to reduce effects from redshift biases in \texttt{COSMOS2015} photometric redshifts. Subject to this caveat, we attribute the differences between our uncertainty found here and their reported values to increased statistical and systematic uncertainty of their method when applied to few-band data, as indicated by systematic offsets of 0.01–0.03 found in their MICE2 simulated analysis (see their appendix), with spectroscopic selection effects primarily responsible. In this work, we mitigate these effects with the inclusion of multi-band data from the Deep Fields and by creating redshift samples that are complete. The use of a larger number of bands on its own is likely to significantly reduce the systematic error due to spectroscopic selection effects \citep{Masters2015,GruenBrimioulle2017,KidsSOM20}.

\subsection{Photometric Calibration Uncertainty}
\label{sec:zeropointunc}
We now turn to testing the sensitivity of our measured $n(z)$ to the uncertainty in the photometric zero-points of the deep fields. Note that our \sompz formalism inherently assumes consistent colours across the four deep fields to assign galaxies to deep SOM cells $c$ and to set the fluxes of their artificial wide field renderings in the \balrog procedure. There are in reality, however, field-to-field variations in photometric calibration (for more detail see \S 6.4 of \citealt*{y3-deepfields}), encoded in the zero-point uncertainty in each field of [0.0548, 0.0039, 0.0039, 0.0039, 0.0039, 0.0054, 0.0054, 0.0054] in the $ugrizJHK_s$ bands, respectively. We propagate this zero-point uncertainty to variations in our resulting $n(z)$. The key physical effect these uncertainties relate to is the interpretation of the 4000\angstrom~break of the Deep Field galaxies in a particular deep band. As shown in Appendix~\ref{app:pit}, we find empirically that the uncertainty in $n(z)$ due to this effect is most pronounced at redshifts corresponding to transitions of the 4000\angstrom~break between the deep photometric filters, as expected, and that the $u$ band zero-point uncertainty dominates, increasing the uncertainty at lower redshift. We summarize briefly the method for measuring and propagating this uncertainty here and present greater detail in Appendix~\ref{app:pit}.

We draw samples of deep-field magnitude zero-point offsets from a Gaussian with standard deviation equal to the photometric zero point uncertainty in the Y3 deep fields catalogue in the relevant band as measured by \citet*{y3-deepfields}.
For each zero-point-error realization, we perturb all magnitudes in the mock \balrog catalogue with these zero-points and re-run the \sompz procedures to generate a perturbed $n(z).$ In this way we generate a full ensemble of $n(z)$s reflecting the uncertainty of our redshift calibration due to the photometric calibration. 

\begin{figure*}
\centering
\includegraphics[width=\linewidth]{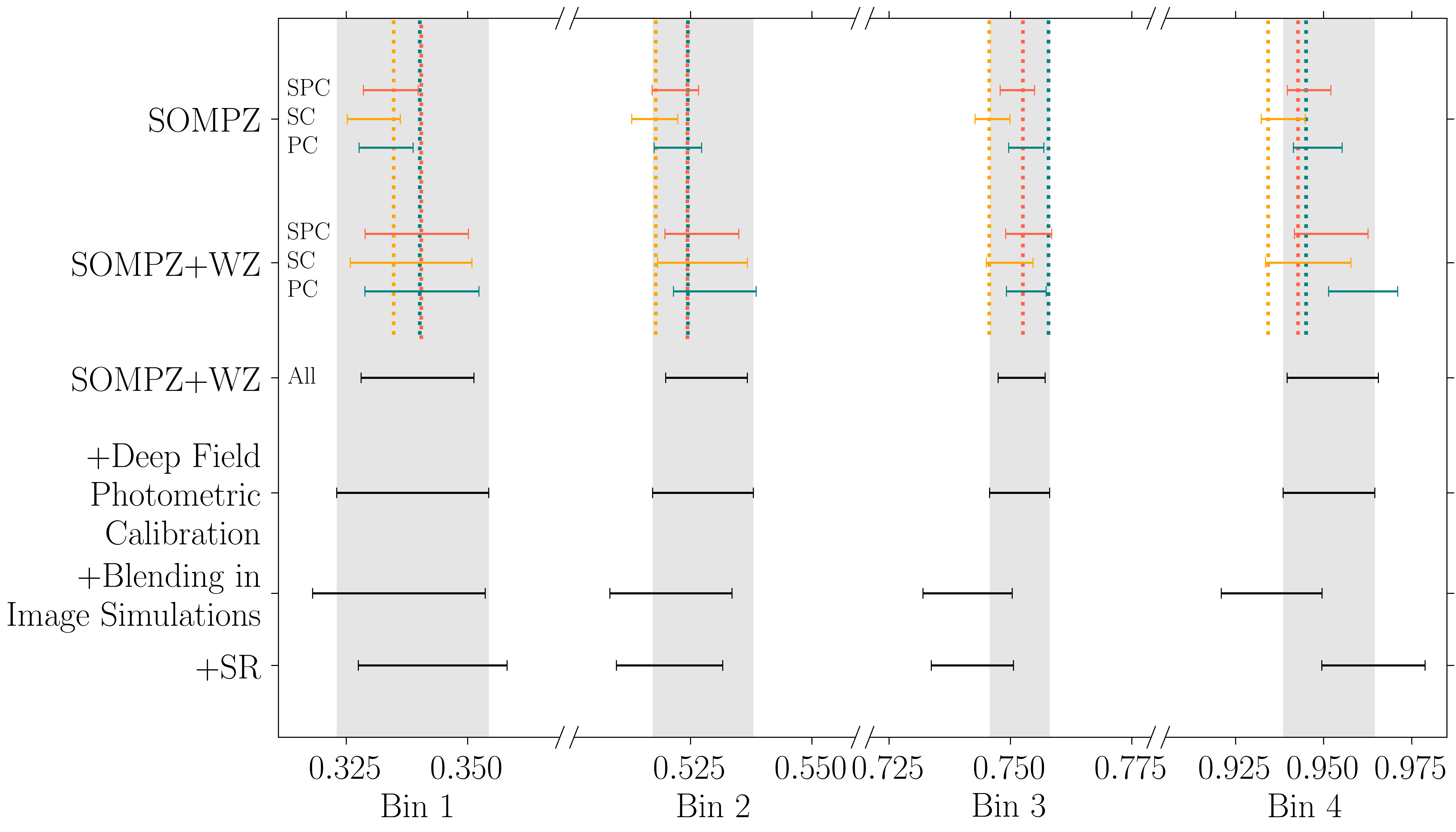}
\caption{Mean redshifts of each tomographic bin for each of the fiducial redshift samples at each stage of the analysis. Vertical dotted lines indicate the mean redshift in each bin from the $n(z)$ output of \sompz, given a particular Redshift Sample. The horizontal intervals indicate the 68 per cent confidence intervals on the mean as estimated according to the methods described in \S \ref{sec:uncertainty}, some of which shift the mean redshift. The larger uncertainty on the mean from the \sompz+WZ ensemble relative to the \sompz ensemble can be attributed to the different sample variance model used to combine \sompz with WZ (\sdirmfwz, rather than \sdir, see \S \ref{sec:3sdir_mfwz}). For details on the modification to incorporate the effect of blending as measured by image simulations see \citet{y3-imagesims}.} \label{fig:redshift_range}
\end{figure*}

It then remains to transfer the variation among the $n(z)$s in this simulation-based ensemble to a corresponding data-based ensemble of $n(z)$ distributions. We implement a novel application of Probability Integral Transforms (PITs) to achieve this. This PIT method transfers the variation encoded in the ensemble from simulated $n(z)$ (ensemble A) to our fiducial data result to ultimately yield a second ensemble (ensemble B). In brief, we achieve this by transferring the difference between the values of the quantile function of each realisation. For the details of this implementation, see Appendix \ref{app:pit}. The impact of this source of uncertainty is shown in Fig. \ref{fig:redshift_range} and \ref{fig:barsummary} and documented in Table \ref{tab:uncertainty}.

\subsection{\balrog Uncertainty} \label{sec:balrogunc}

Recall that we use the \balrog software (see \S~\ref{sec:data_DF_and_balrog}) to empirically estimate the relation between wide and deep field colours, $p^{\B}(c,\chat)$. The marginal distributions $p^{\B}(\chat)$ and $p^{\B}(c)$ from \balrog are not important, (they are measured from the Deep and Wide Samples), but the transfer function, $p^{\B}(c,\chat)/(p^{\B}(c)p^{\B}(\chat))$, is a potentially important source of uncertainty. The probability of observing certain wide colours $\chat$ given deep colours $c$ depends in general on the observing conditions present in the wide field. Observing conditions vary across the DES Y3 wide-field footprint, but for our cosmic shear analysis we are interested in the \textit{average} $n(z)$ across the footprint. Since \balrog injects galaxies with tiles placed at random across the DES Y3 wide-field footprint (covering about $\sim20$ per cent of it), we are fairly sampling the distribution of observing conditions present in the wide field.

To verify that the average transfer function from \balrog is well-estimated, we bootstrap the \balrog galaxies by their injected position in the wide field. First, we create 100 subsamples by grouping the injected position using the \textsc{kmeans\_radec}\footnote{\url{https://github.com/esheldon/kmeans_radec}} software. Then, we draw the same number of subsamples with replacement, use them to recompute the average transfer function and calculate the \sompz $n(z)$. We repeat this process 1000 times and find the dispersion in mean redshift to be smaller than $10^{-3}$ in all tomographic bins. Therefore, we conclude that the internal noise in the average \balrog transfer function is negligible, and consider $f^\B_{c\chat}=N^\B_{c\chat}/N^\B$ to be true (with $f^\B_{c\chat}$ from Equation~\ref{eqn:redshift_coeffs}).

Three of the DES deep fields (C3, E2, X3) overlap with the DES Y3 wide field, which we can use to construct a galaxy sample of position-matched wide-deep photometry pairs. We refer to this galaxy sample as \textsc{wide-deep}. We can empirically estimate the transfer function using the deep and wide colours observed in this catalogue. We do not use this transfer function for our fiducial result because it is computed from one realisation of the deep and wide mapping that happens with the particular wide-field observing conditions found in the deep fields, which are a much smaller area than the overall wide-field footprint. However, we can compare the \balrog and \textsc{wide-deep} transfer functions and their impact on the mean redshift to see if they are reasonably in agreement, subject to the limitations just mentioned. 

For this test, we estimate the \balrog transfer function using only injected deep field galaxies that are also present in the \textsc{wide-deep} Sample. We can simulate the uncertainty due to varying observing conditions of the \textsc{wide-deep} transfer function using \balrog subsamples similar to the \textsc{wide-deep} Sample. However, \balrog galaxies are injected at one-fifth of the density of real galaxies, so we can either reproduce a \textsc{wide-deep}-like Sample with the same number of objects and five times the area, or the same area but one fifth of the number of objects. The uncertainty of the first will be smaller than the real uncertainty of the \textsc{wide-deep}, while the uncertainty of the second case would be larger. We choose the former, which yields a lower limit on the uncertainty due to variable observing conditions. 

We find the difference in mean redshift $\Delta_{\zbar}$ between using the \balrog or the \textsc{wide-deep} transfer functions to be within $\sim2\sigma_{\zbar}$ of the distribution of simulated \textsc{wide-deep} Samples: $(\Delta_{\zbar}\pm\sigma_{\zbar})\times 10^3=[-2.8\pm1.8;\,  3.6\pm1.4;\,  3.1\pm1.4;\,  8.2\pm4.8]$. Since the estimated value of $\sigma_{\zbar}$ is a lower limit, we conclude the difference is consistent with the expected variance from observing conditions.

\subsection{\sompz Method Uncertainty}
\label{sec:sompz_method_unc}
As shown in \S~\ref{sec:method_validation}, we find an intrinsic error on the mean redshift predicted by \sompz when we compare it to the true mean redshift across 300 \buzzard realisations. This inherent method uncertainty, like our zero-point calibration uncertainty, is incorporated into our $n(z)$ ensemble using the PIT method, albeit in a much simpler way: we can incorporate this uncertainty by shifting each probability integral transform by a value drawn from a Gaussian with zero mean and a standard deviation equal to the root-mean-square of these mean offset values, $0.003$.

We note that this ensemble is made with an assignment of wide SOM cells to tomographic bins that is fixed for all realizations. Additionally, this method uncertainty is necessarily produced from runs with finite sample sizes, meaning there is some statistical contribution to the resulting estimate of systematic uncertainty.

\subsection{Summary of sources of uncertainty}

\begin{figure}
\centering
\includegraphics[width=\linewidth]{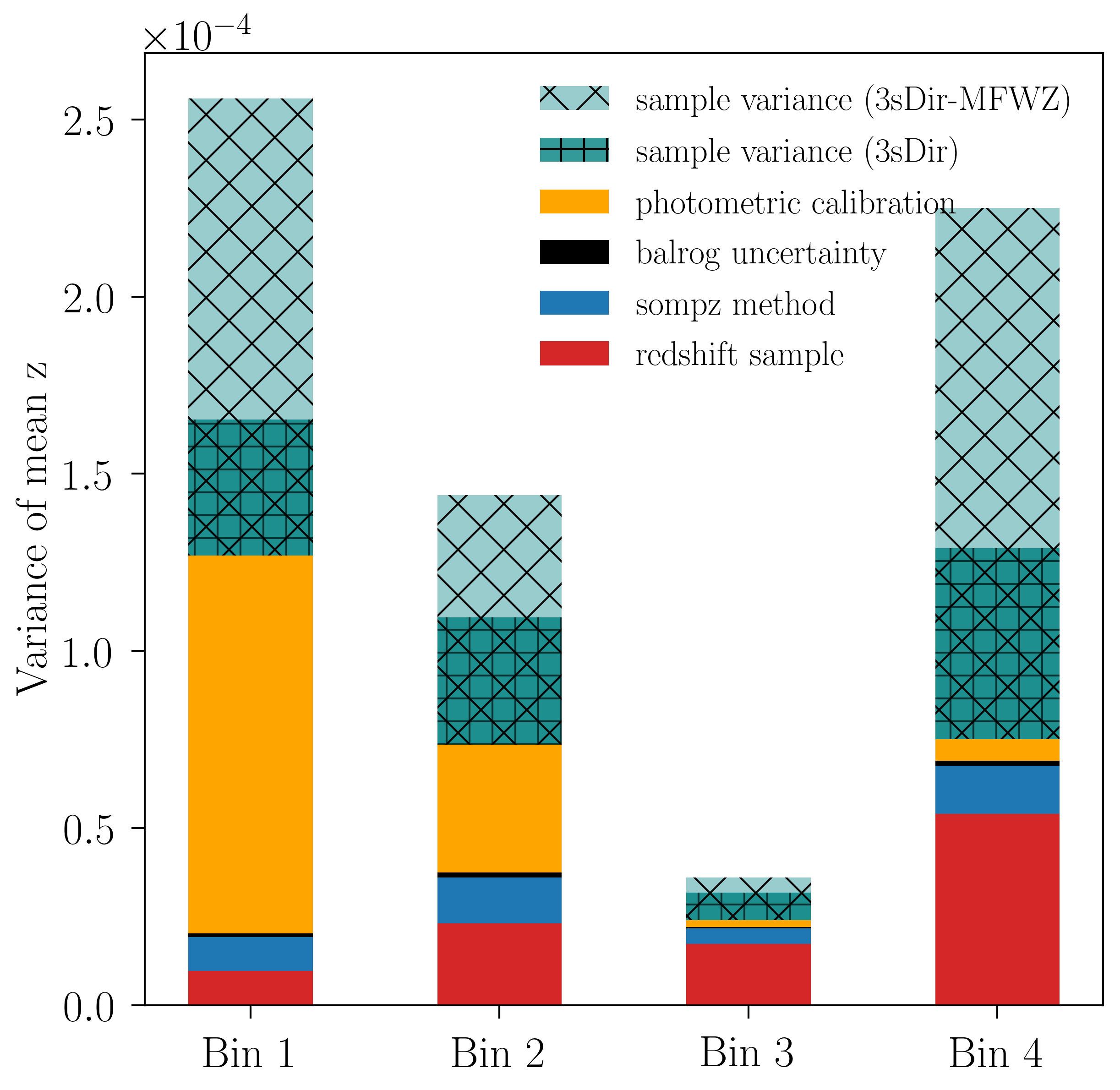}
\caption{Variance of each source of uncertainty in each tomographic bin. Note, the bar symbols indicating contributions from \sdir and \sdir-MFWZ start at the same value for each bin, but \sdir-MFWZ extends to a larger total uncertainty. The larger uncertainty estimated by \sdir-MFWZ is an artefact of the likelihood we must use to combine $n(z)$ constraints from \sompz and WZ (see \S \ref{sec:3sdir_mfwz}). As shown here the Redshift Sample uncertainty becomes a larger contributor to the uncertainty for higher redshift tomographic bins. Note, the contributing sources of uncertainty combine non-linearly. As a result, to illustrate the relative magnitude of each source of uncertainty in each bin, and the relative importance of each contributing source of uncertainty as a function of redshift, we rescale the total variance in this figure to match the combined uncertainty (see Table \ref{tab:uncertainty}).}
\label{fig:barsummary}
\end{figure}

\begin{table*}
\begin{threeparttable}[t]
    \centering
    \begin{tabular}{l c c c c c}
          & & Bin 1 & Bin 2 & Bin 3 & Bin 4\\
         
         $z^{\text{PZ}}$ range & & 0.0---0.358 & 0.358---0.631 & 0.631---0.872 & 0.872---2.0 \\
         \hline
         \hline
         $\langle z \rangle$ \sompz  & & 0.332 & 0.520 & 0.750 & 0.944\\
         $\langle z \rangle$ \sompz + WZ & & 0.339 & 0.528 & 0.752 & 0.952\\
        Effective $\langle z \rangle$ \sompz + WZ + Blending \tnote{a} & & 0.336 & 0.521 & 0.741 & 0.935\\
        Effective $\langle z \rangle$ \sompz + WZ + SR + Blending \tnote{b} & & 0.343 & 0.521 & 0.742 & 0.964\\
         \hline
         \hline
         \textbf{Uncertainty} & \textbf{Method} & & & & \\
         \hline
         \hline
         Shot Noise \& Sample Variance & \sdir & 0.006& 0.005 & 0.004& 0.006\\
         Redshift Sample Uncertainty & Sampling & 0.003& 0.004& 0.006& 0.006\\
         
         \balrog Uncertainty  & None & <0.001 & <0.001 & <0.001 & <0.001 \\
         
         Photometric Calibration Uncertainty  & PIT & 0.010 & 0.005 & 0.002 & 0.002\\
         
         Inherent \sompz Method Uncertainty & PIT & 0.003 & 0.003 & 0.003 & 0.003 \\
         Combined Uncertainty: \sompz (from \sdir)& - & 0.012 & 0.008 & 0.006 & 0.009\\
         \hline
         \hline
         Shot Noise \& Sample Variance & \sdir MFWZ & 0.011 & 0.007 & 0.005 & 0.010\\
         Combined Uncertainty: \sompz (from \sdirmfwz) & - & 0.015 & 0.010 & 0.007 &  0.012\\
         Combined Uncertainty: \sompz + WZ & - & 0.016 & 0.012 & 0.006 & 0.015 \\
         \hline
         \hline
         Effective Combined Uncertainty: \sompz + WZ +  Blending \tnote{a} & - & 0.018 & 0.015 & 0.011 & 0.017\\
         Effective Combined Uncertainty: \sompz + WZ + SR + Blending \tnote{b} & - & 0.015 & 0.011 & 0.008 & 0.015
    \end{tabular}
             \begin{tablenotes}
     \item[a] These values correspond to the $n(z)$ prior used in subsequent cosmological analyses.
     \item[b] These values correspond to the $n(z)$ posterior from a SR-only chain with fixed cosmology parameters. SR information is included in the cosmology analysis as an additional modelled data vector (see \S \ref{sec:shearratio} for more details).
     \end{tablenotes}
    \end{threeparttable}%
    \caption{Values of and approximate error contributions to the mean redshift of each tomographic bin at each stage of the analysis. We find that Sample Variance in the deep fields is the greatest contributor to our overall uncertainty for our fiducial result. The Shot Noise \& Sample Variance term here is computed with the \SPC sample. At low redshifts, the photometric calibration uncertainty is also significant, motivating improved work on the deep field photometric calibration. As expected, the uncertainty due to choice in Redshift Sample is a leading source of uncertainty for the third and fourth bins, motivating follow-up spectroscopic and narrow-band photometric observations. Note, the uncertainties combine non-linearly, so the combined uncertainties are not necessarily the quadrature sum of the contributing factors. Note, we label all results that incorporate blending as `Effective' because we expect non-zero shifts on the mean redshift due to blending (as discussed in \S \ref{sec:imagesims}), but we do not expect non-zero shifts on the mean redshift between \sompz and WZ.}
    \label{tab:uncertainty}
\end{table*}

In summary, we incorporate uncertainties due to the following sources into a final ensemble of redshift distributions. These results are summarized in Table \ref{tab:uncertainty} and illustrated visually in Fig. \ref{fig:redshift_range} and \ref{fig:barsummary}. We note that the individual contributing sources of uncertainty do not combine linearly. We report our best estimate of the uncertainty due to each factor considered in this section in Table \ref{tab:uncertainty}, but note that the combined uncertainty is less than the quadrature sum of these individual approximations. Fig. \ref{fig:barsummary} illustrates the relative magnitude of each source of uncertainty for each bin, and the relative importance of each source of uncertainty as a function of redshift. 

\begin{enumerate}
\item{Sample Variance}: This uncertainty is estimated and incorporated into our result as part of the \sdir formalism. This uncertainty is a main contributor to the uncertainty budget in all of our tomographic bins.
\item{Shot Noise}: This uncertainty is estimated and incorporated into our result as part of the \sdir formalism. 
\item{Redshift Sample Uncertainty}: This uncertainty is estimated by performing our inference with multiple different underlying redshift samples, and marginalizing over these choices by compiling their resultant $n(z)$ samples into a single ensemble. The uncertainty added by this marginalization is non-negligible in the third and fourth tomographic bins and dominant in the third tomographic bin.
\item{Photometric Calibration Uncertainty}: This uncertainty is estimated by running many times in simulations with offsets introduced to the galaxy photometry, and is incorporated into our result using PIT. This uncertainty is non-negligible in the first tomographic bin. 
\item{\balrog Uncertainty}: This uncertainty is estimated by replacing the transfer function $p(c|\hat{c})$ with an equivalent term estimated directly from galaxies for which we have independent deep and wide photometry, rather than using \balrog. This uncertainty is found to be negligible in all bins and is thus not propagated into our final resulting ensemble.
\item{\sompz method uncertainty}: This uncertainty is estimated by running many times in simulations, and is incorporated into our result using PIT. This uncertainty is found to be negligible in all tomographic bins but it nevertheless propagated into our final resulting ensemble.
\end{enumerate}

\section{Results}
\label{sec:results}
\subsection{Redshift Distribution Ensembles}

\begin{figure*}
\centering
\includegraphics[width=\linewidth]{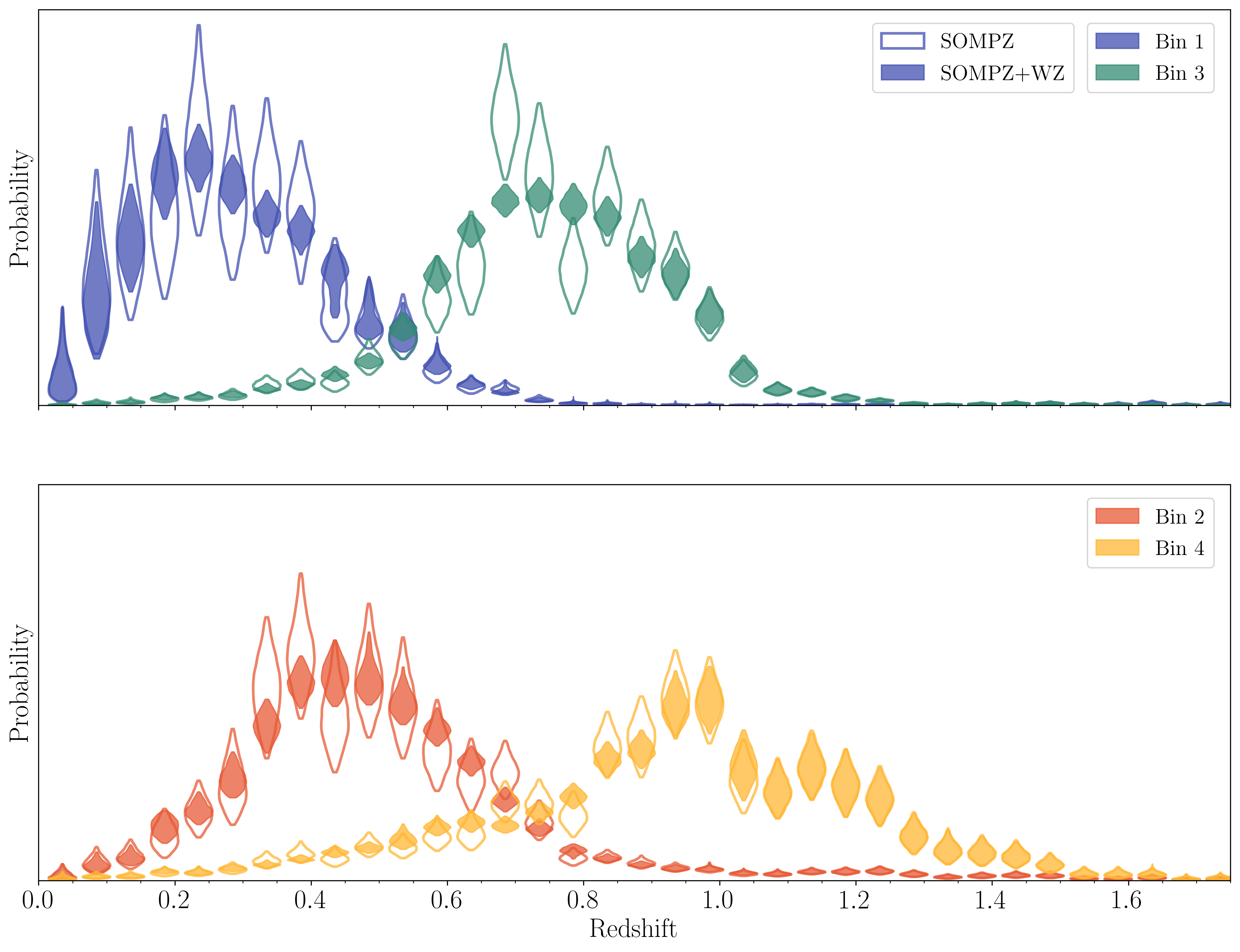}\caption{Visualization of the ensemble of redshift distributions in four tomographic bins, as inferred from \sompz only (open), and from \sompz combined with WZ (filled). Each violin symbol shows the 95 per cent credible interval of the probability of a galaxy in the weak lensing source sample and assigned to a given tomographic bin to have redshift $z$. The width at any part of a violin indicates the relative likelihood of $p(z)$ in that histogram bin. The uncertainty on p(z) is due to biases in the secure redshifts used in the analysis, sample variance and shot noise in the galaxies in the DES deep fields, photometric calibration uncertainty for the DES deep fields, and the inherent uncertainty of the methods applied. SR information is included in the cosmology analysis as an additional modelled data vector whose effect on the $n(z)$ can be quantified in terms of shifts on the mean redshift (see \S \ref{sec:shearratio} for more details). The low probability region of \sompz-only near $z\sim 0.75$ is due to an imprint of large-scale structure in the COSMOS field, as illustrated by the abundance of spectroscopic and photometric redshifts available in that region in Fig. \ref{fig:RS_deep_imaghist}.} \label{fig:nz_ensemble}
\end{figure*}

The results of the combined redshift calibration techniques are shown in Fig. \ref{fig:nz_ensemble}. We show the ensemble produced by \sompz as well as the ensemble constrained by the addition of WZ. Notably, our knowledge of the uncertainty on our measurement is not limited to the mean redshift, or any other finite set of moments of the distributions. Rather, the ensemble of redshift distributions effectively defines a full probability distribution function for the $p(z)$ of each histogram bin, as illustrated by the violin plots of Fig \ref{fig:nz_ensemble}.
Visual inspection of the \sompz-only distributions show that they are often not smooth functions of $z$.  This is expected because the \sdir likelihood (and similar \sdirmfwz likelihood) aims to raise the uncertainties to the levels expected from sample variance, but does not force the resultant distributions to be smooth.
 The filled violins include WZ information, which heavily favors smooth $n(z)$ in the $0.1<z<1.0$ region where WZ data are available.  The smoother nature of the ensemble after incorporating WZ demonstrates the valuable independence of that probe and its lesser reliance on biased redshift samples. SR information is included in the cosmology analysis as an additional modelled data vector whose effect on the $n(z)$ can be quantified in terms of  shifts on the mean redshift (see \S \ref{sec:shearratio} for more details).

\subsection{Consistency Of Independent Redshift Distribution Measures}
\begin{figure*}
\centering
\includegraphics[width=\linewidth]{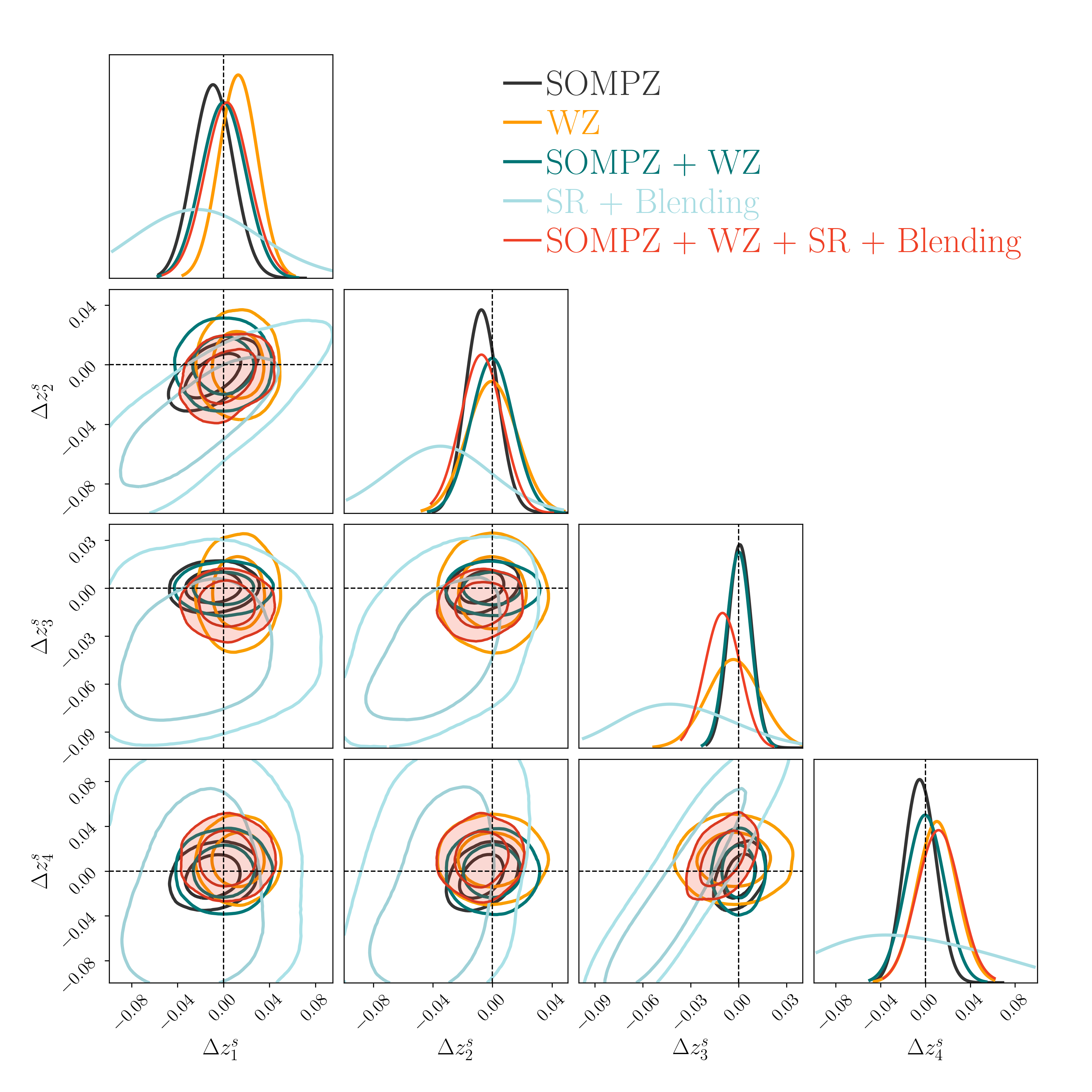}
\caption{Consistency of the measured mean redshift in each tomographic bin from the three inference likelihoods on data. Each axis represents the difference $\Delta \bar z$ in mean redshift $\bar z$ for a particular bin relative to the mean value of $\bar z$ in the \sompz+WZ ensemble. As noted in the text, $\bar z$ can only be calculated from WZ and SR information using a windowed (or weighted) average over $z,$ so this plot makes use of such windows where necessary. As shown by the light-blue contour, the inclusion of information from the ratios of the shear-position correlation functions at small scales significantly reduces the uncertainty on the mean redshift in each tomographic bin. Note the contours including SR information have additional uncertainty due to incorporating the effect of blending, thus leading to the false appearance that our combined \sompz+WZ+SR constraint is less constraining than \sompz, WZ, and SR individually. \label{fig:consistency_data}}
\end{figure*}

Fig. \ref{fig:consistency_data} demonstrates consistency among the distinct sources of information used to determine $n(z)$, namely \sompz (colour-magnitude), WZ (clustering), and SR (shear ratios).  A formal consistency check is complicated by the fact that the methods do not constrain common directions in the space of all possible $n(z)$'s.  We choose to compare $\bar z,$ the mean of $n(z)$, and define $\Delta z$ here as the shift of $\bar z$ relative to the mean $\bar z$ of the \sompz+WZ ensemble.  Even with this simplification there are complications, e.g. WZ can only constrain $n(z)$ (and hence its mean) as restricted to the range in $z$ where adequate reference samples exist.  Similarly, SR data measure redshift with an implicit weight related to lensing efficiency functions.  The $\Delta z$ values are plotted by always applying matching redshift windows to both \sompz and the sample under study.

On this basis, we find consistency between the three methods, as well as the combinations thereof. While the constraints on the mean redshift in each tomographic bin from shear ratios are broader than from \sompz, the relative independence of this information yields significantly more precise combined constraints on these means. The WZ constraints on $\bar z$ are weaker than those from \sompz, but as detailed in \citet*{y3-sourcewz}, the WZ data are much more powerful in constraining the shape and smoothness of $n(z)$ than in constraining the mean. This is illustrated directly by comparing the \sompz ensemble to the \sompz+WZ ensemble in Fig. \ref{fig:nz_ensemble}.

Further, because the shear signals measured in the SR analysis are subject to systematic observational effects described in \citet{y3-imagesims}, we expect a certain degree of inconsistency between SR and \sompz. Overall, however, within the reported uncertainties we find that these three likelihood functions can be combined. As described in \S \ref{sec:shearratio}, the SR information is included in the cosmology chains as an additional data vector, where the SR model is evaluated alongside the cosmological and nuisance parameters of the Y3 lensing analyses. As a result, the uncertainty we report in Table \ref{tab:uncertainty} is not a prior on the uncertainty in the mean directly used in the cosmological Markov chains, but the posterior from a SR-only chain where SOMPZ+WZ is used as the $n(z)$ prior, the cosmological parameters are fixed, and the nuisance parameters are varied within their priors.

\section{Discussion}
\label{sec:discussion}

We derive constraints on the redshift distributions of the DES Y3 lensing source sample from the combination of wide field photometry \citep*{y3-gold, y3-shapecatalog}, deep field photometry \citep*{y3-deepfields}, artificial DES wide field photometry \citep{y3-balrog}, and high quality photometric and spectrocopic redshifts, using and updating the methodology of \citet*{y3-sompzbuzzard}. When quantifying the full uncertainty, including sample variance, the choice of Redshift Sample, calibration uncertainty of the photometric deep fields, and necessary assumptions made in the method, we find small errors ($\sigma_{\langle z \rangle} \sim 0.01$) on the mean redshift of each of the four tomographic bins. Within their joint errors, these redshift distributions are consistent with estimates from cross-correlation of galaxies with high-quality redshift reference samples \citep*{y3-sourcewz} and with the ratios of small-scale galaxy-galaxy lensing signals \citep*{y3-shearratio}, which we incorporate for a joint estimate of $n(z)$s. Similar to \cite{Hildebrandt2017}, we also quantify the  full uncertainty in $n(z)$ shape which, while for many applications being subdominant to the uncertainty in  mean redshift, can be fully propagated to parameter constraints from DES Y3 lensing analyses \citep{y3-hyperrank}. We note in this context that \sdir is the first analytic model whose samples are full-shape $n(z)$.

While these results are encouraging, it is useful to consider the limiting factors of our analysis to inform future work. There is not one single effect. Rather, we find that our uncertainty is dominated by photometric calibration uncertainty of the deep  fields at the low redshift end of our sample, and that sample variance in the deep fields and biases in the redshift samples dominate at higher redshifts. Future work should address these sources of uncertainties with targeted observing campaigns and development of new methods. In particular, the LSST science requirement specification for error on the mean redshift below 0.003 will require improvements on all counts. As discussed by \citet*{euclid_photoz_i, euclid_photoz_ii}, the overlap of LSST photometry with NIR photometry from the \textit{Euclid} survey \citep{euclid}, especially over joint deep fields, will enable methods like those used in our work for future lensing surveys (see also \citet*{euclid_capak, euclid_rhoades}). We enumerate several opportunities for improving weak lensing redshift calibration below:

\begin{enumerate}
    \item \textbf{Spectroscopic follow-up targeting SOM cells}: The deep SOM constructed for this work defines a map of 8-band colour space which can be used to design future spectroscopic surveys. Many, but not all, cells defined by this SOM are populated with spectroscopic redshifts.  Particularly for the 8-band or 9-band (including the NIR $Y$ band, for example) colour space spanned by deeper lensing samples, the fraction of cells covered by spectroscopy is expected to decrease, the number of spectroscopic redshifts per cell is expected to decrease, and the magnitude range, at fixed colour, spanned by spectroscopic observations is expected to not match the magnitude range of the lensing sources. Follow-up observations should prioritize deep SOM cells with few redshifts, or with highly discrepant redshifts, as done by \cite{C3R2_DR2}. Larger samples per cell will be required to address any degeneracies remaining at fixed colour, and to calibrate the effects of magnitude-dependent incompleteness at fixed colour. 
    \item \textbf{Narrow-band imaging}: Narrow-band imaging can serve as a valuable complement to broadband imaging and spectroscopic surveys. Narrow-band imaging offers the benefit of measuring relatively high wavelength resolution data for surveys of large fields without selection biases. Given the intractable nature of selection biases in spectroscopic redshift samples, narrow-band imaging can serve a key role in breaking degeneracies of the colour-redshift relation for the large regions of colour-magnitude space sampled by weak lensing surveys. Given the dominance of redshift sample uncertainty in the most cosmologically constraining bins, redshifts informed by narrow-band imaging may prove key to meeting the LSST redshift calibration requirements. (see e.g. \cite{Alarcon2020, Benitez2014})
    \item \textbf{Improved transfer function}: A key innovation of this work is the construction of a transfer function encoding the probabilistic relation between deep and wide field photometry. While this transfer function was validated to be a negligible contribution to our uncertainty, it could be improved by injecting across a larger fraction of the wide-field survey footprint to probe more variation in survey properties. Further, as described in \cite{y3-balrog}, the \balrog injection procedure itself could be improved by using galaxy image cutouts, rather than CModel fits, to account for the full diversity in galaxy image properties that exceeds what the CModel galaxy profile is able to describe.
    \item \textbf{Photometric calibration} uncertainty leads to redshift uncertainty that, at low redshift, is dominated by the DES deep field $u$-band calibration. Reducing the uncertainty on the $u$-band zero point can significantly aid redshift calibration. Additional $u$-band data collected after the DES Y3 Deep Fields effort will enable an improved photometric calibration in future work.
    \item \textbf{Improved optimization schemes for incorporating magnitude} to the photometric information used in the Deep SOM: We construct the deep SOM with colour only, rather than colour-magnitude, following the finding by \citep*{y3-sompzbuzzard} that the addition of total flux (or magnitude) to the deep SOM does not improve the performance of \sompz (see their section 5.1). Depending on the survey photometric noise, it is in principle possible for there to be residual correlation between redshift and magnitude at fixed 8-band colour, as shown in fig. 4 of \citet{speagle2019}, but also possible for the addition of total flux (or magnitude) to worsen results because magnitude correlates more weakly with redshift than colour. We leave it to future work to perform additional tests including magnitude in the information used in the Deep SOM.
    \item In our analysis, the sample variance on the abundance of an 8-band colour in our Deep Field Sample is propagated throughout, but the abundance is not updated from wide-field information. A \textbf{hierarchical Bayesian model} can significantly reduce the Sample Variance on $p(c)$ by using $p(\hat{c})$ and the transfer function $p(c|\hat{c})$ to update and constrain $p(c)$ \citep{Leistedt2016,Sanchez2018}. Likewise, $p(z|c)$ can be further constrained with a hierarchical Bayesian model that includes clustering information from wide field galaxies \citep*{HBM_clustering}.
    \item Modeling $n(z)$ with \textbf{dependence on observing conditions}: variations in observing conditions over surveys remains a barrier to using the full non-homogeneous photometric data set collected by a given galaxy survey. To enable analysis of cosmic shear two-point functions in a survey with non-uniform depth, future work may require modeling lensing survey $n(z)$ from non-uniform catalogues (see e.g. \citealt*[][appendix B]{Hoyle2018}; \citealt{1910.11327}). Our formalism, by explicitly evaluating the observing-condition-dependent transfer function $p(c|\hat{c})$ naturally extends toward this goal. Future work can use \balrog to mock galaxies at varying levels of survey depth to match non-uniform surveys.
\end{enumerate}

DES Y3 has developed several new redshift calibration methods to facilitate advanced quantification of our uncertainties. Given our results, we conclude that future work for deeper lensing surveys such as DES Y6 and Stage-IV experiments such as the Legacy Survey of Space and Time (LSST) \citep{lsst_desc} must address these challenges to achieve the stated LSST science goal of uncertainty on the mean redshift below $~0.003$. In particular, we highlight the need for targeted spectroscopic and narrow-band photometric observations overlapping the LSST footprint. We emphasize the utility of our constructed SOMs to facilitate effective experimental design for such observations. Work to achieve these goals is underway, see e.g. \citet{euclid_photoz_specz_preparation, Masters2017}.
\section*{Acknowledgements}
This work was supported by the Department of Energy, Laboratory Directed Research and Development program at SLAC National Accelerator Laboratory, under contract DE-AC02-76SF00515 and as part of the Panofsky Fellowship awarded to DG. JM thanks the LSSTC Data Science Fellowship Program, which is funded by LSSTC, NSF Cybertraining Grant \#1829740, the Brinson Foundation, and the Moore Foundation; his participation in the program has benefited this work. Argonne National Laboratory's work was supported by the U.S. Department of Energy, Office of High Energy Physics. Argonne, a U.S. Department of Energy Office of Science Laboratory, is operated by UChicago Argonne LLC under contract no. DE-AC02-06CH11357.

Funding for the DES Projects has been provided by the U.S. Department of Energy, the U.S. National Science Foundation, the Ministry of Science and Education of Spain, 
the Science and Technology Facilities Council of the United Kingdom, the Higher Education Funding Council for England, the National Center for Supercomputing 
Applications at the University of Illinois at Urbana-Champaign, the Kavli Institute of Cosmological Physics at the University of Chicago, 
the Center for Cosmology and Astro-Particle Physics at the Ohio State University,
the Mitchell Institute for Fundamental Physics and Astronomy at Texas A\&M University, Financiadora de Estudos e Projetos, 
Funda{\c c}{\~a}o Carlos Chagas Filho de Amparo {\`a} Pesquisa do Estado do Rio de Janeiro, Conselho Nacional de Desenvolvimento Cient{\'i}fico e Tecnol{\'o}gico and 
the Minist{\'e}rio da Ci{\^e}ncia, Tecnologia e Inova{\c c}{\~a}o, the Deutsche Forschungsgemeinschaft and the Collaborating Institutions in the Dark Energy Survey. 

The Collaborating Institutions are Argonne National Laboratory, the University of California at Santa Cruz, the University of Cambridge, Centro de Investigaciones Energ{\'e}ticas, 
Medioambientales y Tecnol{\'o}gicas-Madrid, the University of Chicago, University College London, the DES-Brazil Consortium, the University of Edinburgh, 
the Eidgen{\"o}ssische Technische Hochschule (ETH) Z{\"u}rich, 
Fermi National Accelerator Laboratory, the University of Illinois at Urbana-Champaign, the Institut de Ci{\`e}ncies de l'Espai (IEEC/CSIC), 
the Institut de F{\'i}sica d'Altes Energies, Lawrence Berkeley National Laboratory, the Ludwig-Maximilians Universit{\"a}t M{\"u}nchen and the associated Excellence Cluster Universe, 
the University of Michigan, NFS's NOIRLab, the University of Nottingham, The Ohio State University, the University of Pennsylvania, the University of Portsmouth, 
SLAC National Accelerator Laboratory, Stanford University, the University of Sussex, Texas A\&M University, and the OzDES Membership Consortium.

Based in part on observations at Cerro Tololo Inter-American Observatory at NSF's NOIRLab (NOIRLab Prop. ID 2012B-0001; PI: J. Frieman), which is managed by the Association of Universities for Research in Astronomy (AURA) under a cooperative agreement with the National Science Foundation.

The DES data management system is supported by the National Science Foundation under Grant Numbers AST-1138766 and AST-1536171.
The DES participants from Spanish institutions are partially supported by MICINN under grants ESP2017-89838, PGC2018-094773, PGC2018-102021, SEV-2016-0588, SEV-2016-0597, and MDM-2015-0509, some of which include ERDF funds from the European Union. IFAE is partially funded by the CERCA program of the Generalitat de Catalunya.
Research leading to these results has received funding from the European Research
Council under the European Union's Seventh Framework Program (FP7/2007-2013) including ERC grant agreements 240672, 291329, and 306478.
We  acknowledge support from the Brazilian Instituto Nacional de Ci\^encia
e Tecnologia (INCT) do e-Universo (CNPq grant 465376/2014-2).

This manuscript has been authored by Fermi Research Alliance, LLC under Contract No. DE-AC02-07CH11359 with the U.S. Department of Energy, Office of Science, Office of High Energy Physics.

\section{Data availability} 
\label{sec:release}
The DES Y3 data products used in this work, as well as the full ensemble of DES Y3 source galaxy redshift distributions described by this work, will be made publicly available following publication, at the URL \url{https://des.ncsa.illinois.edu/releases}. 



\bibliographystyle{mnras_2author}

\bibliography{export-bibtex,y3kp}



\appendix

\section{Appendix on SOM}
\label{app:som}
\begin{figure*}
\centering
\includegraphics[width=\linewidth]{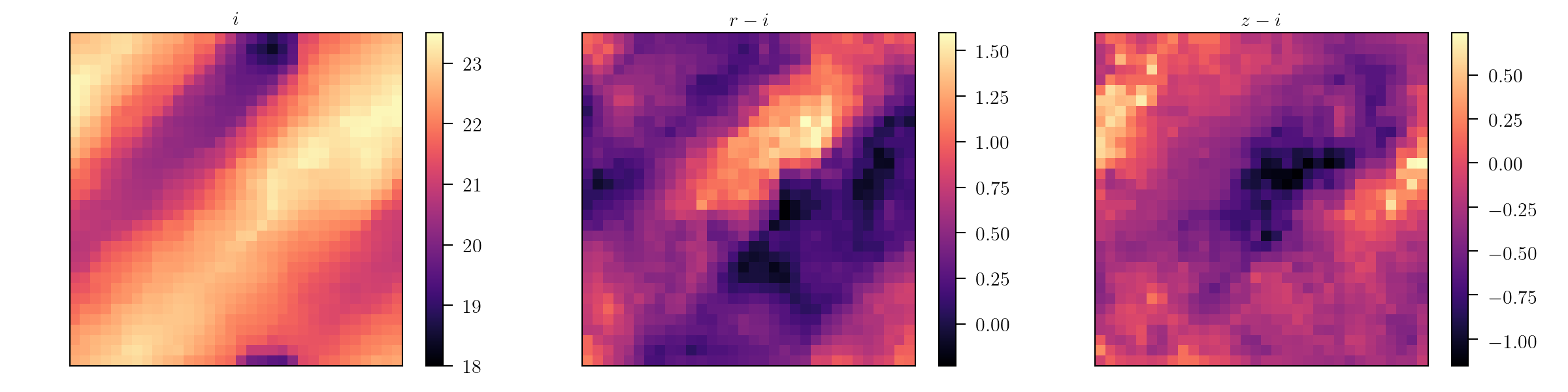}
\caption{Visualization of the wide Self-Organizing Map. Shown here are the mean $i$-band magnitude (left), the mean $r-i$ colour (middle), and the mean $z-i$ colour (right) of each cell in the wide SOM. The implementation of SOMs used in our analysis generates a toroidal map; in other words, the left and right edges of each map correspond to the same region of colour-magnitude space, as do the upper and lower edges. \label{fig:wide_som_colors}}
\end{figure*}

\begin{figure*}
\centering
\includegraphics[width=\linewidth]{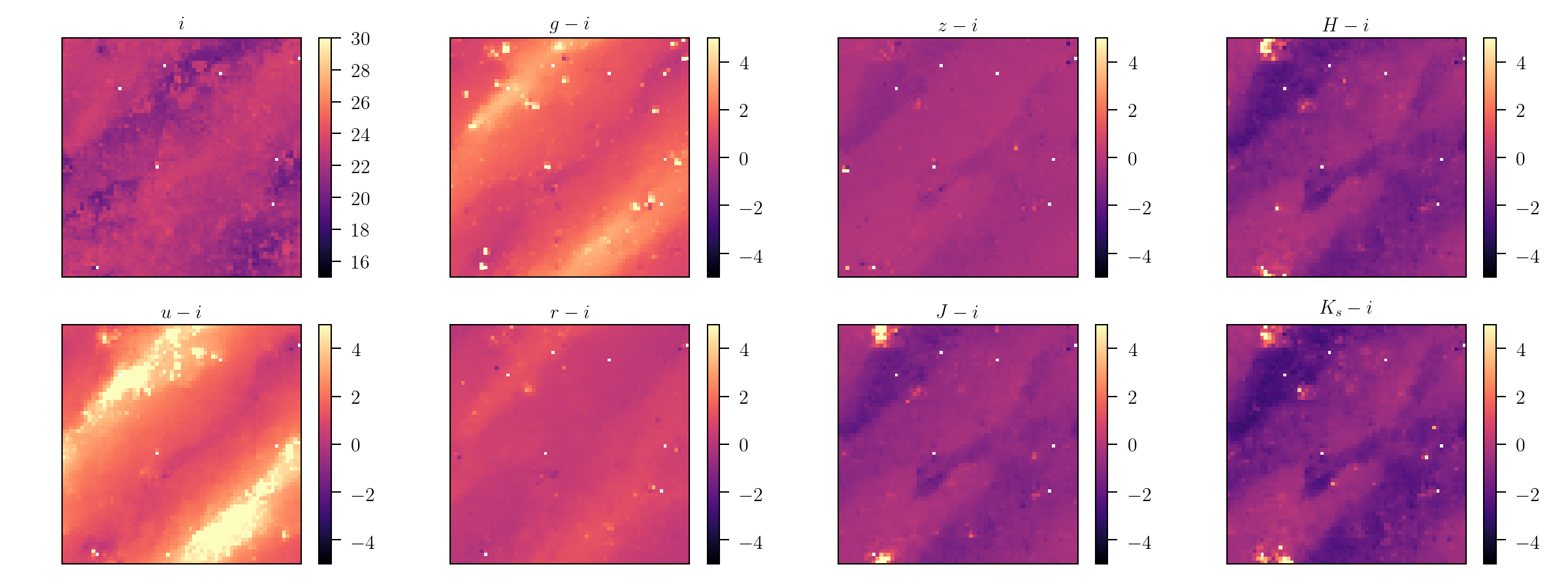}
\caption{Visualization of the deep field Self-Organizing Map. Shown here are the mean $i$-band magnitude (upper left) of each cell of the deep SOM, as well as each colour used in the deep SOM training. The implementation of SOMs used in our analysis generates a toroidal map; in other words, the left and right edges of each map correspond to the same region of colour-magnitude space, as do the upper and lower edges.}  \label{fig:deep_som_colors}
\end{figure*}

Fig. \ref{fig:wide_som_colors} and Fig. \ref{fig:deep_som_colors} show the $i$-band magnitude and colours of each wide and deep SOM cell, respectively. Given that the SOM training algorithm attempts to construct a smooth map in the full parameter space of the training inputs, we can interpret stark differences in adjacent cells as indirect indicators of degeneracies in the colour-redshift relation. Comparison with the upper right panel of Fig. \ref{fig:soms} indicates that the wide SOM cells with the broadest $p(z|\hat{c})$ tend to be cells with overall fainter galaxies, supporting the intuitive conclusion that our redshift constraints are weaker for fainter galaxies.
 
\section{SOMPZ Implementation Details}
\label{app:sompz_tech_details}
We enumerate here several technical details about the implementation of SOMPZ:

\subsection{SOM Training}

We note a few details about the SOM training algorithm here and refer the reader to \citet*{y3-sompzbuzzard} for a full treatment. We use the magnitude scale defined by \citet{Lupton1999} for our SOM training, which we call `luptitude'. The input vector of the Deep SOM is chosen to be a list of lupticolors with respect to the luptitude in $i$-band:

\begin{equation*}
\vec{x} = (\mu_{x_1}{-}\mu_i, ..., \mu_{x_7}{-}\mu_i),
\end{equation*}

where the bands $x_1$ to $x_7$ are $ugrzJHK$.  For the input vector of the Wide SOM, we also use lupticolors with respect to the luptitude in $i$ band, and we add the luptitude in $i$ band:

\begin{equation*}
\vec{\hat{x}} = (\mu_i, \mu_r{-}\mu_i, \mu_z{-}\mu_i).
\end{equation*}

In the case of the wide field, where only few colors are measured, \citet{y3-sompzbuzzard} find empirically that addition of the luptitude improves the performance of the scheme.

\subsection{Deep SOM Training Sample}

We find that training the deep SOM only on deep galaxies whose \balrog realisations are detected and selected by the weak lensing source selection function leads to a SOM with more precise $p(z)$.

\subsection{High Redshift Pile-up}

The redshift samples used contain galaxies with $p(3 < z < 6) > 0$. Although the resulting SOMPZ $n(z)$ with probability density at redshifts greater than three accurately reflect our estimate of the $n(z)$ given the information available, the relatively small probability in this high redshift region inconveniently increases the computation time needed to integrate over the $n(z)$ in cosmological likelihood Markov chains. To mitigate this effect, we shift all probability greater than a cut-off value of three to the final redshift bin at 3. The amount of probability beyond redshift 2 is less than one per cent in all cases: [0.0096, 0.0062, 0.0021, 0.0077].

\subsection{Ramping}
The DES Y3 $3\times2$pt. cosmological analyses sample over this ensemble in cosmological likelihood inference Markov chains. Importantly, we find that non-zero probability density near zero redshift ($p(z\approx0)> 0$) significantly increases the computation time necessary to efficiently sample parameter space due to high sensitivity of the intrinsic galaxy alignment (IA) model. This effect is most pronounced for the lowest redshift bin because this bin has the greatest $p(z\approx0)$. We alter the ensemble of redshift distributions post hoc to manually reduce $p(z\approx0)$ by multiplying the $p(z)$ up to $z = 0.055$ with a linear function. This choice is justified on the grounds of definitive prior knowledge that the source galaxy number density approaches zero as redshift approaches zero. Given that our analytic sample variance model does not account for this prior knowledge, we enforce this prior on the output $n(z)$. We additionally note that the ramping procedure is verified to preserve the mean redshift in the tomographic bin.

\subsection{Deep Field Noise Differentiation}
We note a test run on simulations in which the deep field photometric noise is set to different levels in COSMOS and the other DES deep fields. As for all runs in simulations, we set the noise levels by measuring the median noise levels in the corresponding data catalogues. We find no measurable difference on the mean redshift with this added realism relative to previous work.

\section{PIT Implementation}
\label{app:pit}
This subsection (\ref{sec:zeropointunc}) is dedicated to describing this novel method for transferring the variation in $n(z)$ as a result of our photometric calibration uncertainty and how we implement this method in practice.

The conceptual procedure for achieving this is described below and described in greater detail in Myles et al. in prep. We begin by computing the inverse cumulative distribution function (i.e. quantile function) $F^{-1}_i$ for each simulated realisation $n_i(z)$ in the ensemble labelled A. For a tomographic bin $\hat{b}$, this can be written as
\begin{equation}
    F^{-1}_{i,\hat{b}}(p) = \{z :  F_{i,\hat{b}}(z) = p \} \quad \text{with} \quad F_{i,\hat{b}}(z) = \int_{-\infty}^{z} n_{i,\hat{b}}(z^{\prime}) dz^{\prime} \; . 
\end{equation}

We construct each PIT by computing the difference of the inverse CDF of a given realisation $F^{-1}_{i,\hat{b}}$ with the average inverse CDF of the ensemble:
\begin{equation}
    \label{eqn:pit}
    \text{PIT}_{i, \hat{b}} = F^{-1}_{i,\hat{b}} - \langle F^{-1}_{\hat{b}} \rangle.
\end{equation}

Subtracting the average inverse CDF ensures that the mean redshift is not changed by the PIT. This is necessary because each realisation, in addition to having some zero-point offset introduced, is drawn from a noisy distribution due to (i) deep field photometric noise and (ii) mock-\balrog realisation noise in simulations. As a result of (i) and (ii), there would be a non-zero mean shift of the mean $z$ shifts of the ensemble of PITs if not for subtracting $\langle F^{-1}_{\hat{b}} \rangle$.

We apply these transformations to the data by simply adding each PIT to the inverse CDF of the fiducial data $n(z)$, $F^{-1}_{\text{data, fiducial}}$. The PIT due to one draw of zero-point offsets is determined and applied jointly to all tomographic bins:
\begin{equation}
    \label{eqn:pit_apply}
    F^{-1}_{i, \hat{b}, \text{data}} = F^{-1}_{\hat{b}, \text{data, fiducial}} + \text{PIT}_{i, \hat{b}}.
\end{equation}
Given this ensemble of inverse CDFs of the data $n(z)$, we construct the corresponding ensemble of data $n(z)$ by taking the inverse to yield CDFs, then differentiating:
\begin{equation}
\label{eqn:pit_differentiate}
    n_{i,\hat{b}}(z) = \dfrac{\mathrm{d}}{\mathrm{d}z} \left(F_{i, \hat{b}, \text{data}}\right).
\end{equation}

\begin{figure}
\centering
\includegraphics[width=\linewidth]{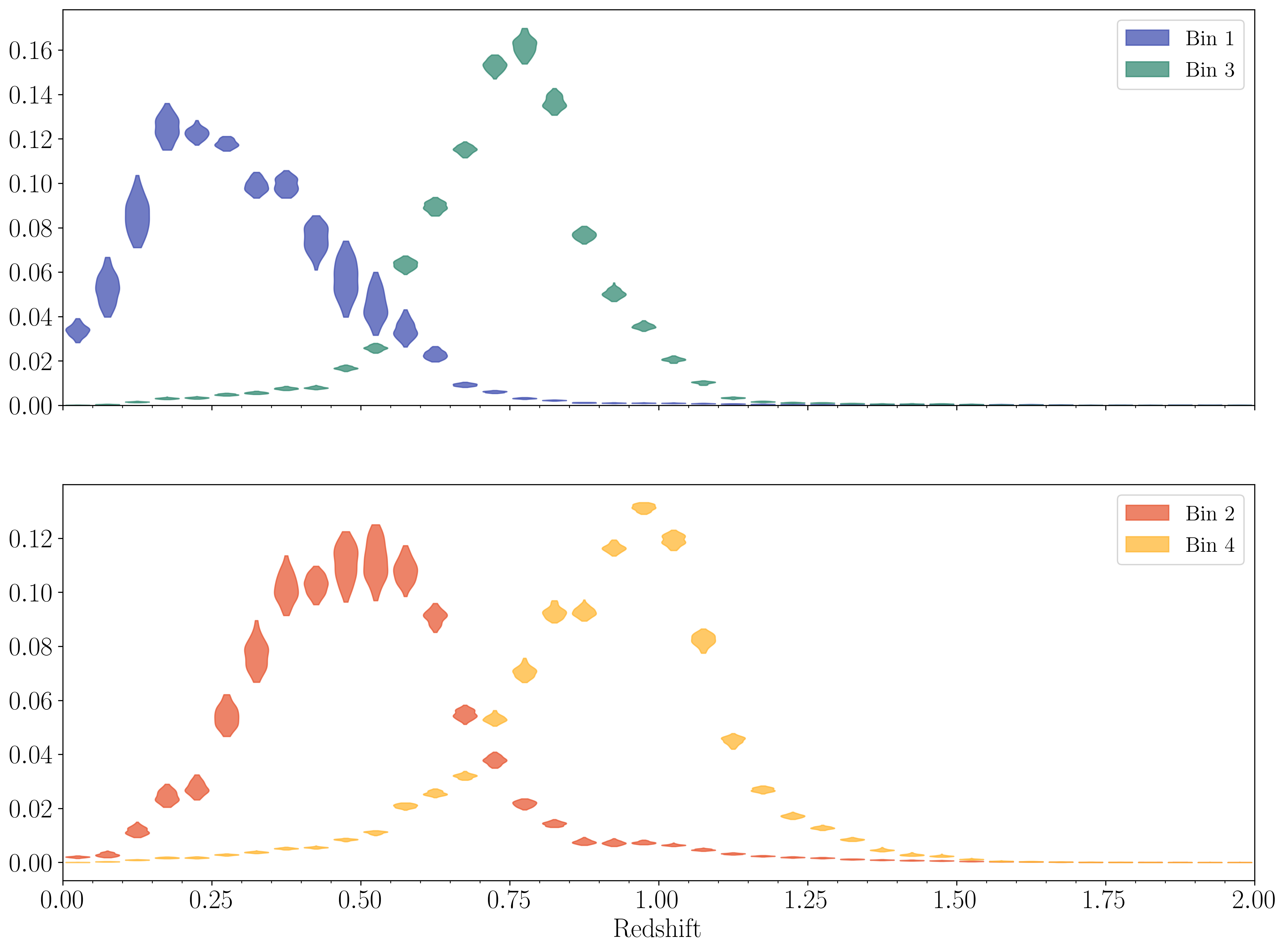}
\caption{Impact of the deep field photometric zero-point offset on estimated $n(z)$. The spread in values of $n(z)$ for any given histogram bin here reflect the propagated impact of the photometric zero-point uncertainty on $n(z)$. We determine the offset used in each band for each realisation by sampling a multivariate Normal distribution with standard deviations set to the zero-point uncertainty in each band. As shown here, some redshifts have much larger spread in $n(z)$ than others. The uncertain region in the interval $0<z<0.2$ corresponds to a redshifted $4000\angstrom$~break between 400 and 480nm, which is in the DES $g$-band filter. Likewise, the interval $0.4<z<0.6$ corresponds to a redshifted $4000\angstrom$~break between 560 and 640nm, which is in the DES $r$-band filter, and the transition to the $i$-band filter occurs at $z\sim0.75$.}\label{fig:pit_filter_transitions}
\end{figure}

Implementing the PIT offers two insights into our calibration uncertainty: first, our uncertainty is driven by the $u$-band calibration, and second, we find n(z) uncertainty increases at wavelengths corresponding to photometric filter transitions of the $4000\angstrom$ break, as shown in Fig. \ref{fig:pit_filter_transitions}. 

\section{\sdir model}
\label{app:3sdir_details}

Here we describe the formalism we use to model shot noise and sample variance in the redshift-colour probability from the deep fields to propagate it through to the redshift distribution of a tomographic bin.

Regarding the notation for the probability of redshift and colour space, note that $c$ and $\hat{c}$ are discrete variables, denoting regions in a partitioning of colour space. Following \citet[][see Eq.1-2 and \S 2]{Leistedt2016}, we will also adopt a piece-wise constant representation of probabilities in redshift space (essentially a probability histogram). In other words, we define any probability of a galaxy in our sample having redshift $z$, $p(z)$, with a finite set of coefficients $f_i$ of step functions $\Theta$,
\begin{equation}\label{eqn:basic_pz_as_f}
    p(z) \equiv  \sum_{i} \frac{f_i}{z_i-z_{i-1}}
    \times\Theta(z-z_{i-1})\Theta(z_i-z).
\end{equation}
In the remainder of the work, we will use the symbol $z$ to represent the discrete index of redshift bins divided at the $z_i$ bin edge values. Given this notation, we can represent the joint probability of colour and redshift with the set of coefficients $\{f_{zc}\}$.

We denote each dataset $D$ as: $\W$ for Wide, $\D$ for Deep, $\B$ for \balrog and $\R$ for Redshift. Note that $\R\subset\D$, and that $B$ contains several mock realisations of galaxies from $\D$ which have been injected and measured as in $\W$ using \balrog (see \S~\ref{sec:data}). We denote a set of coefficients $f$ that has been inferred from a dataset $D$ as $f^{D}$. For example, the coefficients of the joint redshift and colour distribution inferred from the Redshift Sample is denoted by $f^{\R}_{zc}$.

\subsection{Shot Noise} \label{sec:shotnoise}

We begin by rewriting Equation~\ref{eqn:redshift_prob_samples} using the $f$ coefficients notation, indicating which sample is used to infer each of the coefficients: 
\begin{equation} \label{eqn:redshift_coeffs}
    p(z|\hat{b}, \hat{s}) \approx \sum_{\hat{c} \in \hat{b}} \sum_{c} \frac{f^{\R}_{zc}}{f^{\R}_{c}} f^{\D}_{c} \frac{f^{\B}_{c\chat}}{f^{\B}_{c}f^{\B}_{\chat}} f^{\W}_{\chat},
\end{equation}
where $f^{\R}_{c}=\sum_{z} f^{\R}_{zc}$, $f^{\B}_{c}=\sum_{\chat} f^{\B}_{c\chat}$ and $f^{\B}_{\chat}=\sum_{c} f^{\B}_{c\chat}$. Following \citet[][see Section~3.1]{Leistedt2016}, we want to infer the parameters $(\{f^{\R}_{zc}\},\{f^{\B}_{\chat c}\},\{f^{\D}_{c}\})$ from the following sets of galaxy data
\begin{itemize}
\item $D^\R =  \{z_g,c_g\}^{\R}$ for $g=1\ldots N^{\R}$, 
\item $D^\B =  \{\chat_g,c_g\}^{\B}$ for $g=1\ldots N^{\B}$, 
\item $D^\D =  \{c_g\}^{\D}$ for $g=1\ldots N^{\D}$, and
\item $D^\W =  \{\chat_g\}^{\W}$ for $g=1\ldots N^{\W}$,
\end{itemize}
Let's start by assuming that the properties of these galaxies are known (\textit{i.e.} they are noiseless). Consider a scenario in which we ignore line-of-sight density variance, redshift errors, zero-point errors and other systematic uncertainties. In this scenario, a sufficient statistic for inferring the coefficients $(\{f^{\R}_{zc}\},\{f^{\B}_{\chat c}\},\{f^{\D}_{c}\})$, is the count of galaxies in each of the joint bins of redshift and SOM cells. Let's take for example the Redshift Sample, which we can reduce to the counts $D^\R \rightarrow \{N^\R_{zc}\}$ of galaxies detected in redshift bin $z$ and deep SOM cell $c$.

From Bayes' Theorem, the probability of these coefficients $\vf^{\R}\equiv\{f^\R_{zc}\}$  given the observed galaxy counts can be written as
\begin{equation} \label{pfDR}
\begin{split}
     p(\vf^{\R}|D^\R) &\propto p(D^\R|\vf^{\R})\,p(\vf^{\R}) \\
     &= p(\{N^\R_{zc}\}|\vf^\R) p(\vf^\R) \\
     &= p(\vf^R) \prod_{zc}\left(f^\R_{zc}\right)^{N^\R_{zc}}    ,
\end{split}
\end{equation}
and similarly for the other two sets of coefficients.  The likelihood function of the binned data $p(\{N^{\R}_{zc}\}|\vf^{\R})$ is a multinomial distribution by definition, under the assumption of independent selection of each galaxy. The conjugate prior for a multinomial likelihood function is a Dirichlet distribution, so if we choose our prior $p(\vf^{\R})$ to be a Dirichlet distribution with rate parameters $\alpha^\R_{zc}=\epsilon$,  then the posterior is also a Dirichlet distribution which depends only on the galaxy number counts (which makes the posterior analytical and easy to sample from). The Dirichlet distribution is also a natural prior because it enforces the constraints that $f^\R_{zc}>0$ for all $zc,$ $\sum f^\R_{zc}=1$ (required for any probability), and is invariant under any rearrangement of the $f$'s (it is agnostic to the meaning of the bins). The posterior Dirichlet distribution is
\begin{equation} \label{zt_coeff_dist}
\begin{split}
  p(\vf^{\R}|\{N^{\R}_{zc}\}) &= \Dir\left(\vf^\R;\{\alpha_{zc}\}\right) \\
  &=\Dir\left(\vf^\R;\{\alpha_{zc}=N^{\R}_{zc}+\epsilon\}\right) \\
  &\propto \delta\left(\sum_{zc}f^\R_{zc}-1\right) \prod_{zc} (f^{\R}_{zc})^{N^{\R}_{zc}-1+\epsilon},   
\end{split}
\end{equation}
where $\epsilon$ is a positive, small number to ensure that the Dirichlet distribution cannot get zero or negative counts as input (some of the $z,c$ counts will be zero).

For a large number of galaxies, the marginalized mean and variance of $f^{\R}_{zc}$ reduce to $N^{\R}_{zc}/N^{\R}$, which is the classical approximate histogram estimator (note that Dirichlet is the correct posterior distribution for a histogram, but a Gaussian distribution with $N^{\R}_{zc}/N^{\R}$ mean and variance become a good approximation). Equivalent expressions arise for the $f^\B$ and $f^\D$ coefficients. We note that, for the wide sample, we can consider $f^\W_{\chat}=N^\W_{\chat}/N^\W$ to be an exact result (not stochastic), because we are interested in the $p(z)$ for the realisation of the wide-field survey that we have, not the redshift distribution for a hypothetical infinite survey.

\subsection{Sample Variance} \label{sec:samplevariance}

Both Deep and Redshift Samples span a much smaller area than that of the DES Y3 wide source sample. Therefore, the underlying redshift distribution measured in the deep fields -- and since they are correlated, the measured colour distribution -- contain random  large-scale structure fluctuations particular to that volume, commonly referred to as \textit{sample variance}. We can describe the observed redshift distribution in the deep field as
\begin{equation}
    N_z^{\D} = \text{Poisson}\left[ N^{\D} f^{\W}_z (1+\Delta^{\D}_z) \right],
\end{equation}
with $f^{\W}_z$ as the underlying redshift distribution in the wide field, and $\Delta^{\D}_z$ the particular redshift fluctuation found in the deep field with respect to the wide field, with $\text{Var}(\Delta^{\D}_z)$ the sample variance. 

The Dirichlet sampling (Equation~\ref{zt_coeff_dist}, $\Dir(\vf^\R;\alpha_{zc}=N^\R_{zc}+\epsilon)$) described in \S~\ref{sec:shotnoise} only reproduces the variance expected from Poisson noise, but does not account of the additional uncertainty due to sample variance. In order to increase the variance of the Dirichlet sample, one can perform the transformation $\alpha_i\rightarrow  \alpha_i/\lambda$, which does not change the expected value of $f_i$ in the Dirichlet distribution, but does change its variance roughly as $\text{Var}(f_i)\rightarrow \lambda \text{Var}(f_i)$, for those coefficient indices $i$ for which $\alpha_i\ll \sum_i \alpha_i$. When the value of $\lambda$ is the ratio between total noise (sample variance and shot noise) over shot noise we obtain samples of $f_i$ with the larger, correct variance. However, for equally spaced redshift bins $\text{Var}(\Delta^{\D}_z)$ is a function of redshift (\textit{i.e.} sample variance becomes larger at lower redshift where the volume is smaller). A constant value $\lambda$ cannot increase the variance as a function of redshift as needed, but a value of $\lambda$ that changes as a function of redshift would bias the expected value of $f_i$.

However, we do not sample in the $f_z$ space, but in $f_{zc}$ space. Two phenotypes (different deep SOM cells $c$) that overlap in redshift have correlations due to the same underlying large-scale structure fluctuations. We will work under the assumption that different phenotypes at the same redshift have the same sample variance. As phenotypes are defined in observed colour space we do not expect large differences in their galaxy bias when they overlap in redshift, and we defer a more detailed study to future work. \citet[][S20 hereafter]{Sanchez2020} introduced a three-step Dirichlet sampling method (\sdir) which produces samples of $f_{zc}$ that can incorporate the correct level of sample variance as a function of redshift and correlate phenotypes that overlap in redshift. S20 validated in simulations that \sdir correctly reproduces the amount of sample variance expected in both the mean redshift and width of the redshift distribution of a wide field estimated from a smaller patch of the sky.

\subsubsection{\sdir method}
\label{sec:3sdir_original}

The \sdir method from S20 assumes the coefficients $f_{zc}$ are inferred from a single Redshift Sample, $f^{\R}_{zc}$. The \sdir method introduces the concept of a \textit{superphenotype} $T$ as a group of deep SOM cells that are close in redshift, such that the superphenotypes become nearly disjoint in redshift space. This allows us to introduce a redshift dependent parameter $\lambda$ (one $\lambda$ for each $T$) and correlate phenotypes that are close in redshift (different $c$ in the same $T$ are correlated). Following S20, we can write the probability of redshift and colour, with the superphenotypes $T$, as
\begin{align}
    p(z,c) &= \sum_T p(c|z,T)p(z|T)p(T)\\
    \label{eq:fzc_sample}
    f_{zc} &= \sum_T f_{c}^{zT}f_{z}^{T} f_T.
\end{align}
To produce a sample of the coefficients $\{f_{zc}\}$ we need to produce a sample of the coefficients $(\{f_{c}^{zT}\},\{f_{z}^{T}\},\{f_T\})$, which we infer from the observed Redshift Sample number counts in each $zcT$ bin, $N^{\R}_{zcT}$. Note that $\{f_{z}^{T}\}$ are independent from $\{f_T\}$, since the former is conditioned on $T$ (indicated by the superscript). Similarly, $\{f_{c}^{zT}\}$ are independent from both $\{f_{z}^{T}\}$ and $\{f_T\})$. Therefore, we can sample them separately from the observed counts. \sdir consists of drawing, in sequence, values of $\{f_T\}$ , then of $\{f_{z}^{T}\}$ , then of $\{f_{c}^{zT}\}$ with individual Dirichlet distributions from the appropriate galaxy counts, $\{N^{\R}_{T}\}$, $\{N^{\R}_{zT}\}$ and $\{N^{\R}_{zcT}\}$, respectively. However, we will rescale the counts used to infer the samples of both $\{f_T\}$ and $\{f_{z}^{T}\}$. This process increases the variance of the final $\{f_{zc}\}$ sample to the level expected for the sum of shot noise and sample variance, while keeping its expected value. In other words, we draw from the following distributions
\begin{subequations}\label{eq:threesteps}
\begin{align}
    \label{eq:step1}
    p(\{f_T\}|\{N^{\R}_{T}\}) \sim& \Dir\left(\{f_T\};\{\alpha_{T}=\frac{N^{\R}_{T}}{\bar{\lambda}}+\epsilon\}\right);\\
    \label{eq:step2}
    p(\{f_{z}^{T}\}|\{N^{\R}_{zT}\}) \sim& \Dir\left(\{f_{z}^{T}\};\{\alpha_{z}=\frac{N^{\R}_{zT}}{\lambda_T}+\epsilon\}\right) \quad \text{for each $T$};\\
    \label{eq:step3}
    p(\{f_{c}^{zT}\}|\{N^{\R}_{zcT}\}) \sim& \Dir\left(\{f_{c}^{zT}\};\{\alpha_{c}=N^{\R}_{zcT}+\epsilon\}\right)\quad \text{for each $z,T$}.
\end{align}
\end{subequations}
where
\begin{align}
    \label{eq:lambda_bar}
    \bar{\lambda} &\equiv \sum_z \lambda_z \frac{N^{\R}_{z}}{ N^{\R}},\\
    \label{eq:lambda_T}
    \lambda_T &\equiv \sum_z \lambda_z \frac{N^{\R}_{zT}}{N^{\R}_{T}},\\
    \label{eq:lambda_z}
    \lambda_z &\equiv \frac{\text{Var}(N^{\R}_{z})}{N^{\R}_{z}} = 1+N^{\R}_{z}\text{Var}(\Delta^{\R}_z).
\end{align}
Equation~\ref{eq:lambda_z}, $\lambda_z$, is the ratio of the total variance (shot noise and sample variance) to only the shot noise variance. When we infer $\{f_{z}^{T}\}$, the redshift counts for each superphenotype, $\{N^{\R}_{zT}\}$, are rescaled by a constant value equal to the average $\lambda_z$ ratio weighted by the superphenotype's redshift distribution: $\lambda_T$ (Equation~\ref{eq:lambda_T}). When we infer $\{f_T\}$, the counts $\{N^{\R}_{T}\}$ get rescaled by the average  $\lambda_z$ weighted by the sample redshift distribution, $\bar{\lambda}$ (Equation~\ref{eq:lambda_bar}). Overall, this noise-inflated Dirichlet sampling scheme (Equation~\ref{eq:threesteps}) is an approximate model of how sample variance affects the joint redshift and colour redshift distribution, which allows one to increase the variance as a function of redshift without introducing any bias (as noted in S20).

Finally, we estimate the sample variance term, $\text{Var}(\Delta^{\R}_z)$ (Equation~\ref{eq:lambda_z}), from theory following the same assumptions as in S20, which assumed a circular footprint of the same area as the Redshift Sample, which gives a prediction which is good at the $10-20$ per cent level, mostly due to the galaxy bias modeling (see S20 for more details, including small dependence of the prediction on cosmology). S20 validated the method in simulations, and then applied \sdir to the COSMOS field, which is the field of our Redshift Sample, so we directly use the sample variance prediction from S20.

\subsubsection{Sample variance in the Deep Sample}
\label{sec:3sdir_sv_deepsample}

In this analysis the Redshift Sample spans a smaller area than the whole deep field area, which carries additional information of the marginal distribution of colours, $p(c)$. We have four deep fields, $F=\{\text{COSMOS=COS, C3, E2, X3}\}$, so we can write the probability of $f_z$ conditioned on the counts from the four fields as
\begin{equation}\label{eq:4fields_3sdir}
\begin{split}
    p(f_z|N_z^{\mathrm{COS}}, N_z^{\mathrm{C3}}, N_z^{\mathrm{E2}}, N_z^{\mathrm{X3}}) \propto& p(N_z^{\mathrm{COS}}, N_z^{\mathrm{C3}}, N_z^{\mathrm{E2}}, N_z^{\mathrm{X3}}|f_z) p(f_z)\\
    \approx& p(N_z^{\mathrm{COS}}|f_z) p(N_z^{\mathrm{C3}}|f_z) \\
    &\times p(N_z^{\mathrm{E2}}|f_z) p(N_z^{\mathrm{X3}}|f_z) p(f_z)\\
    \propto& \mathrm{Dir}\left( \alpha_z=\sum_{F}\frac{N^{F}_{z}+\epsilon}{1+N^{F}_{z}\mathrm{Var}(\Delta^{F}_z)}\right),
\end{split}
\end{equation}
where in the second line of Equation~\ref{eq:4fields_3sdir} we approximate that the observed redshift number counts of each field $N^F_z$ are independent of each other. However we do not have complete redshift information in all fields: we have complete high-quality photometric redshift information in the COSMOS field, while we have incomplete and inhomogeneous spectroscopic coverage in all fields. For the purpose of modeling sample variance, one limit is to ignore the redshift information in the C3, E2 and X3 fields; and assume that the Redshift Sample is self-contained in the COSMOS field. Then, one can define the redshift number counts in any field by re-weighting the redshift information in the COSMOS field. In other words, we use
\begin{equation} \label{eq:nzfield}
    N_z^{F} \equiv \sum_c \left(\frac{N_{zc}^{\mathrm{COS}}}{\sum_{z^{\prime}} N_{z^{\prime}c}^{\mathrm{COS}}} N_{c}^{F}\right) \qquad \text{for }F\in\{\text{COS, C3, E2, X3}\}.
\end{equation}
The $N^F_z$ are independent from each other in the limit where there is a tight relation between redshift and deep color (\textit{i.e.} $p(z|c)$ is narrow) that is well determined in the Redshift Sample; and when the noise is dominated by the sample variance in the color distribution in each field, $N_{c}^{F}$.

We define the effective ratio of the total variance to only the shot noise in all the deep fields, $\lambda_z^{\mathrm{eff}}$, from Equation~\ref{eq:4fields_3sdir} as
\begin{equation}
\begin{split}
    \frac{\sum_{F} N^{F}_{z}}{\lambda_z^{\mathrm{eff}}} \equiv \sum_{F\in\{\mathrm{COS},\mathrm{C3},\mathrm{E2},\mathrm{X3}\}}\frac{N^{F}_{z}}{1+N^{F}_{z}\mathrm{Var}(\Delta^{F}_z)},
\end{split}
\end{equation}
where $\mathrm{Var}(\Delta^{F}_z)$ is defined by using the correct area of each field. We define $\bar{\lambda}^{\mathrm{eff}}$ as
\begin{equation}\label{lambdaeff}
   \bar{\lambda}^{\mathrm{eff}} \equiv \sum_z \lambda_z^{\mathrm{eff}} \frac{\sum_F N^{F}_{z}}{ \sum_F N^{F}}.
\end{equation}
In practice, the value of $\bar{\lambda}$ and $\bar{\lambda}^{\mathrm{eff}}$ is similar, since the decrease in sample variance (roughly inversely proportional to the area) is in part compensated by the increase in number counts (proportional to the area).

\subsubsection{Application of \sdir to DES Y3}
\label{sec:3sdir_app_desy3}

From Equation~\ref{eqn:redshift_coeffs}, we want to sample the following coefficients
\begin{equation} \label{eq:fzc_3sdiralt}
    f_{zc}\equiv\frac{f^{\R}_{zc}}{\sum_z f^{\R}_{zc}} f^{\D}_{c}.
\end{equation}
First, we sample the coefficients $\{f^{\R}_{zc}\}$ using only the Redshift Sample with the same \sdir formalism from \S~\ref{sec:3sdir_original}. Then, we separately sample the coefficients $\{f^{\D}_{c}\}$ using only the Deep Sample with the formalism that we now describe. Finally, we can compute the sample of coefficients $\{f_{zc}\}$ using Equation~\ref{eq:fzc_3sdiralt}, which replaces the sample from Equation~\ref{eq:fzc_sample}. 

To sample the coefficients $\{f^{\D}_{c}\}$, we write the probability of colour with the superphenotypes $T$ as 
\begin{align}
    p(c) &= \sum_T p(c|T)p(T);\\
    f_{c} &= \sum_T f_{c}^{T} f_T;
\end{align}
similar to Equation~\ref{eq:fzc_sample}. Then, we sample the coefficients $\{f_{c}^{T}\}$ and $\{f_T\}$ with
\begin{subequations}\label{eq:threesteps_2samples}
\begin{align}
    p(\{f_T\}|\{N^{\D}_T\}) \sim& \Dir\left(\{f_T\};\alpha_{T}=\frac{N^{\D}_T}{\bar{\lambda}^{\mathrm{eff}}}+\epsilon\right);\\
    p(\{f_{c}^{T}\}|\{N^{\D}_{cT}\}) \sim& \Dir\left(\{f_{c}^{T}\};\alpha_{c}=N^{\D}_{cT}+\epsilon\right)\quad \text{for each $T$}.
\end{align}
\end{subequations}
with $\bar{\lambda}^{\mathrm{eff}}$ from Equation~\ref{lambdaeff}.

\subsection{Bin conditionalization}
\label{sec:3sdir_bincond}

The sampling process described so far consists of drawing values for $f_{zc}$ (Equation~\ref{eq:fzc_3sdiralt}) which represents the term $f_{zc}=\frac{f^{\R}_{zc}}{f^{\R}_{c}} f^{\D}_{c}$ from Equation~\ref{eqn:redshift_coeffs} because it includes information from both the Deep and Redshift Samples. We already include the probablity that a galaxy is selected into the weak lensing sample in the counts $N^{\R}_{zc}$ and $N^{\D}_{c}$  that we input to the \sdir method (\textit{i.e.} each galaxy counts as a fraction equal to its \balrog detection probability). We draw one sample of $f_{zc}$ for all four tomographic bins, and we add the bin conditionalization (Equation~\ref{eqn:bincond}) by multiplying the fractional probability $g_{zc}$ that each $(z,c)$ bin is assigned to a tomographic bin $\hat{b}$ as measured from the counts:
\begin{equation}
    f^{\R,\bhat}_{zc}\equiv g^{\R,\bhat}_{zc} \times f^{\R}_{zc},
\end{equation}
where $g^{\R,\bhat}_{zc}$ is the fractional probability that galaxies from the Redshift sample end up in each tomographic bin according to \balrog,
\begin{equation} \label{g_zc_redshiftsample}
    g^{\R,\bhat}_{zc} \equiv \frac{\sum\limits_{\chat\in\bhat} N^{\R}_{z,c,\chat}}{\sum\limits_{\chat} N^{\R}_{z,c,\chat}}, \qquad \mathrm{and} \qquad \sum_{\bhat} f^{\R,\bhat}_{zc} = f^{\R}_{zc}.
\end{equation}
Similarly, we can also define for the deep sample
\begin{equation}\label{g_c_deepsample}
    f^{\D,\bhat}_{c}\equiv g^{\D,\bhat}_{c} \times f^{\D}_{c}; \qquad
    g^{\D,\bhat}_{c} \equiv \frac{\sum\limits_{\chat\in\bhat} N^{\D}_{c,\chat}}{\sum\limits_{\chat} N^{\D}_{c,\chat}}; \qquad  \sum_{\bhat} f^{\D,\bhat}_{c} = f^{\D}_{c}.
\end{equation}
We define an effective tomographic bin weight that we can apply to our sample $f_{zc}$, $g^{\bhat}_{zc}$, as
\begin{equation} \label{eqn:fzc_tomobin}
    g^{\bhat}_{zc} \equiv \frac{g^{\R,\bhat}_{zc}}{\sum_z g^{\R,\bhat}_{zc}} g^{\D,\bhat}_{c}  \quad \text{and then} \quad f^{\bhat}_{zc} = g^{\bhat}_{zc}\times f_{zc}.
\end{equation}
Whenever there are no Redshift galaxies measured in a bin and cell, we set the redshift distribution to the non-tomographic one (following Equation~\ref{eqn:nobincond} and the discussion in section~\ref{sec:sompz_formalism}). To summarize, we draw one sample of $f_{zc}$ and use the weight $g^{\bhat}_{zc}$ to compute the four tomographic bin samples $f^{\bhat}_{zc}$ (Equation~\ref{eqn:fzc_tomobin}).

\subsection{Lensing weights}
\label{sec:3sdir_lensingweights}

Similarly, to include the lensing and response weights from \S~\ref{sec:lensingweight_sompz} we define an averaged weight for each $(z,c)$ pair in the Redshift sample:
\begin{equation}
    \langle w^{\mathcal{R}}_{zc} \rangle \propto g^{\R,\bhat}_{zc} \sum_{i\in (z,c)} \left( \frac{1}{M_i} \sum_j w_{ij}   \right),
\end{equation}
where $w_{ij}$ is the lensing weight for the $j\text{-th}$ detection that passes \metacal selection of the $i\text{-th}$ deep field galaxy with redshift information; $M_i$ is the number of times galaxy $i$ has been injected into \balrog; $g^{\R,\bhat}_{zc}$ is the conditioned probability of each tomographic bin (Equation \ref{g_zc_redshiftsample}). We are also interested in the lensing averaged weight for each deep cell in the Redshift Sample: 
\begin{equation}
    \langle w^{\mathcal{R}}_{c} \rangle \propto \left(\sum_z g^{\R,\bhat}_{zc}\right) \sum_{i\in (c)} \left( \frac{1}{M_i} \sum_j w_{ij}   \right).
\end{equation}
Analogously, we define an averaged lensing weight for the Deep Sample:
\begin{equation}
    \langle w^{\mathcal{D}}_{c} \rangle \propto g^{\D,\bhat}_{c} \sum_{i\in (c)} \left( \frac{1}{M_i} \sum_j w_{ij}   \right),
\end{equation}
with $g^{\D,\bhat}_{c}$ from Equation \ref{g_c_deepsample}. Finally, we define the effective weight as
\begin{equation} \label{eq:fzc_lensingweight}
    \langle w^{}_{zc} \rangle \equiv \frac{\langle w^{\mathcal{R}}_{zc} \rangle}{\langle w^{\mathcal{R}}_{c} \rangle} \langle w^{\mathcal{D}}_{c} \rangle, \quad \text{so that} \quad f^{\bhat}_{zc} \rightarrow \langle w^{}_{zc} \rangle f^{\bhat}_{zc}.
\end{equation}
with $f^{\bhat}_{zc}$ from Equation~\ref{eqn:fzc_tomobin}. 

In summary, we obtain a sample of $f_{zc}$ from Equation~\ref{eq:fzc_3sdiralt} from \balrog-weighted counts of the Redshift and Deep fields, to which we apply a tomographic bin selection probability weight to obtain the coefficients for each tomographic bin, $f^{\bhat}_{zc}$, (Equation~\ref{eqn:fzc_tomobin}) and finally apply the lensing and response weight  (Equation~\ref{eq:fzc_lensingweight}).

\subsection{\sdir Modified for WZ (MFWZ)}
\label{sec:3sdir_mfwz}

To jointly sample from the \sdir likelihood from photometry from this paper and the clustering redshifts (WZ) likelihood from \citet*{y3-sourcewz} we have implemented a Hamiltonian Monte Carlo (HMC) algorithm, which is far more efficient than importance sampling \sdir samples with the WZ likelihood (see \citet*{y3-sourcewz} for details). However, we have implemented a modified version of the \sdir likelihood (MFWZ) for the HMC algorithm that we describe here.

The \sdir MFWZ likelihood samples using the equations for the Redshift Sample, Equation~\ref{eq:fzc_sample} and Equation~\ref{eq:threesteps}, and only incorporates the information from the deep-field colour counts during step 1 (Equation~\ref{eq:step1}). Accordingly, we also update the value of $\bar{\lambda}$ in Equation~\ref{eq:lambda_bar} with $\bar{\lambda}^{\mathrm{eff}}$ from Equation~\ref{lambdaeff}.
We sample $\{f_T\}$ from the colour counts from the deep field $\{N^{\D}_T\}$ with
\begin{equation}
    p(\{f_T\}|\{N^{\D}_T\}) \sim \Dir\left(\{f_T\};\alpha_{T}=\frac{N^{\D}_T}{\bar{\lambda}^{\mathrm{eff}}}+\epsilon\right). 
\end{equation}
The samples of $\{f_{c}^{zT}\}$ and $\{f_{z}^{T}\}$ are obtained from Equation~\ref{eq:step2} and Equation~\ref{eq:step3}. Finally one obtains the $\{f_{zc}\}$ sample from $(\{f_{c}^{zT}\},\{f_{z}^{T}\},\{f_T\})$ using Equation~\ref{eq:fzc_sample}.

Although \sdir MFWZ is using less information from the deep fields, we find it easier to implement in a HMC together with the WZ likelihood.

\subsection{ Known errors}
\label{sec:3sdir_errors}

During the processing of the \sdir and \sdirmfwz samples the following error was made. In bin conditionalization, when there is no Redshift galaxy that satisfies both $c$ and $\bhat$, we instead use the redshift information from any tomographic bin in that cell. In other words, we use Equation~\ref{eqn:nobincond} instead of Equation~\ref{eqn:bincond} as discussed in section~\ref{sec:sompz_formalism}. When implementing the lensing responses in \sdir we did not properly implement this last change, and in practice we always used Equation~\ref{eqn:bincond}. This produces a shift in the $n(z)$ average mean redshift equal to $\Delta_z=[0.003, 0.003, \sim 0, -0.004]$ (difference between the correct implementation minus the actual implementation). We note that the effect of this error is small compared to all other uncertainties included in the analysis.
\section{Validation of \sompz and \sdir}
\label{app:validation}

In order to validate the methodology of \sompz and \sdir we use the suite of \buzzard simulations. We note that the \buzzard simulations do not include simulated images, so we cannot test the lensing and response weight methods from \S~\ref{sec:lensingweight_sompz} in \sompz, nor \S~\ref{sec:3sdir_lensingweights} in \sdir. The validation of such weights is explored in \citep{y3-imagesims}, and we have verified that both the \sompz and \sdir weight implementations are consistent: the \sompz weights are applied individually to galaxies, while in \sdir they are applied as averaged quantities to $f_{zc}$. We have verified this change does not introduce biases larger than $10^{-3}$ in the mean redshift in any of the tomographic bins. 

We generate 300 versions of the four DES Deep Samples (where one of the four has perfect redshift information) at different random line-of-sight positions in the \buzzard simulations. For each of the 300 realisations of the deep fields, we run the \sompz algorithm and estimate the different simulated number counts, $N^{\D}_c$, $N^{\R}_{zc}$ and $N^{\B}_{c\chat}$, while the wide field remains constant. Then we obtain an $n(z)$ estimate for each tomographic bin by fixing the probabilities to the observed number counts.

To test the performance of the \sdir method we perform the following procedure in each of the 300 \buzzard realisations of the deep fields. We draw $10^4$ samples from Equation~\ref{eq:fzc_3sdiralt}, $\{f^{i}_{zc};\, i=1,\ldots,10^4\}$ to which we apply the bin conditionalization using Equation~\ref{eqn:fzc_tomobin}, and use Equation~\ref{eqn:redshift_coeffs} to obtain the $10^4$ $\{f^i_z;\, i=1,\ldots,10^4\}$ samples for each tomographic bin. From it, we estimate the mean redshift of each $f^{i}_z$ sample, $\zbar^{i}=\sum_z z f^{i}_z$, and its average value $\zbar^{\sdir}\equiv \langle \zbar^{i}\rangle$ in each \buzzard realisation. We also compute the $\zbar^{\text{SOMPZ}}$ value of the single $n(z)$ from \sompz in each realisation, which we obtain by fixing the probabilities to the number counts. In summary, we have 300 values of $\zbar^{\text{SOMPZ}}$ and $\zbar^{\sdir}$, and a total of $300\times10^4$ values of $\zbar^{i}$ whose variance reflects the uncertainty on the mean redshift per tomographic bin as estimated from \sdir.

\begin{figure}
\centering
\includegraphics[width=\linewidth]{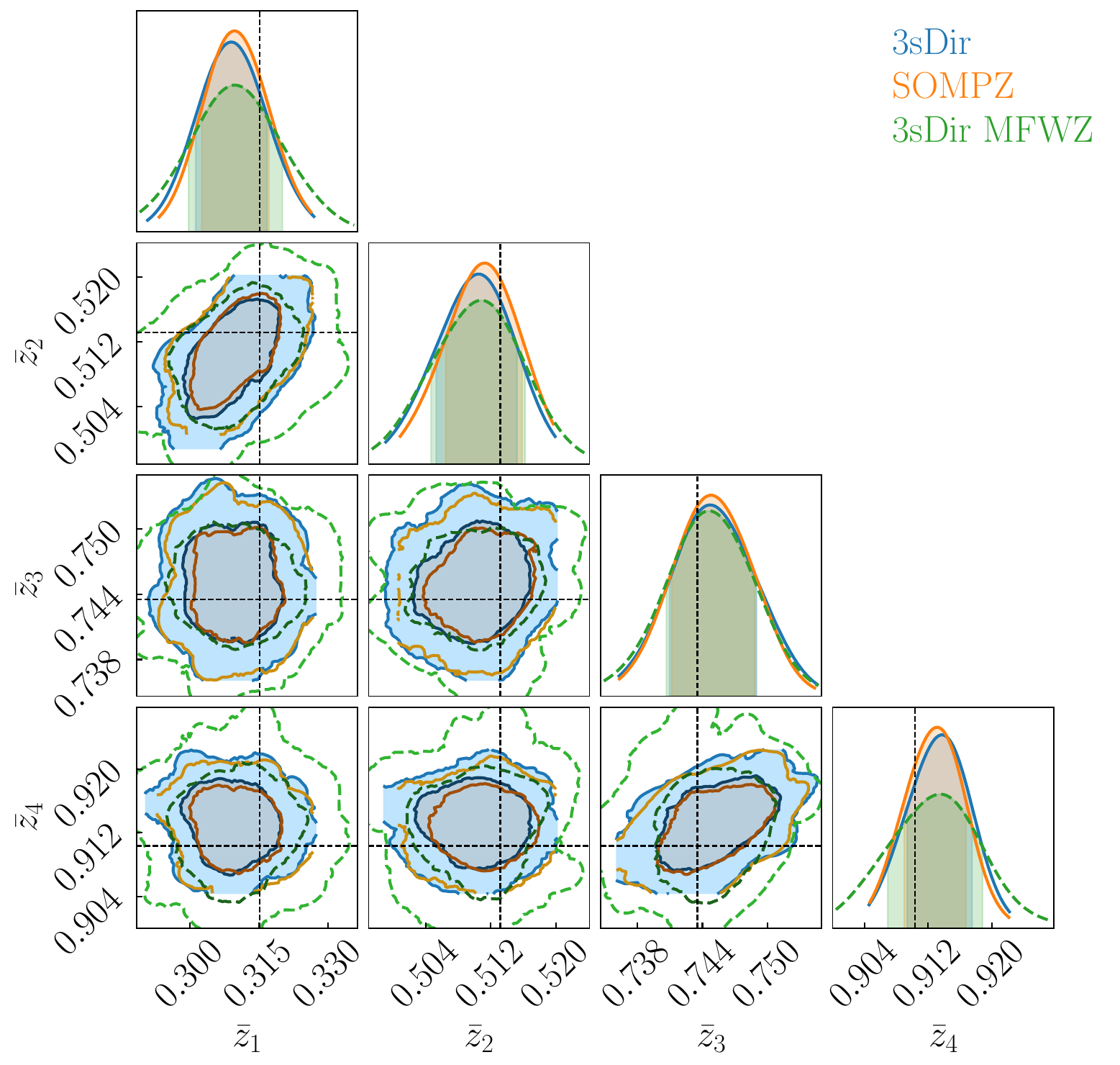}
\caption{Distribution of mean redshift, $\bar{z}$, values in each of the 300 realisations of the deep fields in \buzzard compared to the truth (cross-hatches). In each deep field realisation we run the \sompz code, obtain an $n(z)$ by fixing the probabilities to the number count measurements, and calculate the mean redshift of each tomographic bin. Similarly, we draw $10^{4}$ samples of $f_{zc}$ with the \sdir and \sdiralt method, compute the redshift distribution of each tomographic bin for each sample, their mean redshift, and we finally compute the average $\zbar$ value. The $\zbar$ distribution from \sdir is wider because it is not using all the colour information from the deep fields.}
\label{fig:validation_totruth}
\end{figure}

Figure~\ref{fig:validation_totruth} shows the distribution of the 300 values of $\zbar^{\sompz}$, $\zbar^{\sdir}$ and $\zbar^{\sdirmfwz}$ compared to the true $\zbar^{\text{true}}$ (shown as dotted lines). First, we find the $\zbar^{\sompz}$ distribution to be centred offset from the truth by $\Delta_z=[0.0051, 0.0024, -0.0013, -0.0024]$ in each bin, where $\Delta_z\equiv\langle \zbar^{\sompz} \rangle - \zbar^{\text{true}}$. As discussed in \S~\ref{sec:method_validation}, we expect a nonzero offset due to the bin conditionalization approximation, and we include this nonzero offset as an intrinsic systematic error to the mean redshift (see \S~\ref{sec:sompz_method_unc}). On the other hand, we find the averages over 300 realisations, $\zbaravg$ and $\langle \zbar^{\text{SOMPZ}} \rangle$, to be within 0.001 of each other in redshift in Figure~\ref{fig:validation_totruth}, meaning that \sdir is on average unbiased with respect to the \sompz mean redshift. We also find the width of both distributions to agree. However, we find the distribution of $\zbar^{\sdirmfwz}$ to have more scatter than the $\zbar^{\sompz}$ and $\zbar^{\sdir}$ distributions. This is a consequence of \sdirmfwz not fully exploiting the information available in the Deep Sample on the colour abundance $p(c)$, since we only use it to inform the superphenotype distribution $p(T)$  (\S~\ref{sec:3sdir_mfwz}). The \sdirmfwz likelihood is more suitable to be sampled efficiently together with the clustering redshifts likelihood using an HMC \citep*{y3-sourcewz}.

\begin{figure}
\centering
\includegraphics[width=\linewidth]{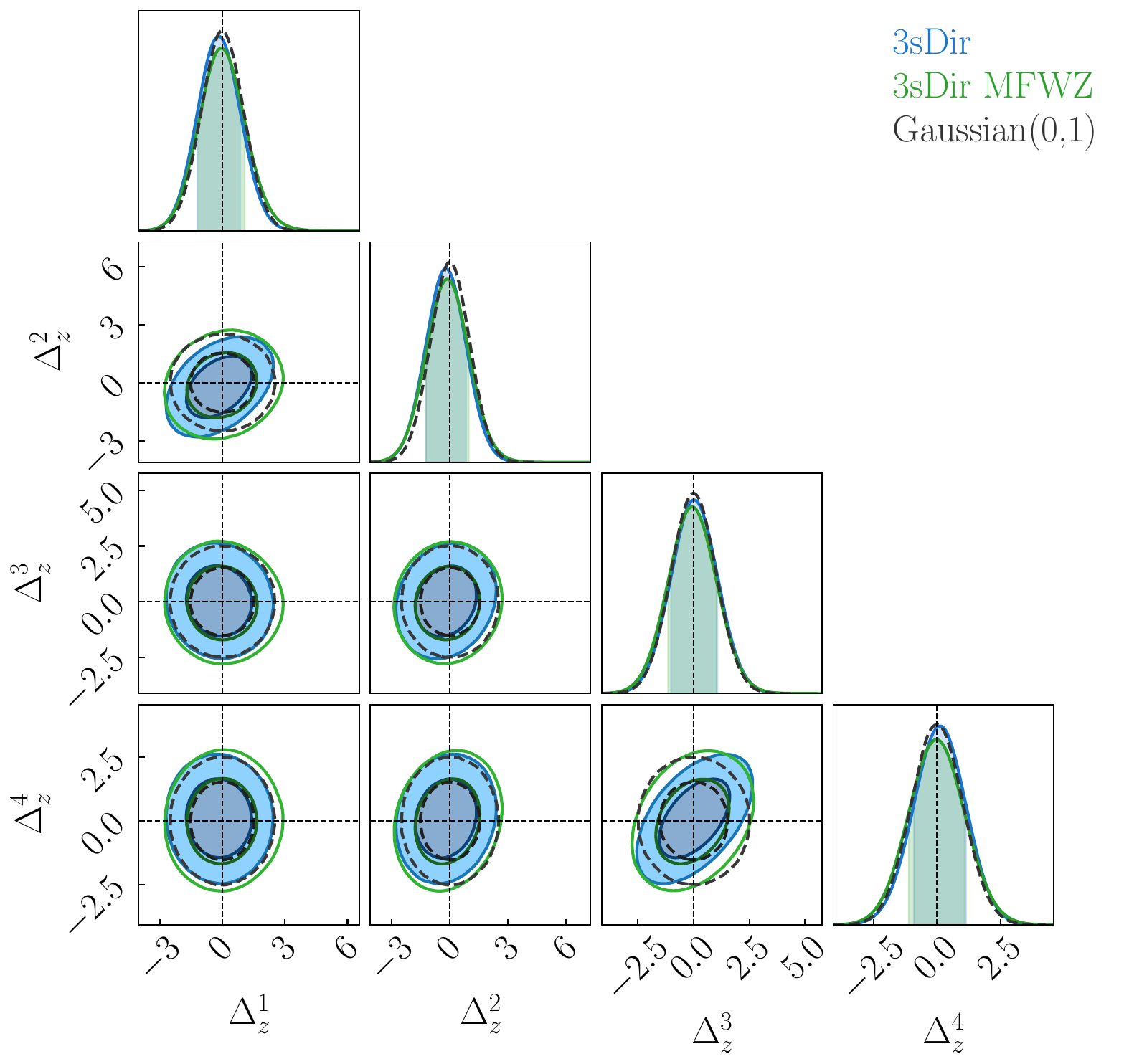}
\caption{Distribution of residual \sdir or \sdiralt samples across 300 realisations of the deep fields on \buzzard. A residual sample is defined as $\Delta^{i}_z = (\zbar^{i}-\zbar^{\text{SOMPZ}})/\sigma(\zbar^{i})$, with $\sigma(\zbar^{i})$ the standard deviation of the $\zbar^{i}$ values from \sdir in each realisation. The distributions agree with a Gaussian distribution with zero mean and unit variance (shown as dashed lines), which shows that the mean redshifts from \sdir and \sdiralt are statistically in agreement with $\zbar^{\text{SOMPZ}}$ across the 300 \buzzard realisations.}
\label{fig:3sdir_validation_2sompz}
\end{figure}

To test if the predicted distribution on $\zbar$ values from \sdir is consistent with $\zbar^{\sompz}$ we compute in each \buzzard realisation the pull distribution as $\Delta^{i}_z = (\zbar^{i}-\zbar^{\sompz})/\sigma(\zbar^{i})$, with $\sigma(\zbar^{i})$ the standard deviation of the $\zbar^{i}$ values from \sdir. Fig.~\ref{fig:3sdir_validation_2sompz} presents the stacked pull distributions from all 300 \buzzard realisations. We find this distribution to be centred at zero and very similar to a Gaussian distribution with zero mean and unit variance, illustrating more rigorously that \sdir predicts samples of $\zbar$ which are fully compatible with the \sompz mean redshift. We also do the same test with \sdirmfwz, finding the same conclusion.

\begin{figure}
\centering
\includegraphics[width=\linewidth]{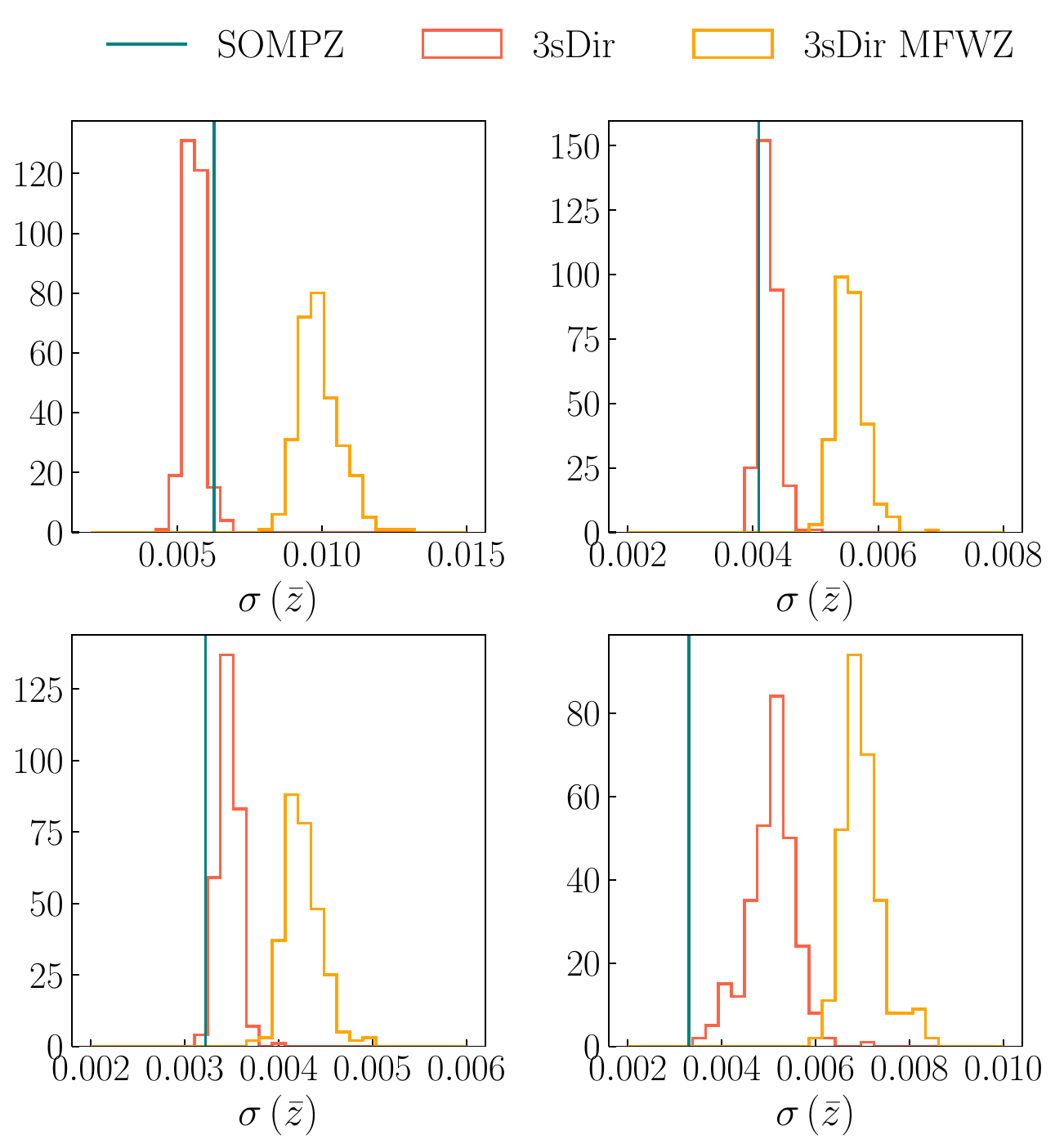}
\caption{Vertical line: Standard deviation of $\zbar^{\sompz}$ across the 300 realisations. Histograms: standard deviation of individual $\zbar$ values drawn with \sdir or \sdiralt in each of the 300 realisations.}
\label{fig:3sdir_std}
\end{figure}

Fig.~\ref{fig:3sdir_std} addresses the width predicted by \sdir or \sdirmfwz in each Buzzard realisation, compared to the scatter in $\zbar$ from \sompz across the 300 Buzzard realisations. The vertical line in each panel shows the spread of $\zbar^{\sompz}$ across the 300 \buzzard realisations (\textit{i.e.} the spread of \sompz in Figure~\ref{fig:validation_totruth}). While \sompz only produces one estimate of $\zbar$ in each realisation, the \sdir and \sdirmfwz models produce a distribution of $\zbar$ values in each \buzzard realisation. In each realisation, we compute the standard deviation of $\zbar$ for both \sdir and \sdirmfwz, and we show the histogram of these 300 values. As expected, the predicted $\sigma(\zbar)$ values from \sdirmfwz are $[78, 31, 23, 39]$ per cent larger than \sdir in each bin, since the former is using less information from the deep fields. We find the $\sigma(\zbar)$ from \sdir to be in reasonable agreement with \sompz, although we find them to be slightly underestimated at lower redshift and overestimated at higher redshift, finding $[-11, 4, 8, 53]$ per cent difference in each bin. This is in agreement with \citet{Sanchez2020} (see their fig.~12) which shows that \sdir tends to underpredict the variance at low redshift, and the opposite at high redshift.

\section*{Affiliations}
\label{sec:affiliations}

$^{1}$ Department of Physics, Stanford University, 382 Via Pueblo Mall, Stanford, CA 94305, USA\\
$^{2}$ Kavli Institute for Particle Astrophysics \& Cosmology, P. O. Box 2450, Stanford University, Stanford, CA 94305, USA\\
$^{3}$ SLAC National Accelerator Laboratory, Menlo Park, CA 94025, USA\\
$^{4}$ Argonne National Laboratory, 9700 South Cass Avenue, Lemont, IL 60439, USA\\
$^{5}$ Department of Physics and Astronomy, University of Pennsylvania, Philadelphia, PA 19104, USA\\
$^{6}$ Santa Cruz Institute for Particle Physics, Santa Cruz, CA 95064, USA\\
$^{7}$ Department of Astronomy, University of California, Berkeley,  501 Campbell Hall, Berkeley, CA 94720, USA\\
$^{8}$ Department of Physics, Duke University Durham, NC 27708, USA\\
$^{9}$ Department of Physics, Carnegie Mellon University, Pittsburgh, Pennsylvania 15312, USA\\
$^{10}$ Department of Applied Mathematics and Theoretical Physics, University of Cambridge, Cambridge CB3 0WA, UK\\
$^{11}$ Kavli Institute for Cosmological Physics, University of Chicago, Chicago, IL 60637, USA\\
$^{12}$ Institute of Cosmology and Gravitation, University of Portsmouth, Portsmouth, PO1 3FX, UK\\
$^{13}$ Center for Cosmology and Astro-Particle Physics, The Ohio State University, Columbus, OH 43210, USA\\
$^{14}$ Jodrell Bank Center for Astrophysics, School of Physics and Astronomy, University of Manchester, Oxford Road, Manchester, M13 9PL, UK\\
$^{15}$ Institut de F\'{\i}sica d'Altes Energies (IFAE), The Barcelona Institute of Science and Technology, Campus UAB, 08193 Bellaterra (Barcelona) Spain\\
$^{16}$ Laborat\'orio Interinstitucional de e-Astronomia - LIneA, Rua Gal. Jos\'e Cristino 77, Rio de Janeiro, RJ - 20921-400, Brazil\\
$^{17}$ Observat\'orio Nacional, Rua Gal. Jos\'e Cristino 77, Rio de Janeiro, RJ - 20921-400, Brazil\\
$^{18}$ Department of Astronomy, University of Illinois at Urbana-Champaign, 1002 W. Green Street, Urbana, IL 61801, USA\\
$^{19}$ National Center for Supercomputing Applications, 1205 West Clark St., Urbana, IL 61801, USA\\
$^{20}$ Department of Physics, University of Oxford, Denys Wilkinson Building, Keble Road, Oxford OX1 3RH, UK\\
$^{21}$ D\'{e}partement de Physique Th\'{e}orique and Center for Astroparticle Physics, Universit\'{e} de Gen\`{e}ve, 24 quai Ernest Ansermet, CH-1211 Geneva, Switzerland\\
$^{22}$ Jet Propulsion Laboratory, California Institute of Technology, 4800 Oak Grove Dr., Pasadena, CA 91109, USA\\
$^{23}$ Fermi National Accelerator Laboratory, P. O. Box 500, Batavia, IL 60510, USA\\
$^{24}$ Jet Propulsion Laboratory, California Institute of Technology, 4800 Oak Grove Drive, Pasadena, CA 91109, USA\\
$^{25}$ Instituci\'o Catalana de Recerca i Estudis Avan\c{c}ats, E-08010 Barcelona, Spain\\
$^{26}$ Department of Astronomy and Astrophysics, University of Chicago, Chicago, IL 60637, USA\\
$^{27}$ Centro de Investigaciones Energ\'eticas, Medioambientales y Tecnol\'ogicas (CIEMAT), Madrid, Spain\\
$^{28}$ Brookhaven National Laboratory, Bldg 510, Upton, NY 11973, USA\\
$^{29}$ Cerro Tololo Inter-American Observatory, NSF's National Optical-Infrared Astronomy Research Laboratory, Casilla 603, La Serena, Chile\\
$^{30}$ Departamento de F\'isica Matem\'atica, Instituto de F\'isica, Universidade de S\~ao Paulo, CP 66318, S\~ao Paulo, SP, 05314-970, Brazil\\
$^{31}$ CNRS, UMR 7095, Institut d'Astrophysique de Paris, F-75014, Paris, France\\
$^{32}$ Sorbonne Universit\'es, UPMC Univ Paris 06, UMR 7095, Institut d'Astrophysique de Paris, F-75014, Paris, France\\
$^{33}$ Department of Physics and Astronomy, Pevensey Building, University of Sussex, Brighton, BN1 9QH, UK\\
$^{34}$ Department of Physics \& Astronomy, University College London, Gower Street, London, WC1E 6BT, UK\\
$^{35}$ Instituto de Astrofisica de Canarias, E-38205 La Laguna, Tenerife, Spain\\
$^{36}$ Universidad de La Laguna, Dpto. Astrofísica, E-38206 La Laguna, Tenerife, Spain\\
$^{37}$ Institut d'Estudis Espacials de Catalunya (IEEC), 08034 Barcelona, Spain\\
$^{38}$ Institute of Space Sciences (ICE, CSIC),  Campus UAB, Carrer de Can Magrans, s/n,  08193 Barcelona, Spain\\
$^{39}$ University of Nottingham, School of Physics and Astronomy, Nottingham NG7 2RD, UK\\
$^{40}$ INAF-Osservatorio Astronomico di Trieste, via G. B. Tiepolo 11, I-34143 Trieste, Italy\\
$^{41}$ Institute for Fundamental Physics of the Universe, Via Beirut 2, 34014 Trieste, Italy\\
$^{42}$ Department of Physics, University of Michigan, Ann Arbor, MI 48109, USA\\
$^{43}$ Department of Physics, IIT Hyderabad, Kandi, Telangana 502285, India\\
$^{44}$ Department of Astronomy/Steward Observatory, University of Arizona, 933 North Cherry Avenue, Tucson, AZ 85721-0065, USA\\
$^{45}$ Department of Physics, The Ohio State University, Columbus, OH 43210, USA\\
$^{46}$ Department of Astronomy, University of Michigan, Ann Arbor, MI 48109, USA\\
$^{47}$ Institute of Theoretical Astrophysics, University of Oslo. P.O. Box 1029 Blindern, NO-0315 Oslo, Norway\\
$^{48}$ Instituto de Fisica Teorica UAM/CSIC, Universidad Autonoma de Madrid, 28049 Madrid, Spain\\
$^{49}$ Institute of Astronomy, University of Cambridge, Madingley Road, Cambridge CB3 0HA, UK\\
$^{50}$ Kavli Institute for Cosmology, University of Cambridge, Madingley Road, Cambridge CB3 0HA, UK\\
$^{51}$ School of Mathematics and Physics, University of Queensland,  Brisbane, QLD 4072, Australia\\
$^{52}$ Faculty of Physics, Ludwig-Maximilians-Universit\"at, Scheinerstr. 1, 81679 Munich, Germany\\
$^{53}$ Max Planck Institute for Extraterrestrial Physics, Giessenbachstrasse, 85748 Garching, Germany\\
$^{54}$ Universit\"ats-Sternwarte, Fakult\"at f\"ur Physik, Ludwig-Maximilians Universit\"at M\"unchen, Scheinerstr. 1, 81679 M\"unchen, Germany\\
$^{55}$ Center for Astrophysics $\vert$ Harvard \& Smithsonian, 60 Garden Street, Cambridge, MA 02138, USA\\
$^{56}$ Australian Astronomical Optics, Macquarie University, North Ryde, NSW 2113, Australia\\
$^{57}$ Lowell Observatory, 1400 Mars Hill Rd, Flagstaff, AZ 86001, USA\\
$^{58}$ George P. and Cynthia Woods Mitchell Institute for Fundamental Physics and Astronomy, and Department of Physics and Astronomy, Texas A\&M University, College Station, TX 77843,  USA\\
$^{59}$ Department of Astronomy, The Ohio State University, Columbus, OH 43210, USA\\
$^{60}$ Radcliffe Institute for Advanced Study, Harvard University, Cambridge, MA 02138\\
$^{61}$ Department of Astrophysical Sciences, Princeton University, Peyton Hall, Princeton, NJ 08544, USA\\
$^{62}$ Physics Department, 2320 Chamberlin Hall, University of Wisconsin-Madison, 1150 University Avenue Madison, WI  53706-1390\\
$^{63}$ School of Physics and Astronomy, University of Southampton,  Southampton, SO17 1BJ, UK\\
$^{64}$ Computer Science and Mathematics Division, Oak Ridge National Laboratory, Oak Ridge, TN 37831\\
$^{65}$Institute of Space Sciences (ICE, CSIC), Campus UAB, Carrer de Can Magrans, s/n, 08193 Barcelona, Spain \\
$^{66}$Institut d'Estudis Espacials de Catalunya (IEEC), E-08034 Barcelona, Spain \\


\bsp	
\label{lastpage}

\end{document}